\documentclass{aa}

\usepackage{graphicx}
\usepackage{txfonts}
\usepackage{color}
\usepackage{stfloats}
\usepackage{amsmath}
\usepackage{multirow}

\usepackage{epsfig}
\usepackage{amsmath}
\usepackage{subfigure}
\usepackage{natbib}

\usepackage{longtable}
\usepackage{graphicx}
\usepackage{epstopdf}
\usepackage{booktabs}
\usepackage{txfonts}
\usepackage[utf8]{inputenc}%
\usepackage{graphicx}
\usepackage{txfonts}
\usepackage{stfloats}

\newcommand{\tablenotea}[1]{\parbox{17.7cm}{\indent \footnotesize{#1}}}
\newcommand{\tablenoteb}[1]{\parbox{8.9cm}{\indent \footnotesize{#1}}}
\newcommand{\tablenotec}[1]{\parbox{17.0cm}{\indent \footnotesize{#1}}}
\newcommand{\tablenoted}[1]{\parbox{15.7cm}{\indent \footnotesize{#1}}}
\newcommand{\tablenotee}[1]{\parbox{17.5cm}{\indent \footnotesize{#1}}}

\newcommand{\cp}{Chem. Phys.}

\newcommand{\fdis}{Faraday Discuss.}

\newcommand{\jms}{J. Mol. Spectr.}
\newcommand{\jmst}{J. Mol. Struct.}

\newcommand{\jpb}{J. Phys. B At. Mol. Phys.}

\newcommand{\pccp}{PCCP}

\begin{document}


\title{Multi-line study of the radial extent of SiO, CS, and SiS in AGB envelopes\thanks{Based on observations carried out with the IRAM 30m and Yebes 40m telescopes. IRAM is supported by INSU/CNRS (France), MPG (Germany) and IGN (Spain). The Yebes 40m telescope at Yebes Observatory is operated by the Spanish Geographic Institute (IGN, Ministerio de Transportes, Movilidad y Agenda Urbana).}}

\titlerunning{Radial extent of SiO, CS, and SiS in AGB envelopes}
\authorrunning{Massalkhi et al.}

\author{S.~Massalkhi\inst{1,2}, M.~Ag\'undez\inst{1}, J.~P.~Fonfr\'ia\inst{3}, J.~R.~Pardo\inst{1}, L.~Velilla-Prieto\inst{1}, \and J.~Cernicharo\inst{1}}

\institute{
Instituto de F\'isica Fundamental, CSIC, C/ Serrano 123, 28006 Madrid, Spain\\
\email{sarah.massalkhi@csic.es, marcelino.agundez@csic.es} \and
Centro de Astrobiolog\'ia (CSIC/INTA), Ctra. de Ajalvir km. 4, Torrej\'on de Ardoz, 28850 Madrid, Spain \and
Observatorio Astron\'omico Nacional (IGN), Alfonso XII 3, 28014 Madrid, Spain
}

\date{Received; accepted}


\abstract
{Circumstellar envelopes around AGB stars contain a rich diversity of molecules whose spatial distribution is regulated by different chemical and physical processes. In the outer circumstellar layers all molecules are efficiently destroyed due to the interaction with interstellar ultraviolet photons. Here we aim to characterize in a coherent and uniform way the radial extent of three molecules (SiO, CS, and SiS) in envelopes around AGB stars of O- and C-rich character, and to study their dependence with mass loss rate. To that purpose, we observed a reduced sample of seven M-type and seven C-type AGB envelopes in multiple lines of SiO, CS, and SiS with the Yebes\,40m and IRAM\,30m telescopes. The selected sources cover a wide range of mass loss rates, from $\sim$\,10$^{-7}$ $M_{\odot}$ yr$^{-1}$ to a few 10$^{-5}$ $M_{\odot}$ yr$^{-1}$, and the observed lines cover a wide range of upper level energies, from 2 K to 130 K. We carried out excitation and radiative transfer calculations over a wide parameter space to characterize the molecular abundance and radial extent. A $\chi^2$ analysis indicates that the abundance is usually well constrained while the radial extent is in some cases more difficult to constrain. Our results indicate that the radial extent increases with increasing envelope density, in agreement with previous observational findings. At high envelope densities, $\dot{M}/v_{\infty}$\,$>$\,10$^{-6}$ $M_{\odot}$ yr$^{-1}$ km$^{-1}$ s, the radial extent of SiO, CS, and SiS are similar, while at low envelope densities, $\dot{M}/v_{\infty}$\,$<$\,10$^{-7}$ $M_{\odot}$ yr$^{-1}$ km$^{-1}$ s, the radial extent differ among the three molecules, in agreement with theoretical expectations based on destruction due to photodissociation. At low envelope densities we find a sequence of increasing radial extent, SiS\,$\rightarrow$\,CS\,$\rightarrow$\,SiO. We also find a tentative dependence of the radial extent with the chemical type (O- or C-rich) of the star for SiO and CS. Interferometric observations and further investigation of the photodissociation of SiO, CS, and SiS should allow to clarify the situation on the relative photodissociation radius of SiO, CS, and SiS in AGB envelopes and the dependence with envelope density and C/O ratio.}


\maketitle

\section{Introduction}

The circumstellar envelopes (CSEs) of Asymptotic Giant Branch (AGB) stars are made of molecules and dust. Some molecules are formed in the warm and dense inner regions, where abundances are thought to be largely controlled by chemical equilibrium \citep{Tsuji1973,Agundez2020}, and are then injected into the expanding envelope. These molecules, some of which contain refractory elements, are simple and stable species, such as CO, H$_2$O, HCN, C$_2$H$_2$, H$_2$S, CS, SiO, SiS, SiC$_2$, NaCl, and AlCl. Other species such as radicals and, in the case of C-rich objects, long carbon chains are formed in the outer layers under the action of photochemistry, which is driven by the penetration of interstellar ultraviolet (UV) photons.

Molecular abundances experience variations as the gas travels away from the star due to interactions with dust grains, gas-phase chemistry, and photodissociation, which eventually destroys all molecules in the tenuous and translucent outer layers. Molecular hydrogen and CO are the two molecules that survive out to larger distances because their high abundances and the fact that they are photodissociated in lines make them to self-shield against interstellar UV photons \citep{Morris1983}. The rest of molecules extend out to shorter distances from the star.

To have a good understanding of circumstellar chemistry it is important to constrain the radial extent of the different molecules. The spatial distribution of CO has been constrained observationally, by mapping low-energy rotational lines \citep{Castro-Carrizo2010,Ramstedt2020}, and theoretically, by modeling its photodissociation \citep{Groenewegen2017,Saberi2019,Groenewegen2021}. For other molecules, the radial extent has been also constrained observationally by modeling multiple rotational lines, as in the cases of SiO \citep{Gonzalez-Delgado2003,Schoier2006a}, SiS and CS \citep{Schoier2007,Danilovich2018}, and HCN \citep{Schoier2013}. For some of these molecules there are also constraints on their abundance distribution in selected objects from large scale mapping of selected rotational lines \citep{Velilla-Prieto2019,Danilovich2019}.

To determine accurate molecular abundances in AGB envelopes from single-dish observations, it is of paramount importance to know how far each molecule extends, especially if only a few lines are available. This was the case of our previous studies in which we determined the abundance of the molecules SiC$_2$, SiO, CS, and SiS in a large sample of AGB envelopes using a few rotational lines \citep{Massalkhi2018,Massalkhi2019,Massalkhi2020}. Here, we aim at using multiple rotational lines to carry out a systematic and coherent determination of the radial extent of SiO, CS, and SiS in a reduced sample of O-rich and C-rich AGB envelopes spanning a wide range of mass loss rates. This valuable information can be used to derive accurate abundances for these three molecules in circumstellar envelopes around other AGB stars. In addition, our comparative study allows to shed light on the different behavior of each molecule against photodissociation. We present the observations carried out in Sect.\,\ref{sec:observations}, describe the adopted model in Sect.\,\ref{sec:model}, discuss the obtained results in Sect.\,\ref{sec:results}, and conclude in Sect.\,\ref{sec:conclusions}.

\section{Observations} \label{sec:observations}

\begin{table*}
\small
\caption{Sample of AGB envelopes and associated parameters.}
\label{table:stars}
\centering
\begin{tabular}{lccccccccccc}
\hline \hline
\multicolumn{1}{l}{Source} & \multicolumn{1}{c}{$V_{sys}$\,$^a$} & \multicolumn{1}{c}{$D$\,$^b$} & \multicolumn{1}{c}{$L_\star$} & \multicolumn{1}{c}{$T_\star$} & \multicolumn{1}{c}{$\tau_{10}$} & \multicolumn{1}{c}{$T_d(r_c)$} & \multicolumn{1}{c}{$r_c$} & \multicolumn{1}{c}{$\dot{M}$\,$^c$} & \multicolumn{1}{c}{$\delta$} & \multicolumn{1}{c}{$v_{\infty}$\,$^a$} & \multicolumn{1}{c}{$\Psi$} \\
& \multicolumn{1}{c}{(km s$^{-1}$)} & \multicolumn{1}{c}{(pc)} & \multicolumn{1}{c}{($L_{\odot}$)} & \multicolumn{1}{c}{(K)} & & \multicolumn{1}{c}{(K)} & \multicolumn{1}{c}{(cm)} & \multicolumn{1}{c}{($M_{\odot}$ yr$^{-1}$)} & & \multicolumn{1}{c}{(km s$^{-1}$)} \\
\hline
\\
\multicolumn{12}{c}{O-rich} \\
\hline
R\,Leo      & $+$0.1  &   111\,$\pm$\,17\,$^d$ &  7000 & 2803\,$^i$ & 0.032 & 1000 & 1.8\,$\times$\,10$^{14}$ & 1.3\,$\times$\,10$^{-7}$ & 0.30 & 5   &  420 \\
R\,Cas      & $+$26.5 &  174\,$\pm$\,6 & 6800 & 3129\,$^j$ & 0.20 & 1400 & 9.3\,$\times$\,10$^{13}$ & 9.5\,$\times$\,10$^{-7}$ & 0.60 & 7.5 &   640 \\
TX\,Cam     & $+$11.5 &  287\,$\pm$\,14 &  4600 & 2778\,$^k$ & 0.63 & 1300 & 9.3\,$\times$\,10$^{13}$ & 3.9\,$\times$\,10$^{-6}$ & 0.50 & 17.5 &  360 \\
IK\,Tau     & $+$34.5 &  261\,$\pm$\,19 &  7900 & 2234\,$^l$ & 1.0 & 1000 & 2.0\,$\times$\,10$^{14}$ & 4.8\,$\times$\,10$^{-6}$ & 0.65 & 17.5 &  130 \\
V1111\,Oph  & $-$31   &  502\,$\pm$\,31 &  5000 & 2667\,$^k$ & 1.3 &  1100 & 1.5\,$\times$\,10$^{14}$ & 8.5\,$\times$\,10$^{-6}$ & 0.70 & 15.5 &  260 \\
GX\,Mon     & $-$9.5  &  569\,$\pm$\,53 &  6300 & 2173\,$^m$ & 1.0 &  900 & 2.2\,$\times$\,10$^{14}$ & 1.6\,$\times$\,10$^{-5}$ & 0.70 & 18   &  380 \\
NV\,Aur     & $+$3    & 1360 &  10700\,$^g$ & 2500\,$^j$ & 2.5 & 1000 & 2.8\,$\times$\,10$^{14}$ & 2.0\,$\times$\,10$^{-5}$ & 0.65 & 17.5 & 150 \\
\\
\multicolumn{12}{c}{C-rich} \\
\hline
Y\,CVn      & $+$22   &  310\,$\pm$\,17 &  8500 & 2760\,$^n$ & 0.0063 & 1000 & 1.9\,$\times$\,10$^{14}$ & 2.1\,$\times$\,10$^{-7}$ & 0.50 &   7   &  1360 \\
R\,Lep      & $+$11.5 &  456\,$\pm$\,13 &  7800 & 2290\,$^n$ & 0.079 & 1300 & 8.9\,$\times$\,10$^{13}$ & 2.1\,$\times$\,10$^{-6}$ & 0.60 & 17.5 &  900 \\
LP\,And     & $-$17   &  526\,$\pm$\,131\,$^e$ &  3700 & 2040\,$^n$ & 0.79 & 1200 & 9.9\,$\times$\,10$^{13}$ & 5.3\,$\times$\,10$^{-6}$ & 0.70 & 14.5 &  250 \\
IRC\,+20370 & $-$0.8  &  753\,$\pm$\,113 &  11300 & 2200\,$^j$ & 0.40 & 1300 & 1.3\,$\times$\,10$^{14}$ & 5.8\,$\times$\,10$^{-6}$ & 0.65 & 14   &  420 \\
IRC\,+30374 & $-$12.5 & 823\,$\pm$\,51 &  6300 & 2500\,$^o$ & 0.79 & 1500 & 8.9\,$\times$\,10$^{13}$ & 9.9\,$\times$\,10$^{-6}$ & 0.75 & 25   & 300 \\
IRC\,+10216 & $-$26.5  &  123\,$\pm$\,14\,$^f$ &  7800 & 2330\,$^p$ & 0.79 & 900 & 2.8\,$\times$\,10$^{14}$ & 2.4\,$\times$\,10$^{-5}$\,$^q$ & 0.95\,$^q$ &   14.5   &  390 \\
CRL\,190    & $-$39.5 & 3370 & 16700\,$^h$ & 2500\,$^o$ & 5.0 & 1400 & 3.1\,$\times$\,10$^{14}$ & 3.9\,$\times$\,10$^{-5}$ & 1.05 & 17   &  80 \\
\hline
\end{tabular}
\tablenotea{Unless otherwise stated, the parameters $L_\star$, $\tau_{10}$, $T_d(r_c)$, $r_c$, $\dot{M}$, $\delta$, and $\Psi$ are determined in this study.\\
$^a$\,From \cite{Massalkhi2020} for O-rich stars and \cite{Massalkhi2018} for C-rich stars. $^b$\,From Gaia DR3 \citep{Gaia2023} unless otherwise stated. Uncertainties in the distances from Gaia DR3 are likely underestimated \citep{Andriantsaralaza2022}. The distances of NV\,Aur and CRL\,190 are derived in this study from the luminosity and a fit to the SED. $^c$\,The estimated uncertainty in the mass loss rate is 20\,\% (see Sect.\,\ref{sec:model_co}). $^d$\,From combined parallax measured by Allegheny Observatory and Hipparcos \citep{Whitelock2008}. $^e$\,Value from Nearby Evolved Stars Survey (\texttt{https://evolvedstars.space}; \citealt{Scicluna2022}) is preferred over Gaia DR3 value (413\,$\pm$\,41) because the resulting luminosity agrees better with the period-luminosity relation \citep{Groenewegen1996}. $^f$\,\cite{Groenewegen2012}. $^g$\,From period-luminosity relation \citep{Olivier2001}. $^h$\,From period-luminosity relation \citep{Groenewegen2002}. $^i$\,\cite{Perrin1999}. $^j$\,\cite{DeBeck2010}. $^k$\,\cite{vanLoon2005}. $^l$\,\cite{Adam2019}. $^m$\,\cite{Karovicova2013}. $^n$\,\cite{Bergeat2001}. $^o$\,Rough estimate by \cite{Groenewegen1998}. $^p$\,\cite{Ridgway1988}. $^q$\,Mass loss rate for IRC\,+10216 fixed to 2.4\,$\times$\,10$^{-5}$ $M_{\odot}$ yr$^{-1}$ \citep{Fonfria2022}. The gas temperature radial profile adopted consists of 3 components with $\delta$\,=\,1.00 between 10$^{15}$ and 10$^{16}$ cm, $\delta$\,=\,0.45 inward of 10$^{15}$ cm and $\delta$\,=\,1.55 beyond 10$^{16}$ cm.}
\end{table*}

We selected a sample of 14 AGB envelopes, half of them of M-type (oxygen-rich) from the sample of \cite{Massalkhi2020} and the other half of C-type (carbon-rich) from \cite{Massalkhi2019}. The sources, which are listed in Table\,\ref{table:stars}, were chosen to cover a wide range of mass loss rates, from 10$^{-7}$ $M_\odot$ yr$^{-1}$ to a few 10$^{-5}$ $M_\odot$ yr$^{-1}$. We used the IRAM\,30m telescope in several sessions from March 2019 to June 2020 and the Yebes\,40m telescope from December 2019 to September 2020. In the case of IRC\,+10216, the Yebes\,40m observations correspond to the high sensitivity spectral survey described in \cite{Pardo2022}.

The IRAM\,30m observations consisted of specific frequency setups across the different available bands. The covered frequency ranges were 81.3-89.1 GHz and 96.9-104.7 GHz (3 mm band), 237.8-245.6 GHz and 253.4-261.2 GHz (1.3 mm band), 287.1-294.9 GHz and 302.7-310.5 GHz (0.9 mm band). We used the E090, E230, and E330 receivers in dual side band, with image rejections $>$\,10 dB, connected to a FFTS providing a spectral resolution of 195 kHz, which for our target lines corresponds to velocity resolutions in the range 0.2-0.7 km s$^{-1}$. We used the wobbler-switching observing mode, with a throw of 180$''$ in azimuth. The focus of the telescope was regularly checked on a planet. The half power beam width (HPBW) of the 30m telescope ranges between 28$''$ at 87 GHz and 8$''$ at 309 GHz. The pointing was systematically checked every 1 h on a nearby quasar, with errors within 2-3$''$. The intensity scale at the 30m telescope is the antenna temperature corrected for atmospheric absorption and for antenna ohmic and spillover losses, $T_A^*$, for which we estimate a calibration error of 10\,\% at 3 mm, 20\,\% at 2 mm, and 30\,\% in the 1.3 and 0.9 mm bands. The $T_A^*$ scale is converted to $T_{mb}$ (main beam brightness temperature) by dividing by $B_{\rm eff}$/$F_{\rm eff}$, where $B_{\rm eff}$\,=\,0.871\,$\exp$[$-$($\nu$(GHz)/359)$^2$] and $F_{\rm eff}$ takes values of 0.95, 0.93, 0.92, and 0.82 at 3, 2, 1.3, and 0.9 mm, respectively.

The Yebes\,40m observations consisted in a full scan of the Q band, from 31 to 50 GHz. We used a 7\,mm receiver connected to a fast Fourier transform spectrometer (FFTS), which provides an instantaneous coverage of the whole Q band in horizontal and vertical polarizations with a spectral resolution of 38 kHz (see \citealt{Tercero2021}). Spectra were later on smoothed to a spectral resolution of 191 kHz, which corresponds to velocity resolutions in the range 1.1-1.8 km s$^{-1}$ across the Q band. The observations were carried out in the position-switching mode, with the off position shifted by 300$''$ in azimuth with respect to the source position. The HPBW ranges between 35$''$ at 50 GHz and 57$''$ at 31 GHz. Pointing corrections were obtained by observing SiO masers and quasars and were always within 2-3$''$. The intensity scale at the 40m telescope is $T_A^*$, for which we estimate a calibration error of 10\,\%. We converted $T_A^*$ to $T_{mb}$ by dividing by $B_{\rm eff}$/$F_{\rm eff}$, where $B_{\rm eff}$\,=\,0.797\,$\exp$[$-$($\nu$(GHz)/71.1)$^2$] and $F_{\rm eff}$\,=\,0.97.

\begin{table}
\small
\caption{Lines observed in this study.}
\label{table:lines}
\centering
\begin{tabular}{lcrr}
\hline \hline
\multicolumn{1}{l}{Molecule} & \multicolumn{1}{c}{Line} & \multicolumn{1}{c}{Frequency (MHz)} & \multicolumn{1}{c}{$E_{up}$ (K)} \\
\hline
SiO & $J$\,=\,1-0   &  43423.844 &   2.1 \\
    & $J$\,=\,2-1   &  86846.971 &   6.3 \\
    & $J$\,=\,6-5   & 260517.985 &  43.8 \\
    & $J$\,=\,7-6   & 303926.783 &  58.3 \\
CS  & $J$\,=\,1-0   &  48990.957 &   2.4 \\
    & $J$\,=\,2-1   &  97980.952 &   7.1 \\
    & $J$\,=\,5-4   & 244935.554 &  35.3 \\
    & $J$\,=\,6-5   & 293912.089 &  49.4 \\
SiS & $J$\,=\,2-1   &  36309.629 &   2.6 \\
    & $J$\,=\,14-13 & 254103.211 &  91.5 \\
    & $J$\,=\,16-15 & 290380.744 & 118.5 \\
    & $J$\,=\,17-16 & 308516.144 & 133.3 \\
\hline
\end{tabular}
\tablenoteb{See Sect.\,\ref{sec:model_molecules} for references on the spectroscopic data.}
\label{sec:model_molecules}
\end{table}

The Yebes\,40m and IRAM\,30m observations allowed us to cover multiple rotational lines of SiO, CS, and SiS with a wide range of upper level energies (see Table\,\ref{table:lines}). The observed lines are shown in Fig.\,\ref{fig:lines}. The line profiles were fitted using the SHELL method implemented in the program CLASS, within the GILDAS software \citep{Pety2005}\footnote{https://www.iram.fr/IRAMFR/GILDAS/}. The line parameters are given in Table\,\ref{table:line_param}. Some of the targeted lines were not detected in all sources. In some cases the line was detected but the intensity was suspected to be erroneous very likely due to problems in the calibration or the pointing, especially for those lines lying at high frequencies. The non-detections and the problematic observations have been omitted in Fig.\,\ref{fig:lines} and Table\,\ref{table:line_param}. The sources IK\,Tau and IRC\,+10216 were not observed with the IRAM\,30m telescope because there are extensive data available from \cite{Velilla-Prieto2017} and \cite{Agundez2012}. We did not observe the 2 mm band because we have IRAM\,30m data from previous studies \citep{Massalkhi2018,Massalkhi2019,Massalkhi2020}.

We also used the IRAM\,30m telescope in May and September 2022 to observe the CO $J$\,=\,1-0 and $J$\,=\,2-1 lines toward CRL\,190 and IRC\,+30374. These data are needed to model the gaseous envelope, as described in Sect.\,\ref{sec:model_co}, but are not available in the literature. The velocity-integrated line intensities are given in Table\,\ref{table:co}.

\section{Envelope model} \label{sec:model}

Our main objective is to constrain the radial extent of SiO, CS, and SiS in the observed sources. We thus need to build a model of each of the sources and perform excitation and radiative transfer calculations to produce synthetic lines that can be compared to the observed lines. To describe the observed objects we consider an idealized model consisting of a spherical envelope of gas and dust expanding around a central AGB star, which is characterized by a luminosity, $L_\star$, and an effective temperature, $T_\star$. Dust is assumed to be present only beyond the condensation radius, $r_c$, with a constant gas-to-dust mass ratio, $\Psi$. The envelope is characterized by a constant mass loss rate, $\dot{M}$, and a radial expansion velocity, $v_{exp}$, which is assumed to be uniform and equal to $v_0$ between the star surface and the dust condensation radius, while outside $r_c$ it can be expressed as function of radius, $r$, as \citep{Hofner2018}
\begin{equation}
v_{exp}(r) = v_0 + (v_{\infty}-v_0)\bigg(1-\frac{r_c}{r} \bigg)^\beta, \label{eq:vexp}
\end{equation}
where $v_{\infty}$ is the terminal expansion velocity and we assume $v_0$\,=\,$v{_\infty}$/4 and $\beta$\,=\,1 based on observational constraints \citep{Decin2010}. The volume density of gas particles, $n_g$, is given by the law of conservation of mass as
\begin{equation}
n_g(r) = \frac{\dot{M}}{\overline{m}_g\,4 \pi r^2\,v_{exp}}, \label{eq:ng}
\end{equation}
where $\overline{m}_g$ is the average mass of gas particles, assumed to be 2.3 amu after considering H$_2$, He, and CO. Since we are mainly interested in the outer layers, for simplicity we assume $v_{exp}$ to be equal to $v_{\infty}$ in Eq.\,(\ref{eq:ng}). The gas kinetic temperature, $T_g$, is assumed to be given by the expression
\begin{equation}
T_g(r) = T_\star \bigg(\frac{r}{R_\star}\bigg)^{-\delta}, \label{eq:tg}
\end{equation}
where the exponent $\delta$ is determined from a fit to multiple CO lines (see Sect.\,\ref{sec:model_co}) and $T_g$ is not allowed to decrease below 10 K. Although Eq.\,(\ref{eq:tg}) does not account for the underlying processes of heating and cooling of the gas, it provides a reasonable first order approximation to the temperature radial profile (e.g., \citealt{DeBeck2010}). The stellar radius, $R_\star$, is set by the luminosity and effective temperature as
\begin{equation}
R_\star = \sqrt{\frac{L_\star}{4 \pi \sigma_{SB} T_\star^4}}, \label{eq:rstar}
\end{equation}
where $\sigma_{SB}$ is the Stefan-Boltzmann constant. The dust temperature is calculated as a function of radius from a fit to the spectral energy distribution (SED) of the envelope (see Sect.\,\ref{sec:model_sed}).

The parameters needed to describe the envelope model for each of the sources are given in Table\,\ref{table:stars}. The distances, $D$, were mainly taken from the Gaia Data Release 3 (DR3; \citealt{Gaia2023}). We note that the adopted distances are within 20\,\% of the values recommended by
\cite{Andriantsaralaza2022}, except for GX\,Mon, for which these authors recommend a much larger distance of 1430 pc based on the period-luminosity relation. The systemic velocities, $V_{sys}$, in the Local Standard of Rest (LSR) frame and the terminal expansion velocities, $v_{\infty}$, were determined from various intense lines in \cite{Massalkhi2018} for the C-rich stars and in \cite{Massalkhi2020} for the O-rich stars. The stellar effective temperatures, $T_\star$, were taken from different literature sources. The remaining parameters in Table\,\ref{table:stars} were determined in this study. On the one hand, the stellar luminosity, $L_\star$, the optical depth at 10 $\mu$m, $\tau_{10}$, the dust condensation radius, $r_c$, the dust temperature at the dust condensation radius, $T_d(r_c)$, and the gas-to-dust mass ratio, $\Psi$, are determined from a fit to the SED (see details in Sec.\,\ref{sec:model_sed}). On the other hand, the mass loss rate, $\dot{M}$, and the exponent $\delta$ in the gas kinetic temperature radial profile in Eq.\,(\ref{eq:tg}), are determined from a fit to multiple CO lines, as described in Sec.\,\ref{sec:model_co}.

The fitting strategy used to reproduce the SED and the CO lines (and also later on the lines of SiO, CS, and SiS) is based on the minimization of the parameter $\chi^2$, which is defined as
\begin{equation}
\chi^2 = \sum_{i=1}^{N} \Bigg[ \frac{(I_{calc} - I_{obs})}{\sigma} \Bigg]^2, \label{eq:chi2}
\end{equation}
where the sum extends over $N$ independent observations, $I_{calc}$ and $I_{obs}$ are the calculated and observed intensities, and $\sigma$ are the uncertainties in $I_{obs}$. We run many models (typically between 100 and 1000) varying a number, $p$, of free input parameters and adopt as best fit model the one that results in the minimum value of $\chi^2$, $\chi^2_{min}$. To evaluate the goodness of the fit we use the reduced $\chi^2$, defined as
\begin{equation}
\chi^2_{red} = \frac{\chi^2_{min}}{(N-p)}. 
\end{equation}
Typically, a value of $\chi^2_{red}$\,$\lesssim$\,1 indicates a good quality of the fit. The methodology provide best fit values for the $p$ adjustable parameters and uncertainties for them, given as the standard deviation $\sigma$. For $p$\,=\,1 the 1\,$\sigma$ level (68\,\% confidence) is given by $\chi^2$\,+\,1.00, while for $p$\,=\,2 (which it is usually the case here) the 1\,$\sigma$ level is given by $\chi^2$\,+\,2.30.

\subsection{Model of the dusty envelope: fit to SED} \label{sec:model_sed}

In order to calculate the SED of each envelope we used the code DUSTY\,V2 \citep{Ivezic1997}\footnote{\texttt{http://faculty.washington.edu/ivezic/dusty\_web/}}, which solves the continuum radiative transfer including light scattering in a spherically symmetric envelope, where dust is present from an inner condensation radius, $r_c$. For simplicity, we assume that grains are spherical with a single size (radius of 0.1 $\mu$m) and adopt the optical constants of warm silicate \citep{Suh1999} for O-rich stars and of amorphous carbon \citep{Suh2000} for C-rich stars.

The adjustable parameters in the $\chi^2$ analysis are the dust temperature at the condensation radius, $T_d(r_c)$, and the dust optical depth at a reference wavelength of 10\,$\mu$m, $\tau_{10}$. The approach used here is similar to that employed in previous studies by F. L. Sch\"oier and colleagues (e.g., \citealt{Schoier2002,Schoier2006a}). The observed SED consist of photometric fluxes measured mostly by space telescopes at infrared wavelengths, which are obtained from the VizieR database \citep{Ochsenbein2000}\footnote{\texttt{http://vizier.cds.unistra.fr/vizier/sed/}}. The fluxes collected for the 14 studied stars are given in Table\,\ref{table:sed}. Since the observed fluxes at different wavelengths differ by orders of magnitude we use the decimal logarithm of the observed flux as $I_{obs}$ in Eq.\,(\ref{eq:chi2}). The uncertainties of the observed fluxes are not determined in a consistent way in the different catalogs and thus we adopt a uniform uncertainty of 20\,\% for all observed fluxes.

\begin{figure*}
\centering
\includegraphics[angle=0,width=0.99\textwidth]{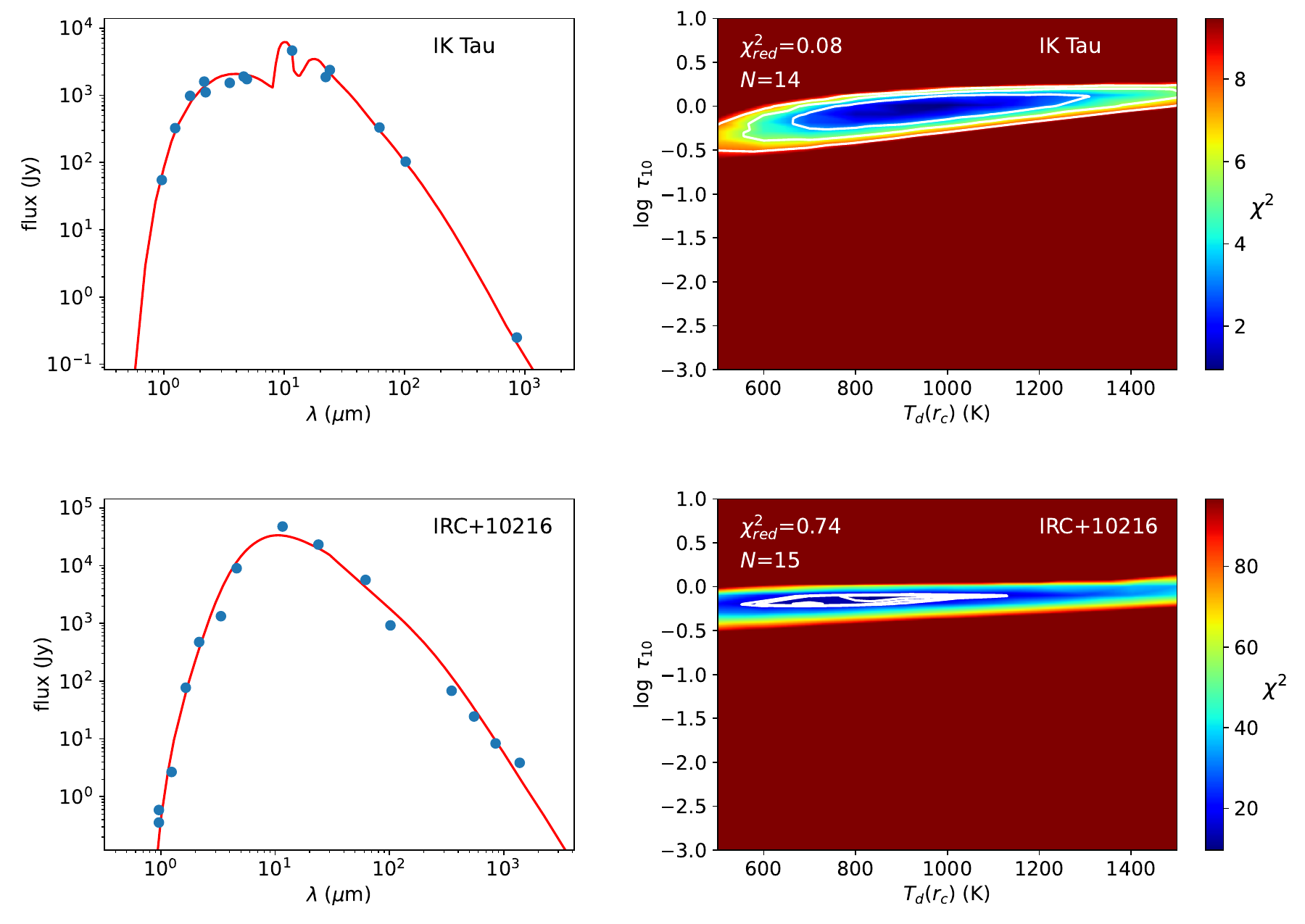}
\caption{Results from SED analysis for IK\,Tau and IRC\,+10216. The left panels show the observed fluxes in blue (see Table\,\ref{table:sed}) and the calculated SED from the best DUSTY model in red. The right panels show $\chi^2$ as a function of the dust temperature at the condensation radius, $T_d(r_c)$, and the logarithm of the dust optical depth at 10 $\mu$m, $\log \tau_{10}$. The white contours correspond to 1, 2, and 3\,$\sigma$ levels. Similar plots for other stars are shown in Fig.\,\ref{fig:sed}.} \label{fig:sed_iktau_irc10216}
\end{figure*}

In Fig.\,\ref{fig:sed_iktau_irc10216} we show the results of the SED analysis for one O-rich envelope, IK\,Tau, and one C-rich envelope, IRC\,+10216. The left panels show the SED of the best fit model overimposed on the observed fluxes, while the right panels show the $\chi^2$ parameter as a function of the two adjustable parameters. The results of the SED analysis for the remaining 12 envelopes are shown in Fig.\,\ref{fig:sed}. The observed SED are reasonably well fitted for all sources, with values of $\chi^2_{red}$ systematically below one. The calculated SED of the O-rich sources show a prominent emission band around 10 $\mu$m due to silicate, while in the case of the C-rich sources the amorphous carbon provides a smooth continuum in the calculated SED. We note that while the optical depth at 10 $\mu$m, $\tau_{10}$, is constrained to relatively narrow ranges, the dust temperature at the condensation radius, $T_d(r_c)$, is poorly constrained in many of the envelopes, in particular in those with low mass loss rates, such as R\,Leo, R\,Cas, Y\,CVn, and R\,Lep (see Fig.\,\ref{fig:sed}). The best fit values of the parameters $T_d(r_c)$ and $\tau_{10}$ are given in Table.\,\ref{table:stars}. The values of $\tau_{10}$ obtained here are similar to those reported by \cite{Schoier2013}, in most cases within a factor of two. However, in the case of $T_d(r_c)$ there are differences as large as 500 K between our values and those of \cite{Schoier2013}, which illustrates the larger error in the determination of this parameter.

In addition to $T_d(r_c)$ and $\tau_{10}$, the dust radiative transfer calculations provide some parameters that are needed to describe the envelope and perform the excitation and radiative transfer calculations of molecules. These parameters are the dust condensation radius, $r_c$, the bolometric luminosity, $L_\star$, and the gas-to-dust mass ratio, $\Psi$, all of which are given in Table\,\ref{table:stars}. The parameter $\Psi$ is computed from the column density of dust, which can be evaluated using $\tau_{10}$ and the optical constants of dust, and the column density of gas beyond the dust condensation radius, which is determined independently from CO through the gas mass loss rate (see Sect.\,\ref{sec:model_co}). In addition, the output of the DUSTY\,V2 code provides the dust temperature as a function of radius.

\subsection{Model of the gaseous envelope: fit to CO} \label{sec:model_co}

\begin{figure*}
\centering
\includegraphics[angle=0,width=0.99\textwidth]{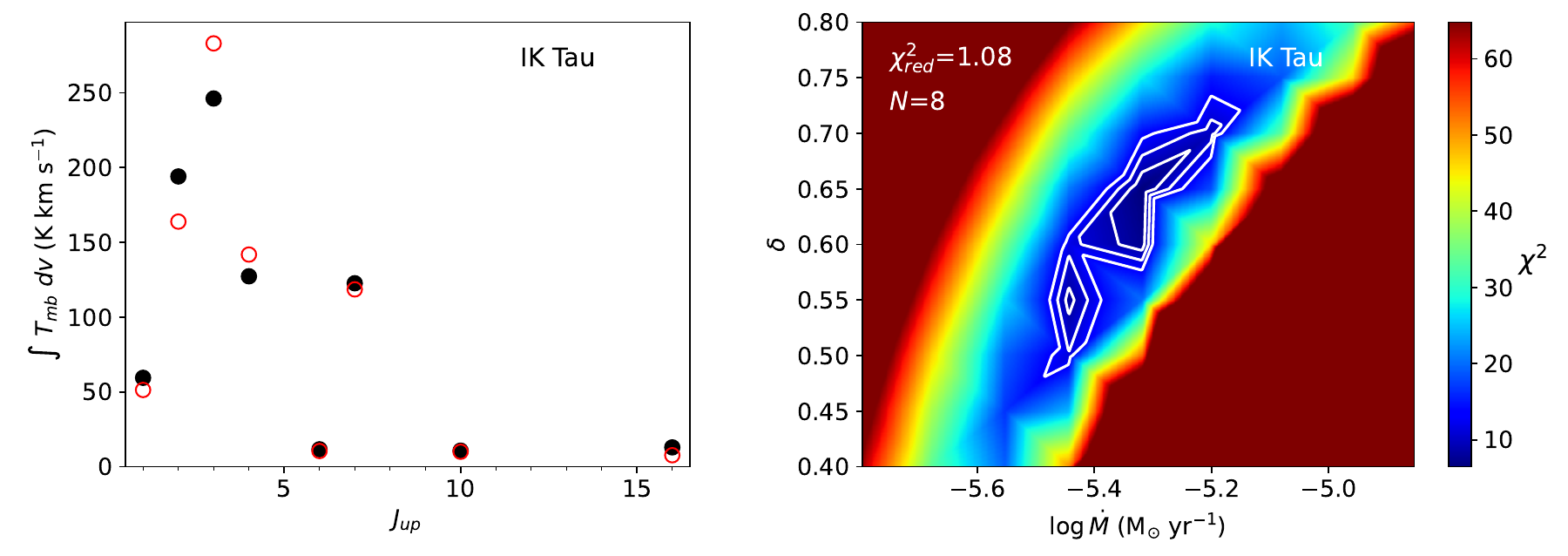}
\caption{Results from CO analysis for IK\,Tau. The left panel shows the observed intensities (see Table\,\ref{table:co}) as black filled circles and the calculated ones as red empty circles. The right panel shows $\chi^2$ as a function of the logarithm of the mass loss rate, $\log \dot{M}$, and the exponent of the gas kinetic temperature radial profile, $\delta$. The white contours correspond to 1, 2, and 3\,$\sigma$ levels. Similar plots for other stars are shown in Fig.\,\ref{fig:co}.} \label{fig:co_iktau}
\end{figure*}

To have a consistent model for each studied source, rather than picking up the mass loss rates from the literature, we determined them by modeling multiple CO lines. The model, which is described in the beginning of Sect.\,\ref{sec:model}, consists of a spherical envelope around an AGB star where various physical quantities may experience variations with radius. Concretely, the expansion velocity, gas volume density, and gas kinetic temperature are given by Eqs.\,(\ref{eq:vexp}-\ref{eq:tg}). The microturbulence velocity is taken as 1 km s$^{-1}$ throughout the envelope (see \citealt{Massalkhi2018}). To be consistent with the dust radiative transfer calculations described in Sec.\,\ref{sec:model_sed}, we assume that dust grains, with a radius of 0.1 $\mu$m and a density of 3.3 g cm$^{-3}$ for silicate and 2.0 g cm$^{-3}$ for amorphous carbon, are present beyond the condensation radius with an uniform gas-to-dust mass ratio and the temperature radial profile resulting from the DUSTY\,V2 calculations. The fractional abundance of CO with respect to H$_2$ at the initial radius (here taken as the stellar photosphere, $R_\star$) is assumed to be 2.0\,$\times$\,10$^{-4}$ for O-rich envelopes, a value typically adopted in previous works (e.g., \citealt{Olofsson2002}), while in the case of C-rich envelopes we adopt a value of 6.7\,$\times$\,10$^{-4}$, as determined from observations of H$_2$ toward IRC\,+10216 by \cite{Fonfria2022}. The CO abundance fall off due to photodissociation in the outer layers is described according to the fitting formula in \cite{Groenewegen2017}. This approach is mainly used for simplicity, although we note that more recent studies \citep{Saberi2019,Groenewegen2021} indicate that CO sizes estimated by \cite{Groenewegen2017} could be too large by 11-60\,\%. An observational validation of the theoretical description given by \cite{Groenewegen2021} would be very useful to shed light on this point.

The excitation and radiative transfer of CO in a circumstellar envelope is solved using two different methods. The first one solves the excitation out of local thermodynamic equilibrium (LTE) and the radiative transfer locally using the large velocity gradient (LVG) formalism. This code has been used previously in \cite{Agundez2012} and \cite{Massalkhi2018,Massalkhi2019,Massalkhi2020}. The second is a non-local non-LTE method based on the Monte Carlo formalism. The code uses the algorithm of RATRAN \citep{Hogerheijde2000}\footnote{\texttt{https://home.strw.leidenuniv.nl/$\sim$michiel/ratran/}}, modified to allow for a layer-dependent number of energy levels, which is chosen depending on the local excitation conditions. This implementation facilitates the convergence when radiative pumping to vibrationally excited states is important in certain but not all circumstellar layers. We include the first 30 rotational levels within the $v$\,=\,0 and $v$\,=\,1 vibrational states of CO. The highest level included ($v$\,=\,1, $J$\,=\,29) has an energy of 5467 K over the ground state ($v$\,=\,0, $J$\,=\,0). In the Monte Carlo calculations each shell is inspected and those levels with a marginal population are removed. The level energies are computed from the rotational constants given by \cite{Winnewiser1997} for the $v$\,=\,0 state and by \cite{Gendriesch2009} for the $v$\,=\,1 state. The line strengths are taken from \cite{Goorvitch1994} for pure rotational transitions and from the HITRAN database \citep{Rothman2005} for the ro-vibrational transitions. We use the rate coefficients for pure rotational transitions induced by inelastic collisions between CO and para and ortho H$_2$ from \cite{Yang2010}, where we assume the statistical ortho-to-para ratio of 3 for H$_2$. We also include collisions between CO and He, the latter assumed to have a solar abundance of 0.17 relative to H$_2$, using the rate coefficients from \cite{Cecchi-Pestellini2002}. For ro-vibrational transitions induced by collisions we adopt the same rate coefficients used for pure rotational transitions but decreased by a factor of 10$^4$ (typically found for similar molecules such as SiO and CS; \citealt{Balanca2017,Lique2007}).

The $\chi^2$ analysis was carried out by running CO models for each source adopting the parameters in Table\,\ref{table:stars}. The mass loss rate, $\dot{M}$, and the exponent, $\delta$, describing the gas kinetic temperature radial profile according to Eq.\,(\ref{eq:tg}), are left as adjustable parameters. Most CO line intensities were collected from the literature (see references in Table\,\ref{table:co}), at the exception of IRC\,+30374 and CRL\,190, in which case we performed new observations (see Sect.\,\ref{sec:observations}). We used the velocity-integrated line intensities in the $T_{mb}$ scale as $I_{obs}$ in Eq.\,(\ref{eq:chi2}). These values are given in Table\,\ref{table:co} for the 14 studied sources. Given the diversity of literature sources and telescopes employed it is difficult to assign a reliable uncertainty to each individual CO observation. We therefore adopt a uniform uncertainty of 20\,\% for all observed CO line intensities. Initially we used the LVG method, which is fast, to explore a broad range of the parameter space [$\dot{M}$, $\delta$], and later on we employed the Monte Carlo method, which is more computationally expensive, to restrict to a narrower range of the parameter space and determine more accurately $\dot{M}$ and $\delta$. All the results shown in this paper correspond to these latter Monte Carlo models.

\begin{figure}
\centering
\includegraphics[angle=0,width=0.99\columnwidth]{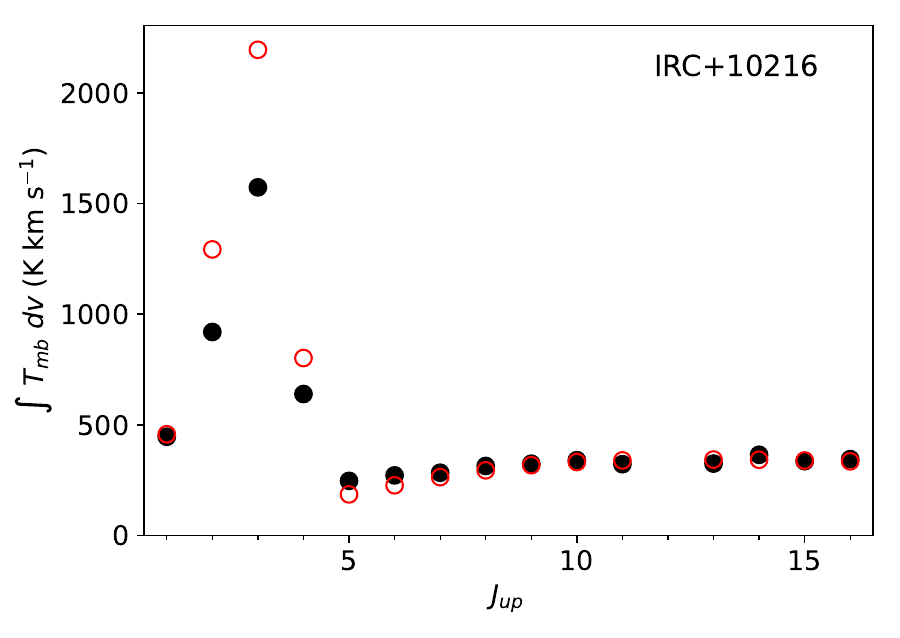}
\caption{Best fit CO model for IRC\,+10216. The observed intensities (see Table\,\ref{table:co}) are represented by black filled circles and the calculated ones by red empty circles.} \label{fig:co_irc10216}
\end{figure}

In Fig.\,\ref{fig:co_iktau} we show the results of the CO analysis for the O-rich star IK\,Tau. The availability of 8 CO lines allows to constrain well the mass loss rate and gas temperature exponent to $\dot{M}$\,=\,(4.8\,$\pm$\,0.7)\,$\times$\,10$^{-6}$ M$_{\odot}$ yr$^{-1}$ and $\delta$\,=\,0.65\,$\pm$\,0.06, with a value of $\chi^2_{red}$ around one. Similar plots for the rest of stars (at the exception of IRC\,+10216, which is discussed below) are shown in Fig.\,\ref{fig:co}. In general, the mass loss rate $\dot{M}$ is well constrained for most of the envelopes, with 1\,$\sigma$ errors in the range 10\,-30\,\%, except for IRC\,+20370, for which the error is $\sim$\,40\,\%. The gas temperature exponent $\delta$ is reasonably well constrained for many of the sources, although for some of them, such as TX\,Cam, R\,Leo, Y\,CVn, and R\,Lep, the uncertainty is high, and in the case of CRL\,190, the parameter $\delta$ is poorly constrained. The best fit values of $\dot{M}$ and $\delta$ are given in Table\,\ref{table:stars}. The mass loss rates derived are consistent with those derived by \cite{Schoier2013} within a factor of two. Larger differences of up to a factor of 3-4 are found for TX\,Cam, R\,Lep, LP\,And, and IRC\,+20370. In most cases changes in $\dot{M}$ can be, at least in part, attributed to the different distances used. We note that the distances adopted here should be more accurate because they are mostly based on Gaia DR3 parallaxes \citep{Gaia2023}, while those in \cite{Schoier2013} are estimated from period-luminosity relationships \citep{Whitelock1994,Groenewegen1996} or from Hipparcos parallaxes when available.

The case of IRC\,+10216 is a special one. This is the only AGB envelope in our sample for which the mass loss rate has been determined directly by observing H$_2$, instead of CO \citep{Fonfria2022}. The mass loss rate derived by these authors is 2.4\,$\times$\,10$^{-5}$ M$_{\odot}$ yr$^{-1}$, in agreement with the range (2-4)\,$\times$\,10$^{-5}$ M$_{\odot}$ yr$^{-1}$ derived in previous studies from maps of low-$J$ CO lines \citep{Cernicharo2015,Guelin2018}. This is also the only source for which we have many high-$J$ CO lines from HIFI. If we perform a $\chi^2$ analysis with $\dot{M}$ and $\delta$ as adjustable parameters, as for the rest of sources, the best fit model has problems to reproduce simultaneously low-$J$ and high-$J$ lines. We therefore decided to fix the mass loss rate to the value derived by \cite{Fonfria2022} and implemented a gas temperature radial profile with three power laws to describe the inner, intermediate and outer envelope. A similar approach was adopted by \cite{Daniel2012} to model the HIFI lines of this source. The best fit model is shown in Fig.\,\ref{fig:co_irc10216}, where the gas temperature is described by $\delta$\,=\,0.45 inside 10$^{15}$ cm, $\delta$\,=\,1.55 beyond 10$^{16}$ cm, and $\delta$\,=\,0.95 in between.

\subsection{Model for SiO, CS, and SiS: abundance and radial extent} \label{sec:model_molecules}

To model the lines of SiO, CS, and SiS we carried out excitation and radiative transfer calculations similar to those described in Sect.\,\ref{sec:model_co} for CO. The main difference is that the abundance fall off in the outer layers, which in the case of CO is described by the fitting formula of \cite{Groenewegen2017}, is described now by the empirical expression
\begin{equation}
f(r) = f_0 \exp \Bigg[-\bigg(\frac{r}{r_e}\bigg)^2\Bigg], \label{eq:f}
\end{equation}
where $f$ is the fractional abundance relative to H$_2$, $f_0$ is the abundance at the initial radius (taken as the stellar photosphere, $R_\star$), and $r_e$ is the $e$-folding radius at which the abundance has decreased by a factor of $e$ with respect to $f_0$. Here, the mass loss rate and the gas temperature radial profile are fixed to the values determined from the CO analysis (see Sect.\,\ref{sec:model_co}), and thus the adjustable parameters in the $\chi^2$ analysis are now $f_0$ and $r_e$. That is, we aim at constraining the abundance and radial extent for SiO, CS, and SiS. In addition to the lines observed in this study we collected SiO, CS, and SiS data from the literature for the 14 sources of our sample to have a number of independent observations as large as possible. The complete number of observations are summarized in Table\,\ref{table:lines_all}. We use the velocity-integrated line intensities in the $T_{mb}$ scale as $I_{obs}$ in Eq.\,(\ref{eq:chi2}). We assume that the uncertainties in $I_{obs}$ are dominated by the calibration error, which usually increases with increasing frequency. Based on the typical calibration errors at the IRAM\,30m telescope, we assume an uncertainty of 10\,\% for frequencies below 120 GHz, of 30\,\% above 190 GHz, and of 20\,\% for frequencies in between. As in the case of the CO models, we started by using the LVG method to explore the parameter space [$f_0$, $r_e$] over a broad range, and later on we switched to the Monte Carlo method to focus on a narrower region of the parameter space.

\begin{figure*}
\centering
\includegraphics[angle=0,width=\textwidth]{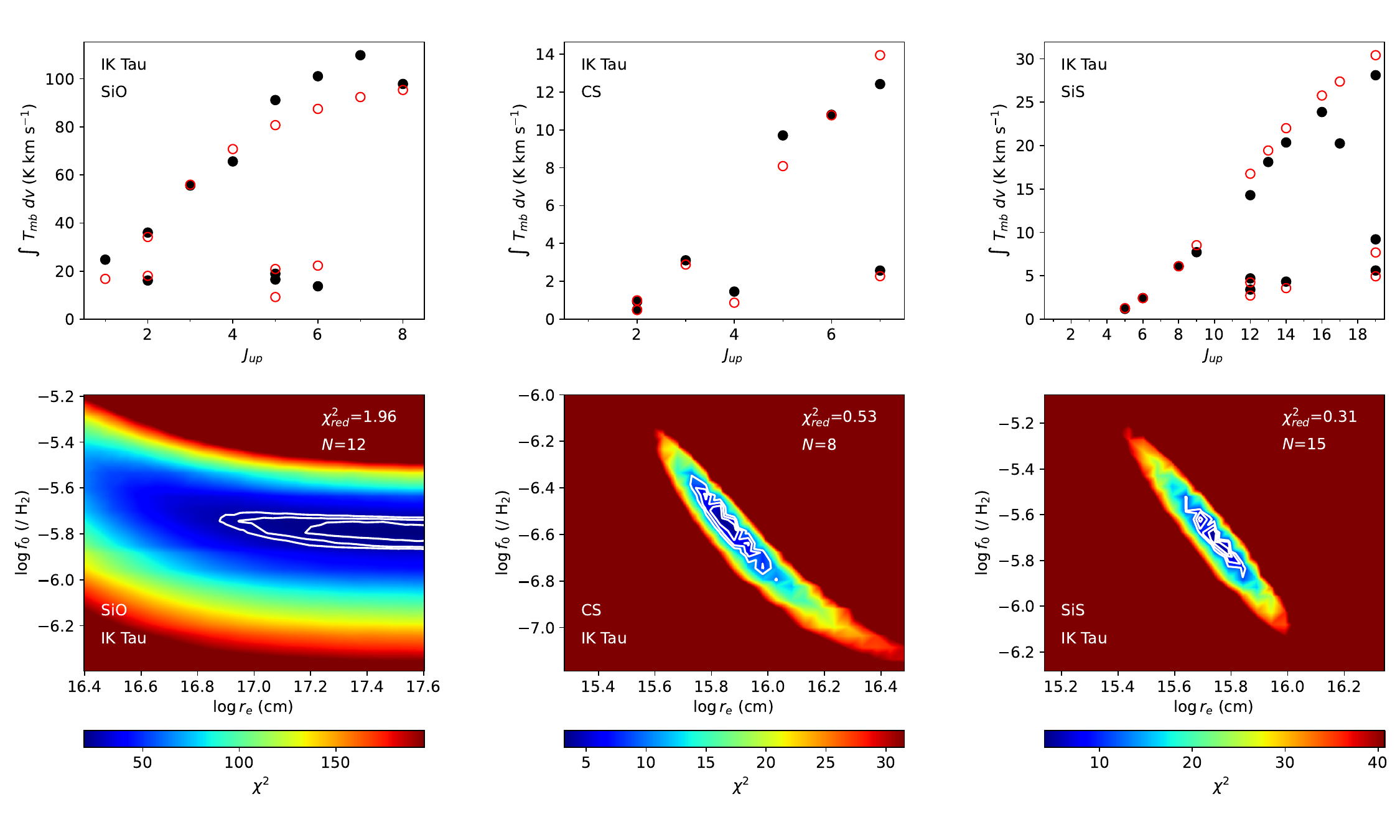}
\caption{Results from SiO, CS, and SiS analysis for IK\,Tau. The top panels show the observed intensities (see Table\,\ref{table:lines_all}) as black filled circles and the calculated ones of the best fit model as red empty circles. The bottom panels show $\chi^2$ as a function of the logarithm of the $e$-folding radius, $\log r_e$, and the logarithm of the fractional abundance relative to H$_2$, $\log f_0$. The white contours correspond to 1, 2, and 3\,$\sigma$ levels. Similar plots for IRC\,+10216 are shown in Fig.\,\ref{fig:irc10216} and for other stars in Fig.\,\ref{fig:sio} (SiO), Fig.\,\ref{fig:cs} (CS), and Fig.\,\ref{fig:sis} (SiS).} \label{fig:iktau}
\end{figure*}

\begin{figure*}
\centering
\includegraphics[angle=0,width=\textwidth]{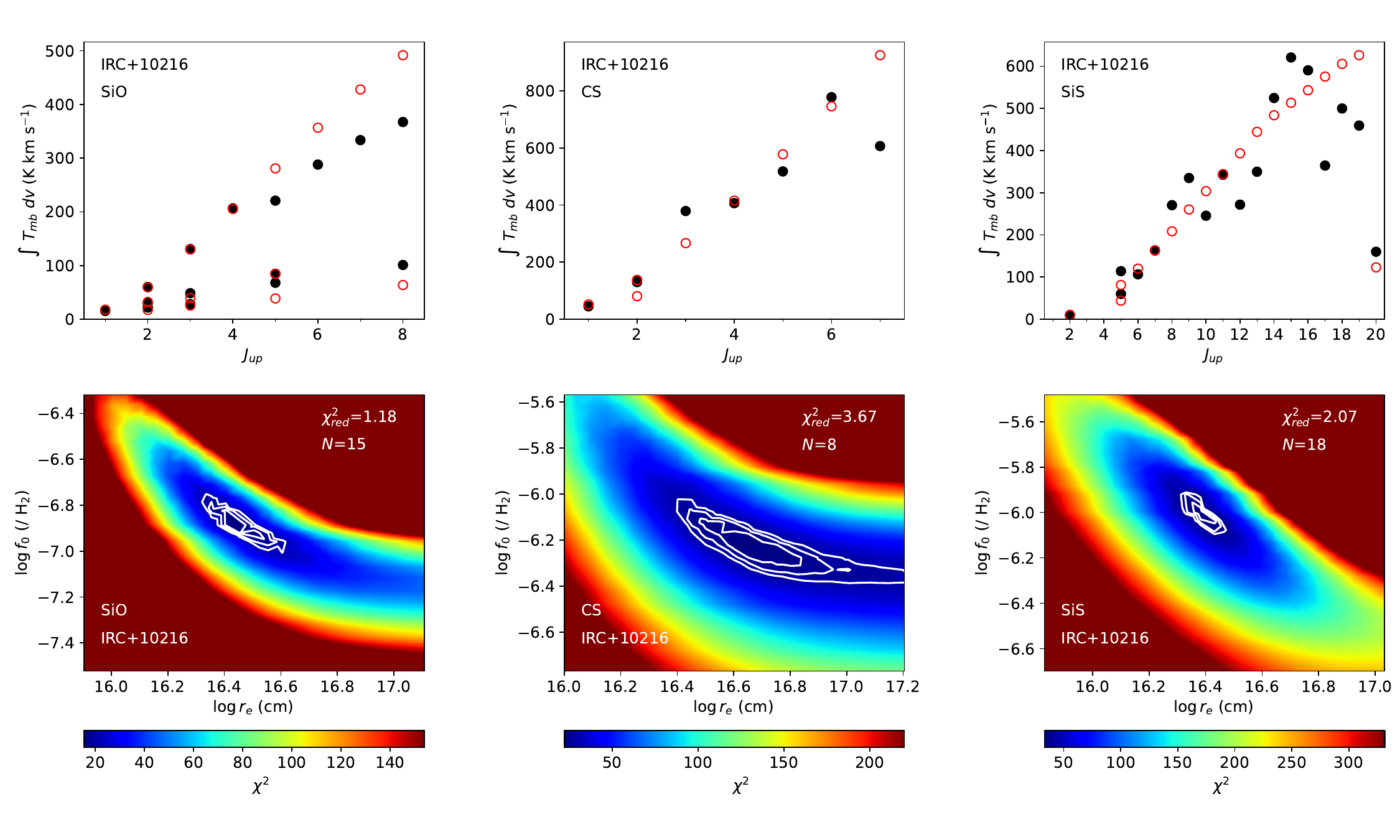}
\caption{Same as Fig.\,\ref{fig:iktau} but for IRC\,+10216.} \label{fig:irc10216}
\end{figure*}

The molecular data used in the excitation and radiative transfer calculations are as follows. For SiO we included the first 50 rotational levels within the $v$\,=\,0 and $v$\,=\,1 vibrational states, with the level energies given by the Dunham coefficients reported by \cite{Sanz2003}. The highest level included ($v$\,=\,1, $J$\,=\,49) has an energy of 4296 K. Line strengths are computed from the dipole moments measured by \cite{Raymonda1970} for pure rotational transitions and from the Einstein coefficients calculated by \cite{Drira1997} for ro-vibrational transitions. For pure rotational transitions induced by inelastic collisions between SiO and para/ortho H$_2$ we use the rate coefficients calculated by \cite{Balanca2018}, adopting an ortho-to-para ratio of 3 for H$_2$. For ro-vibrational transitions induced by collisions with H$_2$ we use the rate coefficients calculated by \cite{Balanca2017}. Since these rate coefficients are originally calculated for He as collider we scale them by multiplying by the square root of the ratio of the reduced masses of the H$_2$ and He colliding systems. We also include collisions with He adopting a solar abundance of 0.17 relative to H$_2$ and the rate coefficients calculated by \cite{Dayou2006} for pure rotational transitions and by \cite{Balanca2017} for ro-vibrational transitions.

For CS we selected the first 50 rotational levels within the first two vibrational states. The highest level included ($v$\,=\,1, $J$\,=\,49) has an energy of 4678 K. The level energies were calculated from the Dunham coefficients reported by \cite{Muller2005}. The line strengths were calculated from the dipole moments measured by \cite{Winnewiser1968} for pure rotational transitions and from the Einstein coefficients calculated by \cite{Chandra1995} for ro-vibrational transitions. The rate coefficients for pure rotational transitions induced by inelastic collisions are taken from \cite{Denis-Alpizar2018} for H$_2$ and \cite{Lique2006} for He, while for ro-vibrational transitions we adopt the values calculated by \cite{Lique2007}.

For SiS we included the first 70 rotational levels within the first two vibrational states. The highest level included ($v$\,=\,1, $J$\,=\,69) has an energy of 3158 K. The level energies were computed from the Dunham coefficients given by \cite{Muller2007} and the line strengths for pure rotational transitions within the $v$\,=\,0 state were calculated from the dipole moment given in the same reference. Line strengths for pure rotational transitions within the $v$\,=\,1 state and for ro-vibrational transitions were computed from the dipole moments calculated by \cite{Pineiro1987}. Regarding the rate coefficients for inelastic collisions, we use the values from \cite{Klos2008} for pure rotational transitions induced by H$_2$ and from \cite{Tobola2008} for He as collider and for ro-vibrational transitions.

\section{Results and discussion} \label{sec:results}

The results from the modeling of SiO, CS, and SiS are shown in Fig.\,\ref{fig:iktau} for IK\,Tau and Fig.\,\ref{fig:irc10216} for IRC\,+10216. The top panels compare the calculated line intensities of the best fit model with the observed ones (see Table\,\ref{table:lines_all}), while the bottom panels show $\chi^2$ as a function of the two adjustable parameters, the initial abundance relative to H$_2$, $f_0$, and the $e$-folding radius, $r_e$. IK\,Tau and IRC\,+10216 are the AGB envelopes with the highest number of observed lines of SiO, CS, and SiS. The observed line intensities are reasonably well reproduced by the best fit model, with $\chi^2_{red}$ values $\lesssim$\,2, at the exception of CS in IRC\,+10216, in which case $\chi^2_{red}$ is larger, probably due to some observational problem in the $J$\,=\,7-6 line. The fractional abundances are relatively well constrained for the three molecules in the two sources. In most cases the $e$-folding radius is well constrained by the $\chi^2$ analysis, although in the case of IK\,Tau only a lower limit to $r_e$(SiO) can be derived (see left-bottom panel in Fig.\,\ref{fig:iktau}).

\begin{table}
\small
\caption{Parameters from the $\chi^2$ analysis for SiO, CS, and SiS.}
\label{table:f0_re}
\centering
\begin{tabular}{l@{\hspace{0.5cm}}c@{\hspace{0.5cm}}c@{\hspace{0.5cm}}cc}
\hline \hline
\multicolumn{1}{l}{Source} & \multicolumn{1}{c}{$f_0$\,$^a$} & \multicolumn{1}{c}{$r_e$ (cm)\,$^a$} & \multicolumn{1}{c}{$N$} & \multicolumn{1}{c}{$\chi^2_{red}$} \\
\hline
\\
\multicolumn{5}{c}{SiO} \\
\hline
R\,Leo & 4.3$\times$\,10$^{-6}$ (0.11) & 1.4$\times$\,10$^{16}$ (0.15) &  8 & 5.5 \\
R\,Cas & 1.8$\times$\,10$^{-6}$ (0.13) & 5.1$\times$\,10$^{16}$ (0.38) &  7 & 3.3 \\
TX\,Cam & 2.0$\times$\,10$^{-6}$ (0.12) & 6.4$\times$\,10$^{16}$ (0.37) &  6 & 1.3 \\
IK\,Tau & 1.6$\times$\,10$^{-6}$ (0.08) & ... & 12 & 2.0 \\
V1111\,Oph & 8.9$\times$\,10$^{-7}$ (0.08) & ... &  8 & 0.8 \\
GX\,Mon & 8.5$\times$\,10$^{-7}$ (0.08) & ... & 10 & 1.2 \\
NV\,Aur & 1.4$\times$\,10$^{-6}$ (0.08) & ... &  6 & 2.1 \\
\vspace{0.2cm}
Y\,CVn & 3.1$\times$\,10$^{-7}$ (0.17) & 8.7$\times$\,10$^{15}$ (0.40) &  3 & 0.1 \\
R\,Lep & 2.4$\times$\,10$^{-6}$ (0.12) & 6.5$\times$\,10$^{15}$ (0.08) &  8 & 2.4 \\
LP\,And & 4.1$\times$\,10$^{-7}$ (0.12) & 3.0$\times$\,10$^{16}$ (0.18) &  7 & 4.0 \\
IRC\,+20370 & 2.5$\times$\,10$^{-6}$ (0.21) & 1.2$\times$\,10$^{16}$ (0.11) &  5 & 3.6 \\
IRC\,+30374 & 1.6$\times$\,10$^{-6}$ (0.23) & 4.3$\times$\,10$^{16}$ (0.51) &  4 & 0.1 \\
IRC\,+10216 & 1.3$\times$\,10$^{-7}$ (0.08) & 2.9$\times$\,10$^{16}$ (0.08) & 15 & 1.2 \\
CRL190 & 2.7$\times$\,10$^{-7}$ (0.48) & 2.9$\times$\,10$^{16}$ (0.50) &  2 & 0.0 \\
\\
\multicolumn{5}{c}{CS} \\
\hline
R\,Leo & 7.3$\times$\,10$^{-8}$ (0.30) & 2.2$\times$\,10$^{15}$ (0.24) &  3 & 0.1 \\
R\,Cas & 4.1$\times$\,10$^{-8}$ (0.23) & 4.6$\times$\,10$^{15}$ (0.18) &  3 & 0.2 \\
TX\,Cam & 1.3$\times$\,10$^{-7}$ (0.08) & ... &  4 & 1.0 \\
IK\,Tau & 2.6$\times$\,10$^{-7}$ (0.11) & 7.6$\times$\,10$^{15}$ (0.08) &  8 & 0.5 \\
V1111\,Oph & 1.0$\times$\,10$^{-7}$ (0.18) & 1.0$\times$\,10$^{16}$ (0.13) &  4 & 0.9 \\
GX\,Mon & 7.9$\times$\,10$^{-8}$ (0.12) & 2.4$\times$\,10$^{16}$ (0.18) &  6 & 1.1 \\
NV\,Aur & 1.6$\times$\,10$^{-7}$ (0.25) & 2.1$\times$\,10$^{16}$ (0.25) &  3 & 0.2 \\
\vspace{0.2cm}
Y\,CVn & 1.3$\times$\,10$^{-5}$ (0.32) & 6.2$\times$\,10$^{15}$ (0.30) &  4 & 3.0 \\
R\,Lep & 1.3$\times$\,10$^{-5}$ (0.41) & 3.8$\times$\,10$^{15}$ (0.20) &  5 & 0.9 \\
LP\,And & 4.5$\times$\,10$^{-6}$ (0.16) & ... &  4 & 3.2 \\
IRC\,+20370 & 1.3$\times$\,10$^{-5}$ (0.44) & 2.0$\times$\,10$^{16}$ (0.35) &  7 & 0.3 \\
IRC\,+30374 & 1.8$\times$\,10$^{-5}$ (0.32) & 5.6$\times$\,10$^{16}$ (0.60) &  4 & 0.0 \\
IRC\,+10216 & 6.0$\times$\,10$^{-7}$ (0.08) & 4.5$\times$\,10$^{16}$ (0.18) &  8 & 3.7 \\
\\
\multicolumn{5}{c}{SiS} \\
\hline
R\,Cas & 4.3$\times$\,10$^{-7}$ (0.53) & 1.4$\times$\,10$^{15}$ (0.34) &  3 & 1.3 \\
TX\,Cam & 3.6$\times$\,10$^{-7}$ (0.14) & 2.6$\times$\,10$^{16}$ (0.26) &  5 & 3.8 \\
IK\,Tau & 2.1$\times$\,10$^{-6}$ (0.09) & 5.5$\times$\,10$^{15}$ (0.08) & 15 & 0.3 \\
V1111\,Oph & 4.8$\times$\,10$^{-7}$ (0.15) & 1.6$\times$\,10$^{16}$ (0.40) &  4 & 0.5 \\
GX\,Mon & 4.0$\times$\,10$^{-7}$ (0.10) & 1.2$\times$\,10$^{16}$ (0.09) &  9 & 0.6 \\
NV\,Aur & 2.5$\times$\,10$^{-6}$ (0.25) & 1.4$\times$\,10$^{16}$ (0.16) &  5 & 0.3 \\
\vspace{0.2cm}
LP\,And & 6.4$\times$\,10$^{-6}$ (0.10) & 1.1$\times$\,10$^{16}$ (0.08) &  7 & 3.4 \\
IRC\,+20370 & 8.6$\times$\,10$^{-6}$ (0.37) & 6.3$\times$\,10$^{15}$ (0.18) & 10 & 1.9 \\
IRC\,+30374 & 2.7$\times$\,10$^{-6}$ (0.15) & 1.2$\times$\,10$^{16}$ (0.36) &  5 & 0.5 \\
IRC\,+10216 & 1.0$\times$\,10$^{-6}$ (0.08) & 2.4$\times$\,10$^{16}$ (0.08) & 18 & 2.1 \\
CRL190 & 1.3$\times$\,10$^{-5}$ (0.44) & 5.5$\times$\,10$^{16}$ (0.43) &  4 & 0.5 \\
\hline
\end{tabular}
\tablenoteb{$^a$ Errors for $f_0$ and $r_e$ are given in parentheses in units of dex. The minimum adopted error is 20\,\% (0.08 dex).}
\end{table}

For the rest of stars, similar plots are shown in Fig.\,\ref{fig:sio} for SiO, Fig.\,\ref{fig:cs} for CS, and Fig.\,\ref{fig:sis} for SiS. The values of $f_0$ and $r_e$ derived from the $\chi^2$ analysis are listed in Table\,\ref{table:f0_re}. The quality of the fit and the ability to constrain the abundance and radial extent vary among the different cases studied. Although this is not shown, in general observed line profiles were well reproduced by the best-fit model (see \citealt{Agundez2012} for the case of IRC\,+10216, where the calculated line profiles are very similar to those obtained in this work). The match between observed and calculated line profile was not particularly good for those lines that show maser emission. This is the case of some SiS lines in C-rich sources, the $J$\,=\,11-10, $J$\,=\,14-13, and $J$\,=\,15-14 lines of SiS in IRC\,+10216 \cite{Fonfria2006,Agundez2012}) and the $J$\,=\,14-13 line of SiS in LP\,And and IRC\,+20370, and the $J$\,=\,1-0 line of SiO in O-rich sources, which shows one or various narrow peaks likely due to maser emission (see Fig.\,\ref{fig:lines}). The only case where we had serious difficulties to reproduce the observed line intensities was that of CS in CRL\,190, and therefore we did not attempt to fit it. In most cases the quality of the fit was good, with the parameter $\chi^2_{red}$ below 2.0 in 3/4 of the cases and going up to 4-6 only in a few cases. In all the cases it was possible to constrain the abundance to a more or less narrow range, while the radial extent was in general less well constrained than the abundance. The average error in $f_0$ is 0.20 dex while for $r_e$ it is somewhat higher, 0.26 dex (see individual values in Table\,\ref{table:f0_re}). Moreover, in a few cases the radial extent could not be constrained and only a lower limit to $r_e$ could be derived. This situation was encountered for SiO in the O-rich sources IK\,Tau, GX\,Mon, NV\,Aur, and V1111\,Oph, and also for CS in TX\,Cam and LP\,And. We note that a similar problem was found by \cite{Gonzalez-Delgado2003} for various O-rich envelopes when studying SiO.

The abundances derived here (see Table\,\ref{table:f0_re}) follow the general behavior found in previous works. That is, the abundance of SiO does not show a marked differentiation between O- and C-rich sources, although it does show a trend in which the denser the envelope the lower the SiO abundance, in agreement with previous studies \citep{Gonzalez-Delgado2003,Schoier2006a,Massalkhi2019,Massalkhi2020}. On the other hand, CS and SiS show a marked differentiation between O- and C-rich sources, being 100 and 10 times, respectively, more abundant in C-rich envelopes than in O-rich ones, in line with previous findings \citep{Schoier2007,Danilovich2018,Massalkhi2019,Massalkhi2020}. The main interest of this study is however the radial extent rather than the abundance.

The radial extent of SiO, CS, and SiS in AGB envelopes has been determined in previous studies from multiple lines observed with single dish telescopes \citep{Gonzalez-Delgado2003,Schoier2006a,Schoier2007,Danilovich2018}. The main difference of this work with respect to previous studies is that here we include a larger number of lines, which ensures a more robust determination of the radial extent, and make a systematic and coherent study for the three molecules. Constraints on the emission size of these molecules is also available from interferometric observations \citep{Lucas1992,Sahai1993,Schoier2004,Danilovich2019,Velilla-Prieto2019,Verbena2019}, although we caution that in some of these cases the brightness distribution has not been converted to an abundance distribution through radiative transfer modeling. It is interesting to compare the sizes derived here with those obtained from interferometric observations. The radial extent of SiO in IK\,Tau has been studied in several works, with conflicting results. Interferometric observations find an emission size of about 5\,$\times$\,10$^{15}$ cm \citep{Lucas1992,Sahai1993,Verbena2019}, while \cite{Gonzalez-Delgado2003} derive a larger size, 2.5\,$\times$\,10$^{16}$ cm, in their study of multiple lines, and our work points to values above 10$^{17}$ cm. A combined model of single dish and interferometric data should allow to better constrain the radial extent of SiO. The size of CS and SiS in IK\,Tau have been also studied through interferometric observations by \cite{Danilovich2019}, who find $e-$folding radii around 8\,$\times$\,10$^{15}$ cm and 4\,$\times$\,10$^{15}$ cm, respectively, in very good agreement with the values derived here. Our results show also very good agreement with the interferometric study of IRC\,+10216 by \cite{Velilla-Prieto2019} regarding the $e$-folding radii of SiO, CS, and SiS, and also the order in which molecules disappear in this particular envelope: SiS $\rightarrow$ SiO $\rightarrow$ CS. Our  value of $r_e$(SiO) in IRC\,+10216 is also similar to that derived by \cite{Schoier2006b}, 2.4\,$\times$\,10$^{16}$ cm, from a modeling study combining single dish and interferometric data.

\begin{figure}
\centering
\includegraphics[angle=0,width=\columnwidth]{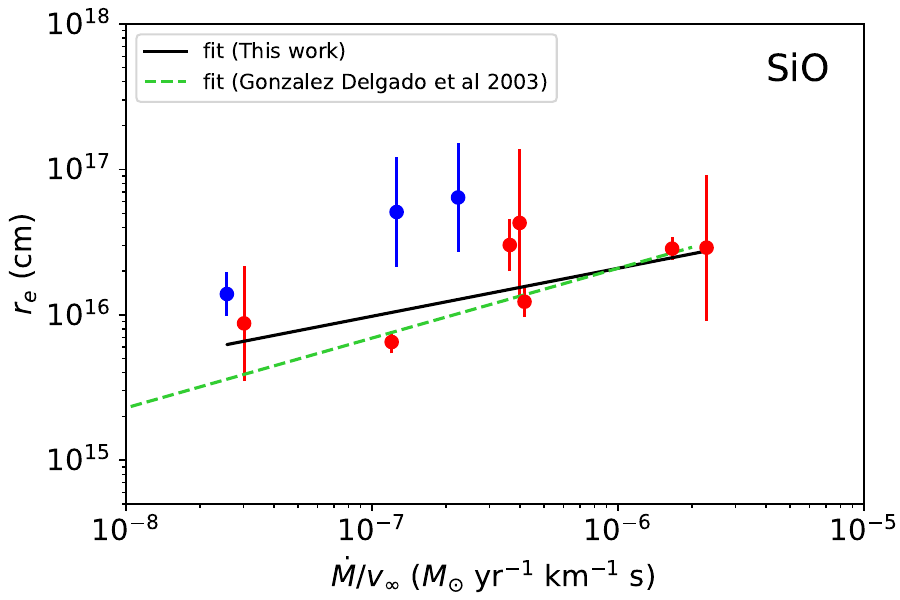} \includegraphics[angle=0,width=\columnwidth]{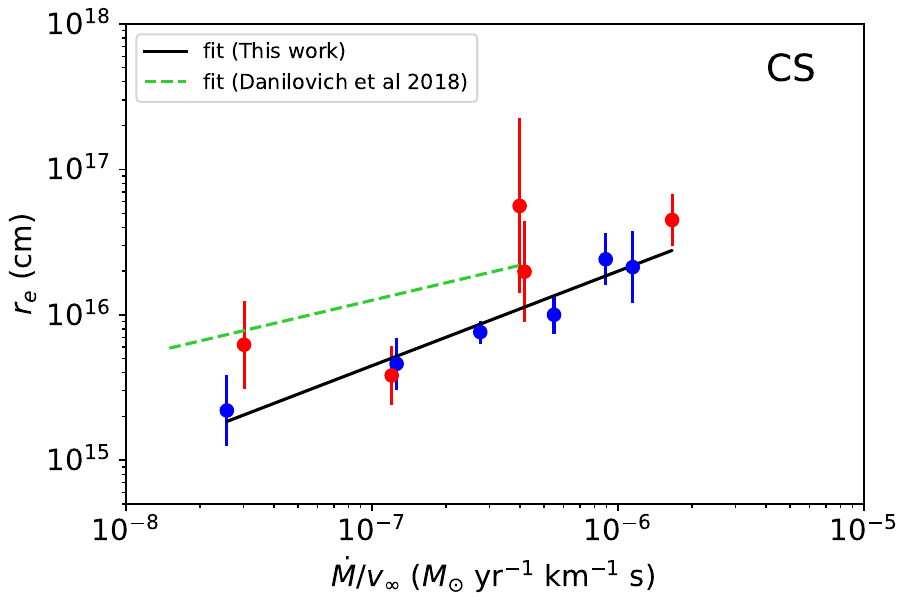} \includegraphics[angle=0,width=\columnwidth]{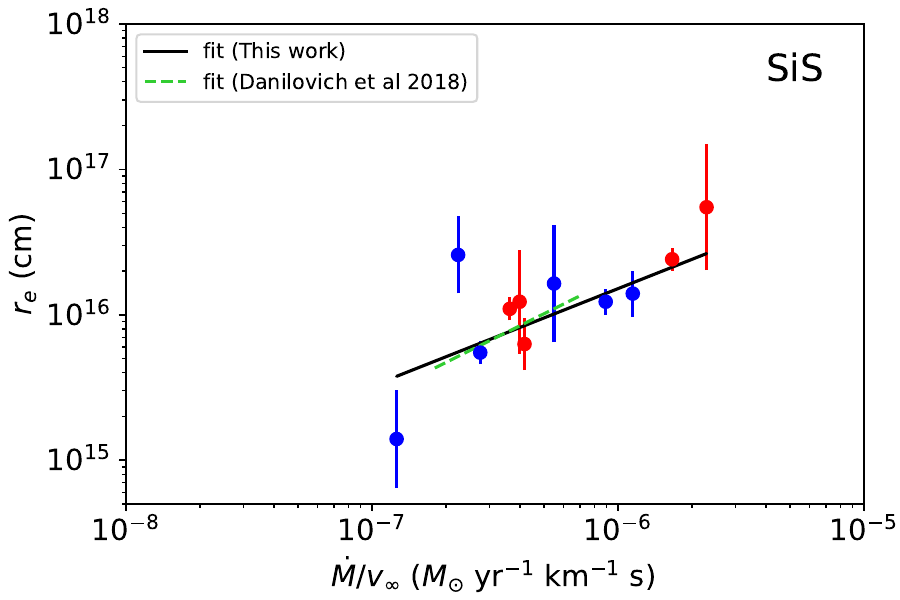}
\caption{$e$-folding radii derived in this study are represented as a function of the envelope density proxy, $\dot{M}/v_{\infty}$ for SiO (upper panel), CS (middle panel), and SiS (lower panel). Blue symbols correspond to O-rich envelopes and red ones to C-rich sources. The resulting linear fits are also plotted and compared with fits from previous studies.} \label{fig:re}
\end{figure}

We are interested in the dependence of the radial extent with the mass loss rate, or more specifically the envelope density, which can be evaluated by the parameter $\dot{M}/v_{\infty}$. There are theoretical grounds to expect a larger radial extent in denser envelopes, occasioned by the enhanced envelope extinction and the corresponding decrease in photodissociation rates. Such dependence has been empirically observed for SiO, CS, and SiS \citep{Gonzalez-Delgado2003,Schoier2006a,Danilovich2018} by constraining the radial extent over a more or less wide range of envelope densities. To shed light on this particular point, in Fig.\,\ref{fig:re} we represent the $e$-folding radii of SiO, CS, and SiS determined here (see Table\,\ref{table:f0_re}) as a function of the envelope density proxy $\dot{M}/v_{\infty}$. The plots show the aforementioned trend in which $r_e$ increases as $\dot{M}/v_{\infty}$ increases. A weighted linear fit in a log-log scale for each of the three molecules yields the following expressions
\begin{equation}
\log r_e ({\rm SiO}) = (18.3\,\pm\,0.5) + (0.33\,\pm\,0.08) \log \bigg( \frac{\dot{M}}{v_\infty} \bigg), \label{eq:re_sio}
\end{equation}
\begin{equation}
\log r_e ({\rm CS}) = (20.2\,\pm\,0.8) + (0.65\,\pm\,0.12) \log \bigg( \frac{\dot{M}}{v_\infty} \bigg), \label{eq:re_cs}
\end{equation}
\begin{equation}
\log r_e ({\rm SiS}) = (20.2\,\pm\,0.7) + (0.67\,\pm\,0.12) \log \bigg( \frac{\dot{M}}{v_\infty} \bigg), \label{eq:re_sis}
\end{equation}
The fits are in line with those obtained previously by \cite{Gonzalez-Delgado2003} for SiO in M-type AGB envelopes and by \cite{Danilovich2018} for CS and SiS in AGB envelopes of M-, C, and S-type over a limited range of envelope densities (see Fig.\,\ref{fig:re}). In the cases of SiO and SiS the agreements with the previously reported fits are almost perfect, while in the case of CS, our fit results in a smaller radial extent, by a factor 2-4, compared to that found by \cite{Danilovich2018}.

Our data suggest that in C-rich envelopes CS could have a larger radial extent than in O-rich sources. Although this conclusion is only tentative, it may be due to the different abundance of CS in each type of source. In C-rich sources, the abundance of CS is on the order of 10$^{-5}$ relative to H$_2$, which may cause some self-shielding effect, and thus a larger radial extent, while the same does not happen in O-rich sources, where the abundance is two orders of magnitude lower \citep{Massalkhi2020}. In the case of SiO, there may be also some differentiation between O- and C-rich sources, which is evident if we consider the four O-rich sources IK\,Tau, GX\,Mon, NV\,Aur, and V1111\,Oph, which are not included in the fit but for which our $\chi^2$ analysis suggest an $e$-folding radius in excess of 10$^{17}$ cm (see Fig.\,\ref{fig:sio}). The reason of the large radial extent inferred for SiO in O-rich sources is unclear, although it may be related to the $J$\,=\,1-0 line. In O-rich sources this line (not included in the analysis of \citealt{Gonzalez-Delgado2003}) shows narrow peaks in the profile (see Fig.\,\ref{fig:lines}). The origin is probably non-thermal maser emission, something that our radiative transfer model does not treat properly. However, the velocity-integrated intensity of this line in O-rich sources is well reproduced by the best fit models shown in Fig.\,\ref{fig:sio} (although the narrow peaks are not reproduced), and if this line is discarded the $e$-folding radius derived does not change significantly. As discussed above for IK\,Tau, including interferometric data could allow to constrain better the radial extent of SiO in O-rich sources.

\begin{figure}
\centering
\includegraphics[angle=0,width=\columnwidth]{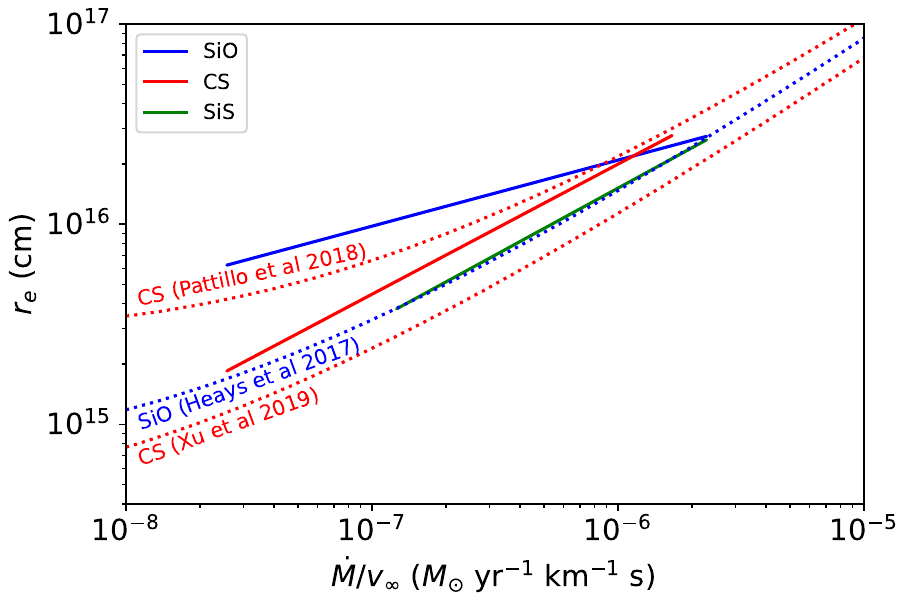}
\caption{Relation between radial extent and envelope density according to the empirical fits and to the photodissociation model. The solid lines are the empirical fits given by Equations\,(\ref{eq:re_sio}), (\ref{eq:re_cs}), and (\ref{eq:re_sis}), and the dotted lines are the predictions from the photodissociation model using different photodissociation rates from the literature (see text).} \label{fig:re_ph}
\end{figure}

Another interesting aspect to discuss is how the sizes of SiO, CS, and SiS compare to each other. In Fig.\,\ref{fig:re_ph} we show the fits derived for the three molecules. It is seen that at high envelope densities, a few 10$^{-6}$ $M_{\odot}$ yr$^{-1}$ km$^{-1}$ s, the radial extent of the three molecules is similar, while at low densities, below 10$^{-7}$ $M_{\odot}$ yr$^{-1}$ km$^{-1}$ s, the radial extent is found to increase in this order SiS $\rightarrow$ CS $\rightarrow$ SiO. Since at large scales the envelope size is expected to be mostly regulated by photodissociation, we have computed the $e$-folding radius adopting the simple photodissociation model described in \cite{Massalkhi2018}. In this model we follow \cite{Agundez2017} and assume that the $N_{\rm H}/A_V$ ratio is 1.5 times lower than the canonical value of 1.87\,$\times$\,10$^{21}$ cm$^{-2}$ mag$^{-1}$ of \cite{Bohlin1978} and adopt a round value for the expansion velocity of 10 km s$^{-1}$. The abundance radial profile from this model is determined by the photodissociation rate, which is parameterized as a function of the visual extinction according to the expression $\alpha$\,$\exp(-\beta A_V)$, where $A_V$ is the visual extinction in mag, $\alpha$ is the unattenuated photodissociation rate, and $\beta$ regulates the effect of dust shielding. For SiO we adopt $\alpha$\,=\,1.6\,$\times$\,10$^{-9}$ s$^{-1}$ and $\beta$\,=\,2.66 \citep{Heays2017}, based on the oscillator strengths calculated by \cite{vanDishoeck2006} for several electronic transitions. For CS, there are conflicting results on its photodissociation rate. \cite{Heays2017} recommends $\alpha$\,=\,9.5\,$\times$\,10$^{-10}$ s$^{-1}$ and $\beta$\,=\,2.77 based on absorption line measurements by \cite{Stark1987}, energies of excited electronic states calculated by \cite{Bruna1975}, plus some oscillator strength guesses. Similar values were found by \cite{Agundez2018}, $\alpha$\,=\,9.5\,$\times$\,10$^{-10}$ s$^{-1}$ and $\beta$\,=\,2.60, based on the same cross section. More recently, a couple of theoretical studies determined quite different photodissociation rates. On the one hand, \cite{Pattillo2018} derive a low photodissociation rate, $\alpha$\,=\,3.7\,$\times$\,10$^{-10}$ s$^{-1}$ and $\beta$\,=\,2.32, while \cite{Xu2019} calculate a higher value, $\alpha$\,=\,2.9\,$\times$\,10$^{-9}$ s$^{-1}$.

The predictions from the photodissociation model are compared to the empirical fits in Fig.\,\ref{fig:re_ph}. The photodissociation model predicts the same behavior seen in the empirical fits in which the radial extents of molecules with different unattenuated photodissociation rates $\alpha$ differ at low envelope densities but converge as the envelope density increases. This occurs because at low envelope densities, and thus low extinction, the radial extent is essentially regulated by the unattenuated photodissociation rate $\alpha$, while at high envelope densities the radial extent depends mostly on dust extinction, i.e., on $\exp(-\beta A_V)$. At high envelope densities empirical fits and photodissociation model agree, but at low envelope densities there are significant discrepancies. The photodissociation model predicts a lower size for SiO compared to the empirical fit. It is unclear whether this results from a too high photodissociation rate on the theoretical side or from an overestimation of the radial extent on the observational side. Concerning CS, there are significant differences depending on whether we use the photodissociation rate of \cite{Pattillo2018} or that of \cite{Xu2019}. The theoretical photodissociation rate by \cite{Pattillo2018} predicts a larger radial extent for CS than our observational fit and CS is predicted to be more extended than SiO, in contrast with our empirical fits. On the other hand, if we favor the photodisociation rate of \cite{Xu2019}, the radial extent predicted for CS would be somewhat lower than given by our empirical fit, although in this case the sequence $r_e$(SiO)\,$>$\,$r_e$(CS) is correctly predicted. In the case of SiS there is no prediction because the photodissociation rate is unknown, although it has been argued that it should be similar to that of SiO \citep{vanDishoeck1988,Wirsich1994}. On the other hand, our empirical fits and the interferometric study by \cite{Velilla-Prieto2019} favor a higher photodissociation rate for SiS compared to SiO. Our empirical fits indicate also that the photodissociation rate of SiS should be somewhat larger than that of CS, something that is also in line with interferometric studies \citep{Velilla-Prieto2019,Danilovich2019}, which find a smaller radial extent for SiS compared to CS.

To shed some light on the different radial extent of the three molecules there seems to be two directions of progress. On the one side, dedicated modeling studies combining single dish and interferometric data should focus on AGB envelopes of O- and C-rich character at the low and high mass loss rate edges. On the other hand, it would be worth to investigate either theoretically or experimentally the photodissociation of SiS, and to revisit also that of SiO and CS. Establishing a sequence in the photodissociation rates of the three molecules would allow to test the empirical fits obtained from observations and ultimately to validate the underlying idea that photodissociation regulates the radial extent of molecules in AGB envelopes. 

\section{Conclusions} \label{sec:conclusions}

We carried out an observational study of SiO, CS, and SiS in AGB envelopes aiming at constraining their radial extent in envelopes of different density and chemical type. We find that for the three molecules, the envelope size increases with increasing envelope density, in agreement with previous observational studies. We also find that at high envelope densities, $\dot{M}/v_{\infty}$\,$>$\,10$^{-6}$ $M_{\odot}$ yr$^{-1}$ km$^{-1}$ s, the three molecules show a similar radial extent, while for decreasing envelope densities SiO, CS, and SiS become more differentiated with respect to their radial extent, which is in line with expectations based on a simple photodissociation model. At low envelope densities we find that molecules extend farther in this order SiS\,$\rightarrow$\,CS\,$\rightarrow$\,SiO. The scarce interferometric studies and the sparsity of data on the photodissociation of the three molecules do not allow for a solid confirmation or refutation of this scheme. We argue that further modeling studies combining single dish and interferometric data, together with investigation on the photodissociation of SiO, CS, and SiS should allow to draw a clear picture on the radial extent of these molecules in AGB envelopes.

\begin{acknowledgements}

We acknowledge funding support from Spanish Ministerio de Ciencia e Innovaci\'on through grants AYA2016-75066-C2-1-P, PID2019-106110GB-I00, PID2019-107115GB-C21, PID2019-105203GB-C21, PID2019-105203GB-C22, PID2020-117034RJ-I00, and PIE 202250I097, and from the European Research Council (ERC Grant 610256: NANOCOSMOS). Calculations in this work were run at the SGAI-CSIC supercomputer DRAGO and the Galicia Supercomputing Center (CESGA). This work has made use of data from the European Space Agency (ESA) mission {\it Gaia} (\url{https://www.cosmos.esa.int/gaia}), processed by the {\it Gaia} Data Processing and Analysis Consortium (DPAC, \url{https://www.cosmos.esa.int/web/gaia/dpac/consortium}). Funding for the DPAC has been provided by national institutions, in particular the institutions participating in the {\it Gaia} Multilateral Agreement. This work made also use of the VizieR catalogue access tool, CDS, Strasbourg, France (DOI: 10.26093/cds/vizier). The original description of the VizieR service was published in \cite{Ochsenbein2000}. This publication makes use of data products from the Wide-field Infrared Survey Explorer, which is a joint project of the University of California, Los Angeles, and the Jet Propulsion Laboratory/California Institute of Technology, and NEOWISE, which is a project of the Jet Propulsion Laboratory/California Institute of Technology. WISE and NEOWISE are funded by the National Aeronautics and Space Administration. The Pan-STARRS1 Surveys (PS1) and the PS1 public science archive have been made possible through contributions by the Institute for Astronomy, the University of Hawaii, the Pan-STARRS Project Office, the Max-Planck Society and its participating institutes, the Max Planck Institute for Astronomy, Heidelberg and the Max Planck Institute for Extraterrestrial Physics, Garching, The Johns Hopkins University, Durham University, the University of Edinburgh, the Queen's University Belfast, the Harvard-Smithsonian Center for Astrophysics, the Las Cumbres Observatory Global Telescope Network Incorporated, the National Central University of Taiwan, the Space Telescope Science Institute, the National Aeronautics and Space Administration under Grant No. NNX08AR22G issued through the Planetary Science Division of the NASA Science Mission Directorate, the National Science Foundation Grant No. AST-1238877, the University of Maryland, Eotvos Lorand University (ELTE), the Los Alamos National Laboratory, and the Gordon and Betty Moore Foundation. This publication makes use of data products from the Two Micron All Sky Survey, which is a joint project of the University of Massachusetts and the Infrared Processing and Analysis Center/California Institute of Technology, funded by the National Aeronautics and Space Administration and the National Science Foundation. We are grateful to the anonymous referee for his/her a constructive report.

\end{acknowledgements}

\appendix
\onecolumn

\section{Supplementary tables and figures}

\small
\begin{longtable}{lllrrrr}
\caption{\label{table:line_param} Lines of SiO, CS, and SiS detected in this study.}\\\hline\hline
Star & Molecule & Transition & \multicolumn{1}{c}{$\nu_{\rm calc}$} & \multicolumn{1}{c}{$\nu_{\rm obs}$} & \multicolumn{1}{c}{$V_{\rm exp}$} & \multicolumn{1}{c}{$\int T_{\rm mb} dv$} \\
        &                &                  & \multicolumn{1}{c}{(MHz)}                 & \multicolumn{1}{c}{(MHz)}                & \multicolumn{1}{c}{(km s$^{-1}$)}   & \multicolumn{1}{c}{(K km s$^{-1}$)} \\
\hline
\endfirsthead
\caption{continued.}\\
\hline\hline
Star & Molecule & Transition & \multicolumn{1}{c}{$\nu_{\rm calc}$} & \multicolumn{1}{c}{$\nu_{\rm obs}$} & \multicolumn{1}{c}{$V_{\rm exp}$} & \multicolumn{1}{c}{$\int T_{\rm mb} dv$} \\
        &                &                  & \multicolumn{1}{c}{(MHz)}                 & \multicolumn{1}{c}{(MHz)}                & \multicolumn{1}{c}{(km s$^{-1}$)}   & \multicolumn{1}{c}{(K km s$^{-1}$)} \\
\hline
\endhead
\hline
\endfoot
     R\,Leo & SiO &   J=1-0 &  43423.844 &  -- &  -- &   3.9\,$^a$ \\
            & SiO &   J=2-1 &  86846.971 &  86847.0( 2) &  5.0( 2) &  16.1(16) \\
            & SiO &   J=6-5 & 260517.985 & 260517.8( 2) &  5.2( 2) &  43( 13) \\
            & SiO &   J=7-6 & 303926.783 & 303926.4( 3) &  4.9( 3) &  49( 15) \\
            &  CS &   J=5-4 & 244935.554 & 244935.2( 3) &  4.8( 3) &   0.41(12) \\
            &  CS &   J=6-5 & 293912.089 & 293911.9( 3) &  4.7( 3) &   0.59(18) \\
     R\,Cas & SiO &   J=1-0 &  43423.844 &  -- &  -- &  10.2\,$^a$ \\
            & SiO &   J=2-1 &  86846.971 &  -- &  -- &  28.3\,$^a$ \\
            & SiO &   J=6-5 & 260517.985 &  -- &  -- &  53.7\,$^a$ \\
            &  CS &   J=2-1 &  97980.952 &  97979.8( 8) &  7.1(10) &   0.10( 1) \\
            &  CS &   J=5-4 & 244935.554 & 244935.1( 6) &  7.8( 7) &   0.84(25) \\
            & SiS & J=14-13 & 254103.211 & 254102.5( 8) &  9.1(10) &   0.92(28) \\
    TX\,Cam & SiO &   J=1-0 &  43423.844 &  -- &  -- &  10.4\,$^a$ \\
            & SiO &   J=2-1 &  86846.971 &  86847.1( 2) & 18.5( 6) &  30.4(30) \\
            &  CS &   J=2-1 &  97980.952 &  97981.0( 1) & 17.1( 2) &   1.69(17) \\
    IK\,Tau & SiO &   J=1-0 &  43423.844 &  -- &  -- &  24.8\,$^a$ \\
 V1111\,Oph & SiO &   J=1-0 &  43423.844 &  -- &  -- &   6.4\,$^a$ \\
            & SiO &   J=2-1 &  86846.971 &  86847.0( 2) & 15.9( 3) &  12.8(13) \\
            & SiO &   J=6-5 & 260517.985 & 260517.7( 2) & 15.5( 3) &  19.4(58) \\
            & SiO &   J=7-6 & 303926.783 & 303926.0( 4) & 15.9( 6) &  24.5(73) \\
            &  CS &   J=2-1 &  97980.952 &  97981.6( 8) & 13.1( 8) &   0.32( 3) \\
            &  CS &   J=5-4 & 244935.554 & 244935.9( 5) & 13.9( 8) &   1.93(58) \\
            &  CS &   J=6-5 & 293912.089 & 293912.9(10) & 14.5(10) &   2.88(86) \\
            & SiS & J=14-13 & 254103.211 & 254103.5( 5) & 13.7( 9) &   3.00(90) \\
            & SiS & J=16-15 & 290380.744 & 290381.6(10) & 14.2(10) &   4.7(14) \\
            & SiS & J=17-16 & 308516.144 & 308515.5(10) & 13.3(10) &   4.3(13) \\
    GX\,Mon & SiO &   J=1-0 &  43423.844 &  -- &  -- &   9.9\,$^a$ \\
            & SiO &   J=2-1 &  86846.971 &  86847.1( 2) & 18.5( 6) &  23.1(23) \\
            & SiO &   J=6-5 & 260517.985 & 260517.6( 3) & 18.0( 5) &  29.1(87) \\
            & SiO &   J=7-6 & 303926.783 & 303926.4( 3) & 17.6( 6) &  31.7(95) \\
            &  CS &   J=2-1 &  97980.952 &  97981.0( 1) & 18.0( 2) &   0.75( 8) \\
            &  CS &   J=5-4 & 244935.554 & 244935.7( 3) & 16.6( 6) &   3.4(10) \\
            &  CS &   J=6-5 & 293912.089 & 293912.8(10) & 17.1(10) &   3.4(10) \\
            & SiS & J=14-13 & 254103.211 & 254103.2( 2) & 18.4( 6) &   4.2(12) \\
            & SiS & J=16-15 & 290380.744 & 290380.6( 3) & 16.3( 8) &   4.4(13) \\
            & SiS & J=17-16 & 308516.144 & 308517.1(10) & 16.4(10) &   4.0(12) \\
    NV\,Aur & SiO &   J=1-0 &  43423.844 &  -- &  -- &   5.1\,$^a$ \\
            & SiO &   J=2-1 &  86846.971 &  86846.9( 1) & 17.9( 3) &   7.61(76) \\
            & SiO &   J=6-5 & 260517.985 & 260517.3( 3) & 17.1( 3) &  12.7(38) \\
            &  CS &   J=2-1 &  97980.952 &  97981.0( 3) & 16.5( 5) &   0.31( 3) \\
            &  CS &   J=5-4 & 244935.554 & 244935.2( 5) & 16.5( 8) &   1.76(53) \\
            & SiS & J=14-13 & 254103.211 & 254102.4( 4) & 18.0( 4) &   5.3(16) \\
     Y\,CVn & SiO &   J=2-1 &  86846.971 &  86847.0( 2) &  6.7( 5) &   0.16( 2) \\
            & SiO &   J=7-6 & 303926.783 & 303928.2(15) &  7.1(10) &   1.08(33) \\
            &  CS &   J=2-1 &  97980.952 &  97981.0( 2) &  8.3( 4) &   1.80(18) \\
            &  CS &   J=6-5 & 293912.089 & 293912.2( 4) &  7.7( 5) &  10.3(31) \\
     R\,Lep & SiO &   J=1-0 &  43423.844 &  43423.9( 3) & 16.8(10) &   0.26( 3) \\
            & SiO &   J=2-1 &  86846.971 &  86846.4( 4) & 21.1(10) &   1.22(12) \\
            & SiO &   J=6-5 & 260517.985 & 260516.5( 8) & 18.6(10) &   8.2(25) \\
            &  CS &   J=2-1 &  97980.952 &  97980.4( 4) & 20.2(10) &   1.46(15) \\
            &  CS &   J=5-4 & 244935.554 & 244934.5( 4) & 19.0(10) &  14.0(42) \\
    LP\,And & SiO &   J=1-0 &  43423.844 &  43423.9( 1) & 13.1( 2) &   0.89( 9) \\
            & SiO &   J=2-1 &  86846.971 &  86846.8( 2) & 12.9( 4) &   3.09(31) \\
            & SiO &   J=6-5 & 260517.985 & 260517.5( 3) & 14.5( 4) &   8.5(26) \\
            &  CS &   J=2-1 &  97980.952 &  97980.9( 2) & 12.9( 4) &  14.2(14) \\
            &  CS &   J=5-4 & 244935.554 & 244935.4( 3) & 13.7( 4) &  31.2(94) \\
            & SiS &   J=2-1 &  36309.629 &  36309.7( 1) & 13.0( 2) &   0.30( 3) \\
            & SiS & J=14-13 & 254103.211 &  -- &  -- &  15.7\,$^a$ \\
IRC\,+20370 & SiO &   J=1-0 &  43423.844 &  43423.8( 2) & 14.1( 5) &   0.57( 6) \\
            & SiO &   J=2-1 &  86846.971 &  86846.9( 2) & 12.9( 5) &   3.08(31) \\
            & SiO &   J=6-5 & 260517.985 & 260517.7( 3) & 13.8( 4) &  17.3(52) \\
            & SiO &   J=7-6 & 303926.783 & 303926.4( 3) & 13.4( 4) &  19.4(58) \\
            &  CS &   J=2-1 &  97980.952 &  97981.0( 2) & 12.6( 8) &   7.74(77) \\
            &  CS &   J=5-4 & 244935.554 & 244935.2( 3) & 13.3( 7) &  45( 13) \\
            &  CS &   J=6-5 & 293912.089 & 293911.7( 4) & 14.1( 7) &  41( 12) \\
            & SiS & J=14-13 & 254103.211 &  -- &  -- &  16.8\,$^a$ \\
            & SiS & J=16-15 & 290380.744 & 290380.4( 4) & 12.9( 8) &  14.3(43) \\
            & SiS & J=17-16 & 308516.144 & 308516.0( 7) & 12.2(10) &  13.0(39) \\
IRC\,+30374 & SiO &   J=2-1 &  86846.971 &  86846.9( 2) & 21.9( 6) &   3.69(37) \\
            & SiO &   J=6-5 & 260517.985 & 260517.6( 4) & 26.1( 8) &  14.2(42) \\
            & SiO &   J=7-6 & 303926.783 & 303925.7( 6) & 26.0(10) &  18.6(56) \\
            &  CS &   J=2-1 &  97980.952 &  97980.6( 3) & 23.0( 8) &  15.6(16) \\
            &  CS &   J=5-4 & 244935.554 & 244935.0( 4) & 23.6( 8) &  42( 13) \\
            &  CS &   J=6-5 & 293912.089 & 293911.2( 5) & 25.1( 8) &  48( 14) \\
            & SiS & J=14-13 & 254103.211 & 254102.5( 7) & 21.7(10) &   5.4(16) \\
            & SiS & J=16-15 & 290380.744 & 290380.2( 6) & 23.8(10) &   7.8(23) \\
            & SiS & J=17-16 & 308516.144 & 308514.8(10) & 24.5(12) &   5.2(16) \\
IRC\,+10216 & SiO &   J=1-0 &  43423.844 &  43423.8( 1) & 14.4( 1) &  15.4(15) \\
            &  CS &   J=1-0 &  48990.957 &  48990.9( 1) & 14.0( 1) &  45.0(45) \\
            & SiS &   J=2-1 &  36309.629 &  36309.6( 1) & 14.1( 1) &   9.08(91) \\
     CRL190 & SiO &   J=2-1 &  86846.971 &  86846.8(10) & 13.1(10) &   0.19( 2) \\
            &  CS &   J=1-0 &  48990.957 &  48990.7( 3) & 13.7(10) &   0.83( 8) \\
            &  CS &   J=2-1 &  97980.952 &  97980.9( 2) & 14.7( 7) &   3.90(39) \\
            &  CS &   J=5-4 & 244935.554 & 244935.1( 5) & 14.1( 7) &   8.7(26) \\
            &  CS &   J=6-5 & 293912.089 & 293911.3( 5) & 14.2( 8) &   7.1(21) \\
            & SiS & J=14-13 & 254103.211 & 254103.3(10) & 12.8(10) &   1.67(50) \\
            & SiS & J=16-15 & 290380.744 & 290380.0( 8) & 13.6(10) &   2.06(62) \\
\end{longtable}
\hspace{2.3cm} \tablenotea{$^a$ Line is not fitted due to a complex line profile. Only velocity-integrated intensity is given.} \\

\clearpage

\begin{figure*}
\centering
\includegraphics[angle=0,width=0.99\textwidth]{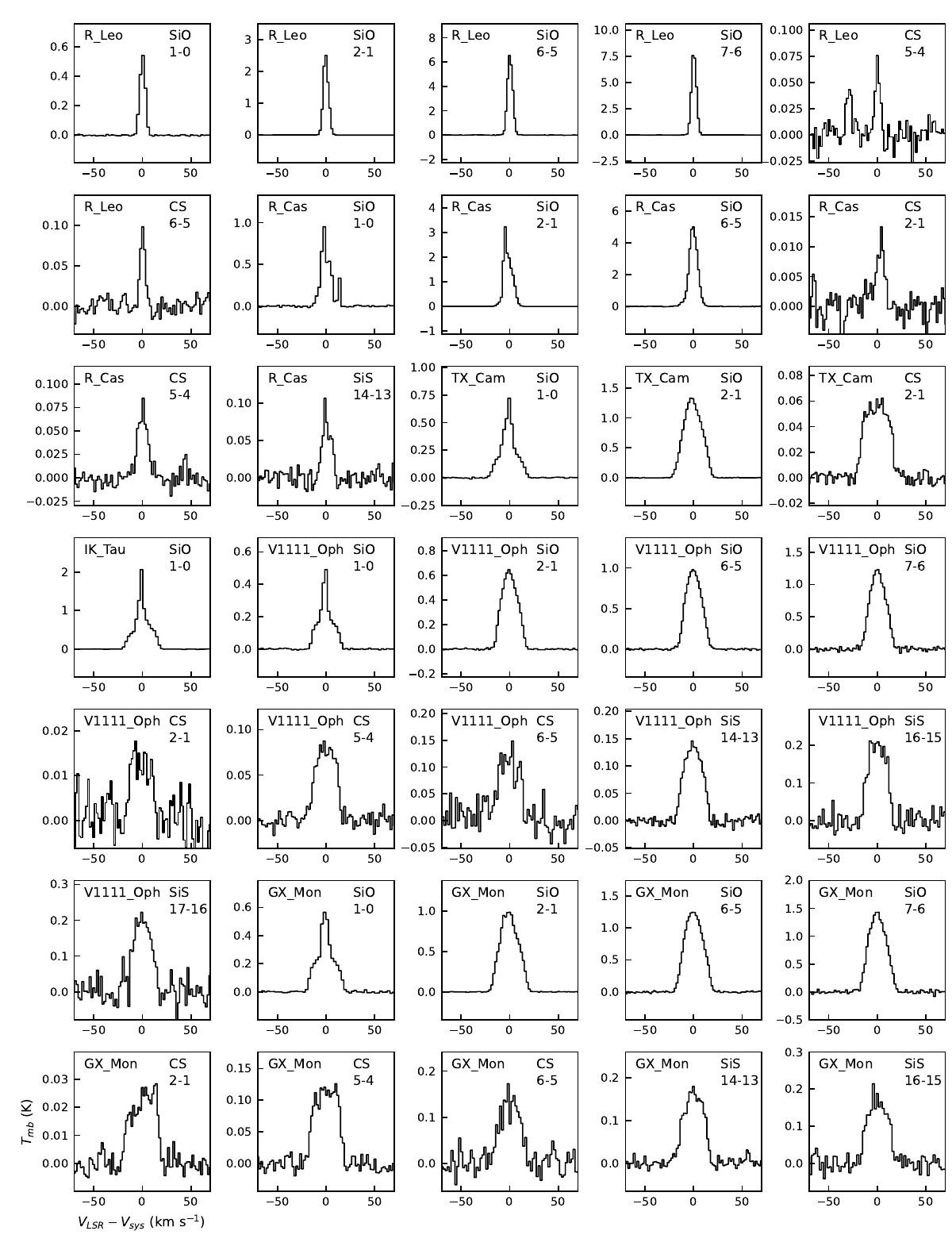}
\caption{Lines of SiO, CS, and SiS detected in this study (see parameters in Table\,\ref{table:line_param}) ordered from top to bottom and from left to right. The velocity resolution has been smoothed to $\sim$\,2 km s$^{-1}$ for a better visualization.} \label{fig:lines}
\end{figure*}

\setcounter{figure}{0}
\begin{figure*}
\centering
\includegraphics[angle=0,width=0.99\textwidth]{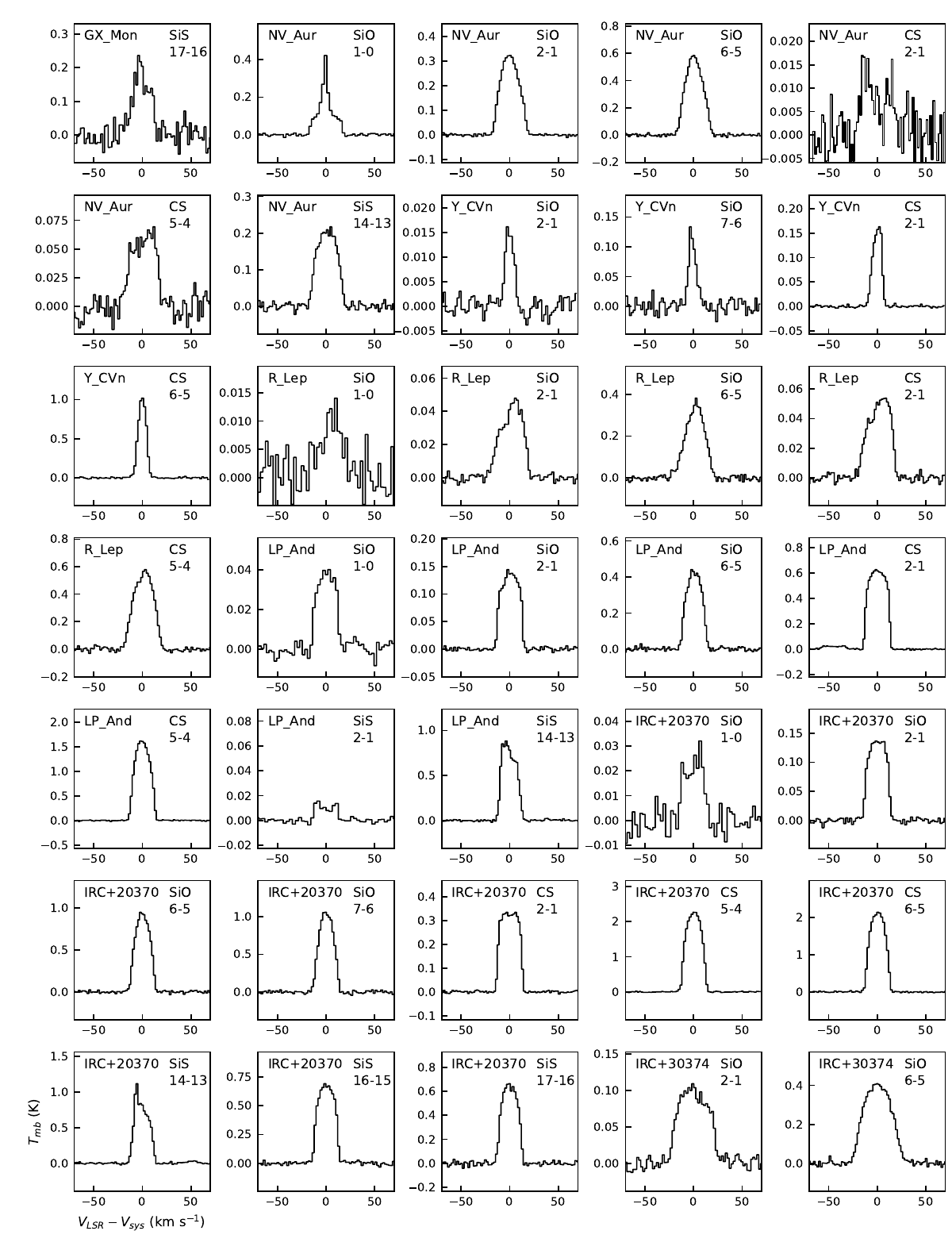}
\caption{continued.}
\end{figure*}

\setcounter{figure}{0}
\begin{figure*}
\centering
\includegraphics[angle=0,width=0.99\textwidth]{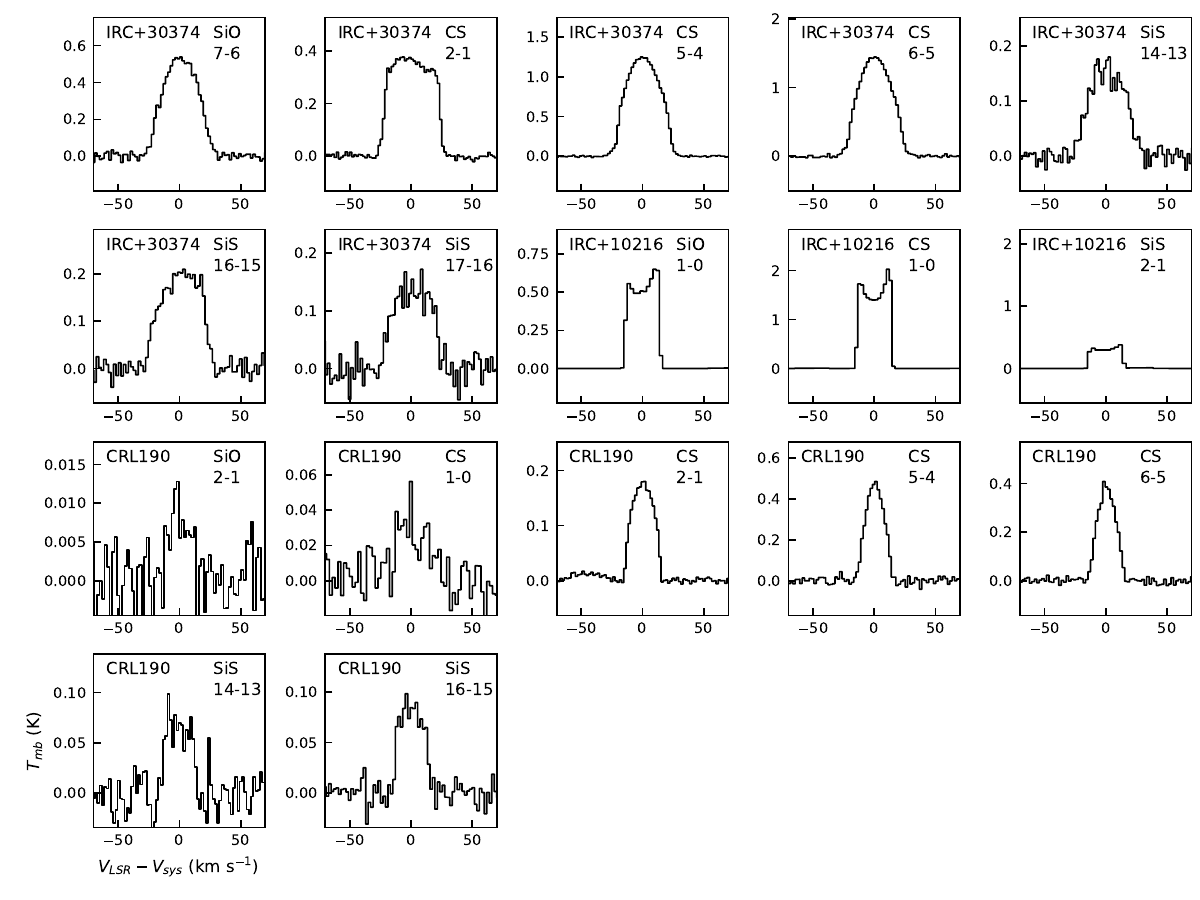}
\caption{continued.}
\end{figure*}

\clearpage

\scriptsize
\begin{longtable}{rrl c rrl c rrl c rrl}
\caption{\label{table:sed} Continuum flux data.}\\
\hline\hline
\multicolumn{1}{c}{$\lambda$} & \multicolumn{1}{c}{Flux} & \multicolumn{1}{l}{Catalog} & & \multicolumn{1}{c}{$\lambda$} & \multicolumn{1}{c}{Flux} & \multicolumn{1}{l}{Catalog} & & \multicolumn{1}{c}{$\lambda$} & \multicolumn{1}{c}{Flux} & \multicolumn{1}{l}{Catalog} & & \multicolumn{1}{c}{$\lambda$} & \multicolumn{1}{c}{Flux} & \multicolumn{1}{l}{Catalog} \\
\multicolumn{1}{c}{($\mu$m)} & \multicolumn{1}{c}{(Jy)} & \multicolumn{1}{c}{} & & \multicolumn{1}{c}{($\mu$m)} & \multicolumn{1}{c}{(Jy)} & \multicolumn{1}{c}{} & & \multicolumn{1}{c}{($\mu$m)} & \multicolumn{1}{c}{(Jy)} & \multicolumn{1}{c}{} & & \multicolumn{1}{c}{($\mu$m)} & \multicolumn{1}{c}{(Jy)} & \multicolumn{1}{c}{} \\ 
\hline
\endfirsthead
\caption{continued.}\\
\hline\hline
\multicolumn{1}{c}{$\lambda$} & \multicolumn{1}{c}{Flux} & \multicolumn{1}{l}{Catalog} & & \multicolumn{1}{c}{$\lambda$} & \multicolumn{1}{c}{Flux} & \multicolumn{1}{l}{Catalog} & & \multicolumn{1}{c}{$\lambda$} & \multicolumn{1}{c}{Flux} & \multicolumn{1}{l}{Catalog} & & \multicolumn{1}{c}{$\lambda$} & \multicolumn{1}{c}{Flux} & \multicolumn{1}{l}{Catalog} \\
\multicolumn{1}{c}{($\mu$m)} & \multicolumn{1}{c}{(Jy)} & \multicolumn{1}{c}{} & & \multicolumn{1}{c}{($\mu$m)} & \multicolumn{1}{c}{(Jy)} & \multicolumn{1}{c}{} & & \multicolumn{1}{c}{($\mu$m)} & \multicolumn{1}{c}{(Jy)} & \multicolumn{1}{c}{} & & \multicolumn{1}{c}{($\mu$m)} & \multicolumn{1}{c}{(Jy)} & \multicolumn{1}{c}{} \\ 
\hline
\endhead
\hline
\endfoot
\multicolumn{15}{c}{} \\
\multicolumn{3}{c}{R\,Leo} & & \multicolumn{3}{c}{R\,Cas} & & \multicolumn{3}{c}{TX\,Cam} & & \multicolumn{3}{c}{IK\,Tau} \\
\cline{1-3} \cline{5-7} \cline{9-11} \cline{13-15}
 1.24       & 7940       & 2MASS      & & 1.24       & 1360       & 2MASS      & & 1.24       & 205        & 2MASS      & & 0.96       & 55         & PAN-STARRS \\
 1.26       & 4280       & DIRBE      & & 1.26       & 2090       & DIRBE      & & 1.26       & 380        & DIRBE      & & 1.24       & 325        & 2MASS      \\
 1.65       & 5280       & 2MASS      & & 1.65       & 2300       & 2MASS      & & 1.65       & 495        & 2MASS      & & 1.65       & 981        & 2MASS      \\
 2.16       & 5610       & 2MASS      & & 2.16       & 2460       & 2MASS      & & 2.16       & 683        & 2MASS      & & 2.16       & 1600       & 2MASS      \\
 2.22       & 7280       & DIRBE      & & 2.22       & 3490       & DIRBE      & & 2.22       & 981        & DIRBE      & & 2.22       & 1110       & DIRBE      \\
 3.35       & 3680       & unWISE     & & 3.35       & 1880       & unWISE     & & 3.35       & 884        & unWISE     & & 3.52       & 1530       & DIRBE      \\
 3.52       & 5640       & DIRBE      & & 3.52       & 2590       & DIRBE      & & 3.52       & 901        & DIRBE      & & 4.60       & 1900       & unWISE     \\
 4.60       & 4020       & unWISE     & & 4.60       & 2250       & unWISE     & & 4.60       & 1160       & unWISE     & & 4.89       & 1740       & DIRBE      \\
 11.6       & 2070       & IRAS       & & 4.89       & 1750       & DIRBE      & & 4.89       & 675        & DIRBE      & & 11.6       & 4630       & IRAS       \\
 22.1       & 673        & WISE       & & 11.6       & 1260       & IRAS       & & 11.6       & 514        & allWISE    & & 22.1       & 1870       & WISE       \\
 23.9       & 625        & IRAS       & & 22.1       & 706        & WISE       & & 18.4       & 508        & AKARI      & & 23.9       & 2380       & IRAS       \\
 61.8       & 114        & IRAS       & & 23.9       & 543        & IRAS       & & 22.1       & 285        & allWISE    & & 61.8       & 332        & IRAS       \\
 102        & 39.3       & IRAS       & & 61.8       & 102        & IRAS       & & 23.9       & 635        & IRAS       & & 102        & 103        & IRAS       \\
            &            &            & & 65         & 82.7       & AKARI      & & 61.8       & 134        & IRAS       & & 855        & 0.25       & SCUBA      \\
            &            &            & & 90         & 55.4       & AKARI      & & 102        & 38.6       & IRAS       & &            &            &            \\
            &            &            & & 102        & 38.8       & IRAS       & &            &            &            & &            &            &            \\
            &            &            & & 140        & 10.8       & AKARI      & &            &            &            & &            &            &            \\
            &            &            & & 160        & 6.78       & AKARI      & &            &            &            & &            &            &            \\
\multicolumn{15}{c}{} \\
\multicolumn{3}{c}{V1111\,Oph} & & \multicolumn{3}{c}{GX\,Mon} & & \multicolumn{3}{c}{NV\,Aur} & & \multicolumn{3}{c}{Y\,CVn} \\
\cline{1-3} \cline{5-7} \cline{9-11} \cline{13-15}
 0.96       & 8.73       & PAN-STARRS & & 0.96       & 10.2       & PAN-STARRS & &  0.96       & 0.32       & PAN-STARRS & &  1.24       & 634        & 2MASS    \\
 1.24       & 35.4       & 2MASS      & & 1.24       & 49.2       & 2MASS      & &  1.24       & 0.833      & 2MASS      & &  1.26       & 704        & DIRBE    \\
 1.65       & 125        & 2MASS      & & 1.65       & 123        & 2MASS      & &  1.26       & 7.0        & DIRBE      & &  1.65       & 1360       & 2MASS    \\
 2.16       & 239        & 2MASS      & & 2.16       & 201        & 2MASS      & &  1.65       & 5.14       & 2MASS      & &  2.16       & 1330       & 2MASS    \\
 3.35       & 370        & unWISE     & & 3.35       & 428        & unWISE     & &  2.16       & 19.0       & 2MASS      & &  2.22       & 1240       & DIRBE    \\
 4.60       & 648        & unWISE     & & 4.60       & 692        & unWISE     & &  2.22       & 32.0       & DIRBE      & &  3.35       & 934        & unWISE   \\
 11.6       & 452        & allWISE    & & 8.61       & 275        & AKARI      & &  3.35       & 70.4       & unWISE     & &  3.52       & 931        & DIRBE    \\
 18.4       & 483        & AKARI      & & 11.6       & 387        & allWISE    & &  3.52       & 117        & DIRBE      & &  4.60       & 835        & unWISE   \\
 22.1       & 388        & WISE       & & 18.4       & 260        & AKARI      & &  4.89       & 168        & DIRBE      & &  4.89       & 336        & DIRBE    \\
 23.9       & 318        & IRAS       & & 22.1       & 172        & allWISE    & &  8.61       & 259        & AKARI      & &  8.61       & 304        & AKARI    \\
 61.8       & 66.0       & IRAS       & & 23.9       & 360        & IRAS       & &  11.6       & 177        & allWISE    & &  11.6       & 193        & WISE     \\
 65         & 65.6       & AKARI      & & 61.8       & 106        & IRAS       & &  18.4       & 349        & AKARI      & &  18.4       & 91.7       & AKARI    \\
 90         & 36.2       & AKARI      & & 65         & 76.5       & AKARI      & &  22.1       & 157        & allWISE    & &  22.1       & 57         & WISE     \\
 102        & 22.7       & IRAS       & & 90         & 43.8       & AKARI      & &  23.9       & 274        & IRAS       & &  23.9       & 70.3       & IRAS     \\
 140        & 9.55       & AKARI      & & 102        & 40.4       & IRAS       & &  61.8       & 72.4       & IRAS       & &  61.8       & 17.2       & IRAS     \\
 160        & 6.14       & AKARI      & & 140        & 16.5       & AKARI      & &  102        & 23.0       & IRAS       & &  102        & 7.82       & IRAS     \\
            &            &            & & 160        & 8.0        & AKARI      & &             &            &            & &             &            &          \\
\multicolumn{15}{c}{} \\
\multicolumn{3}{c}{R\,Lep} & & \multicolumn{3}{c}{LP\,And} & & \multicolumn{3}{c}{IRC\,+20370} & & \multicolumn{3}{c}{IRC\,+30374} \\
\cline{1-3} \cline{5-7} \cline{9-11} \cline{13-15}
 1.24       & 216        & 2MASS      & & 1.24       & 0.223      & 2MASS      & & 1.24       & 8.18       & 2MASS      & & 1.24       & 0.406      & 2MASS      \\
 1.26       & 130        & DIRBE      & & 1.65       & 3.01       & 2MASS      & & 1.65       & 44.0       & 2MASS      & & 1.65       & 3.54       & 2MASS      \\
 1.65       & 465        & 2MASS      & & 2.16       & 19.3       & 2MASS      & & 2.16       & 134        & 2MASS      & & 2.16       & 16.9       & 2MASS      \\
 2.16       & 616        & 2MASS      & & 3.35       & 104        & unWISE     & & 2.22       & 186        & DIRBE      & & 3.35       & 154        & unWISE     \\
 3.35       & 689        & unWISE     & & 4.60       & 603        & unWISE     & & 3.35       & 349        & unWISE     & & 4.29       & 291        & MSX        \\
 3.52       & 549        & DIRBE      & & 11.6       & 445        & allWISE    & & 3.52       & 447        & DIRBE      & & 4.35       & 255        & MSX        \\
 4.60       & 1080       & unWISE     & & 18.4       & 582        & AKARI      & & 4.60       & 846        & allWISE    & & 4.60       & 457        & unWISE     \\
 4.89       & 493        & DIRBE      & & 22.1       & 207        & allWISE    & & 4.89       & 653        & DIRBE      & & 8.61       & 822        & AKARI      \\
 11.6       & 279        & WISE       & & 61.8       & 112        & IRAS       & & 8.61       & 433        & AKARI      & & 11.6       & 325        & DIRBE      \\
 18.4       & 140        & AKARI      & & 102        & 32.4       & IRAS       & & 11.6       & 283        & WISE       & & 12.1       & 361        & MSX        \\
 22.1       & 123        & WISE       & &            &            &            & & 18.4       & 296        & AKARI      & & 14.6       & 240        & MSX        \\
 23.9       & 113        & IRAS       & &            &            &            & & 22.1       & 231        & WISE       & & 18.4       & 186        & AKARI      \\
 61.8       & 24.9       & IRAS       & &            &            &            & & 23.9       & 293        & IRAS       & & 21.3       & 156        & MSX        \\
 102        & 9.69       & IRAS       & &            &            &            & & 61.8       & 54.3       & IRAS       & & 23.9       & 170        & DIRBE      \\
            &            &            & &            &            &            & & 65         & 63.9       & AKARI      & & 61.8       & 39         & DIRBE      \\
            &            &            & &            &            &            & & 90         & 24.4       & AKARI      & & 65         & 30.7       & AKARI      \\
            &            &            & &            &            &            & & 102        & 20.3       & IRAS       & & 90         & 17.1       & AKARI      \\
            &            &            & &            &            &            & & 140        & 8.27       & AKARI      & & 102        & 13.5       & DIRBE      \\
            &            &            & &            &            &            & & 160        & 4.09       & AKARI      & & 140        & 4.91       & AKARI      \\
            &            &            & &            &            &            & &            &            &            & & 160        & 4.48       & AKARI      \\
\multicolumn{3}{c}{IRC\,+10216} & & \multicolumn{3}{c}{CRL\,190} & & \multicolumn{3}{c}{} & & \multicolumn{3}{c}{} \\
\cline{1-3} \cline{5-7}
 0.96       & 0.356      & ATLAS      & & 3.35       & 0.323      & allWISE    & &            &            &            & &            &            &            \\
 0.96       & 0.587      & PAN-STARRS & & 4.60       & 4.34       & unWISE     & &            &            &            & &            &            &            \\
 1.24       & 2.67       & 2MASS      & & 8.28       & 56.5       & MSX        & &            &            &            & &            &            &            \\
 1.65       & 76.9       & 2MASS      & & 8.61       & 75.5       & AKARI      & &            &            &            & &            &            &            \\
 2.16       & 475        & 2MASS      & & 11.6       & 110        & allWISE    & &            &            &            & &            &            &            \\
 3.35       & 1330       & unWISE     & & 12.1       & 115        & MSX        & &            &            &            & &            &            &            \\
 4.60       & 9000       & unWISE     & & 14.6       & 138        & MSX        & &            &            &            & &            &            &            \\
 11.6       & 47500      & DIRBE      & & 18.4       & 197        & AKARI      & &            &            &            & &            &            &            \\
 23.9       & 23100      & DIRBE      & & 21.3       & 146        & MSX        & &            &            &            & &            &            &            \\
 61.8       & 5650       & DIRBE      & & 22.1       & 149        & allWISE    & &            &            &            & &            &            &            \\
 102        & 922        & DIRBE      & & 23.9       & 206        & IRAS       & &            &            &            & &            &            &            \\
 350        & 68         & PLANCK     & & 61.8       & 64.7       & IRAS       & &            &            &            & &            &            &            \\
 550        & 24.4       & PLANCK     & & 102        & 15.9       & IRAS       & &            &            &            & &            &            &            \\
 849        & 8.34       & PLANCK     & &            &            &            & &            &            &            & &            &            &            \\
 1380       & 3.85       & PLANCK     & &            &            &            & &            &            &            & &            &            &            \\
\end{longtable}
\hspace{0.8cm} \tablenoted{\scriptsize{Fluxes obtained from the VizieR database \citep{Ochsenbein2000} at \texttt{http://vizier.cds.unistra.fr/vizier/sed/}.\\ References: IRAS \citep{Beichman1988}, PAN-STARSS \citep{Chambers2019,Flewelling2019}, MSX \citep{Price2001,Egan2003}, AKARI \citep{Ishihara2010}; unWISE \citep{Lang2014}; allWISE \citep{Mainzer2011}; WISE \citep{Mainzer2011,Meisner2016,Meisner2017,Wright2010}, DIRBE \citep{Price2010,Smith2004}, 2MASS \citep{Skrutskie2006}.}}

\clearpage

\begin{figure*}
\centering
\includegraphics[angle=0,width=0.88\textwidth]{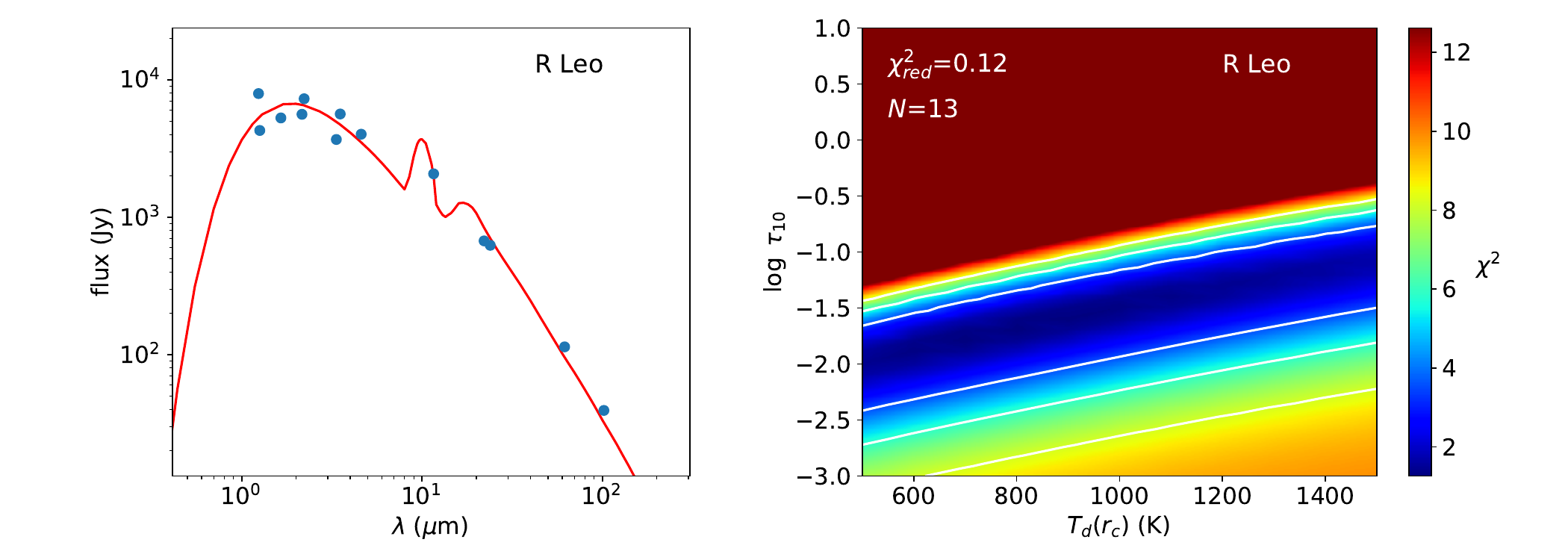} \includegraphics[angle=0,width=0.88\textwidth]{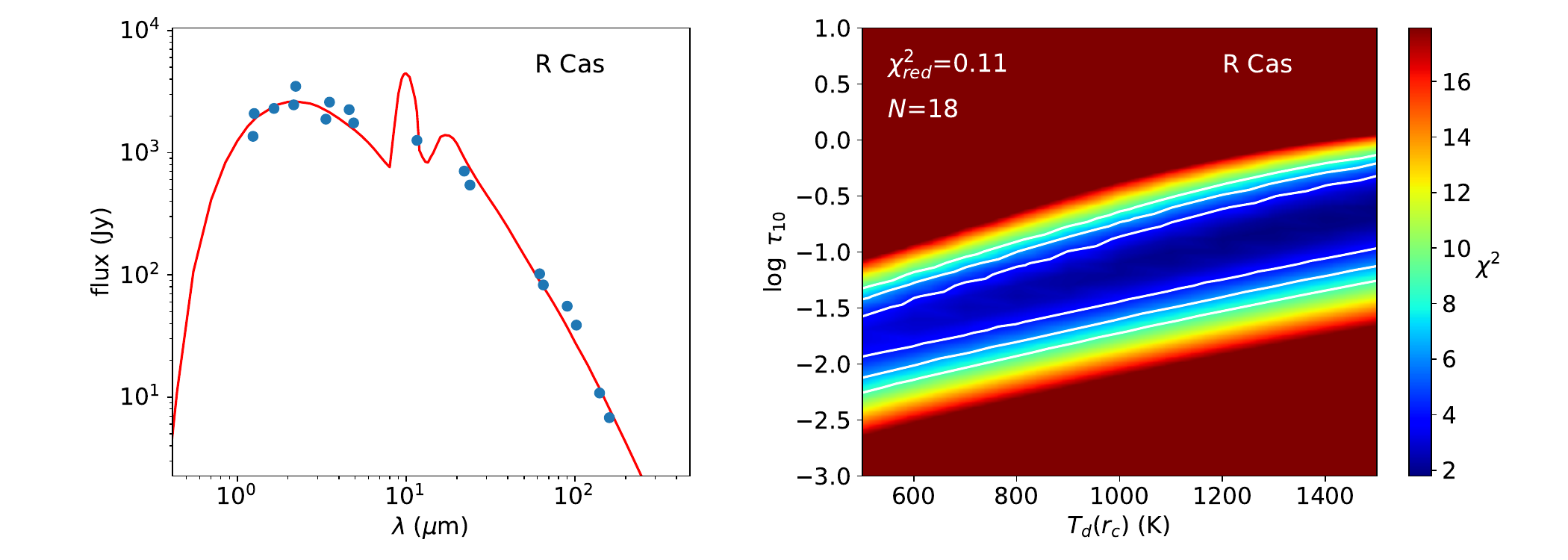} \includegraphics[angle=0,width=0.88\textwidth]{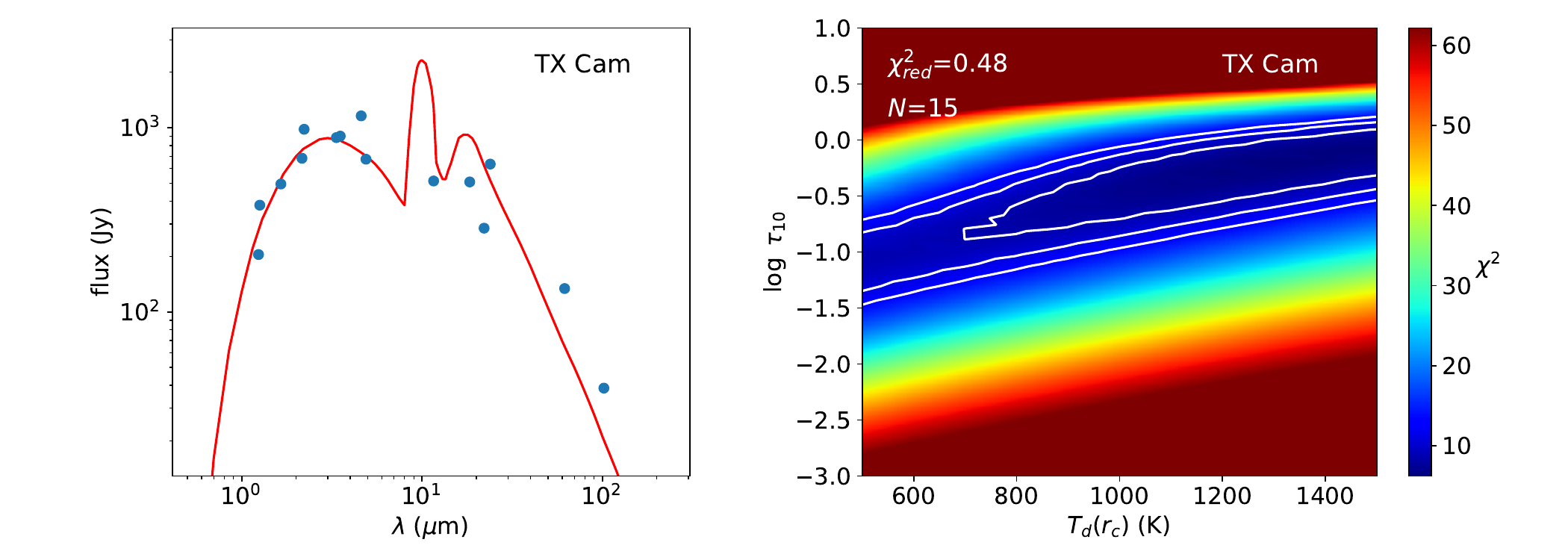} \includegraphics[angle=0,width=0.88\textwidth]{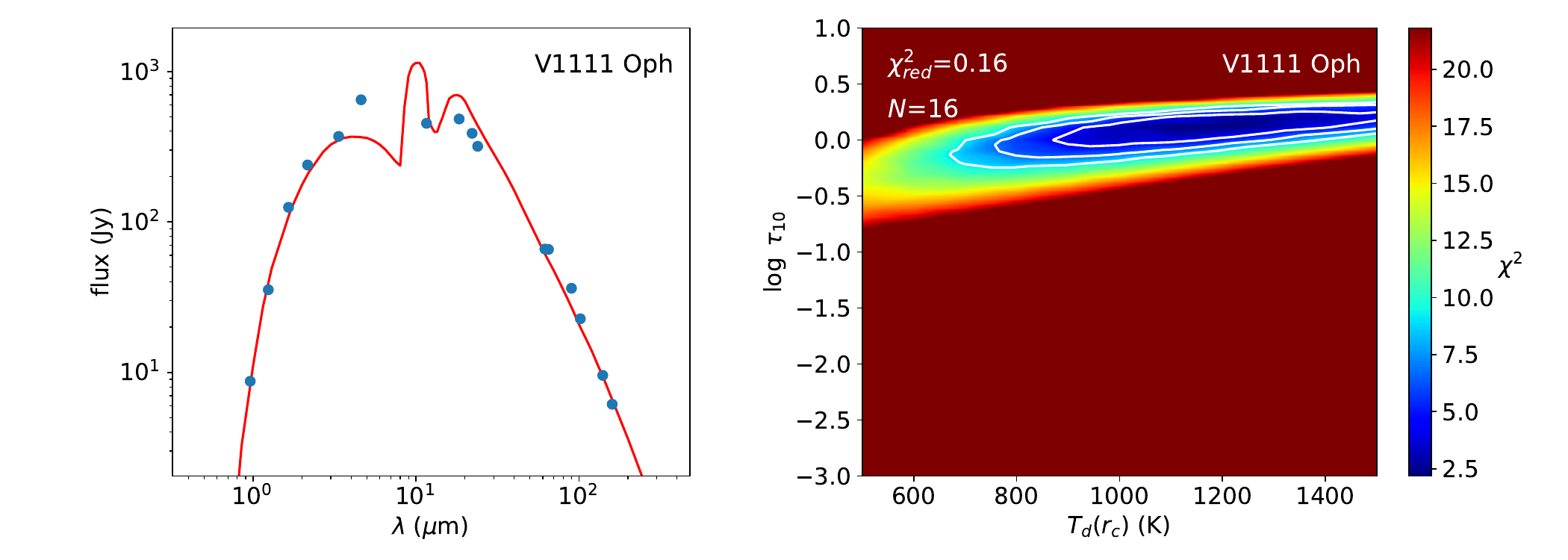}
\caption{Results from SED analysis for all envelopes except IK\,Tau and IRC\,+10216 (shown in Fig.\,\ref{fig:sed_iktau_irc10216}). The left panels show the observed fluxes in blue (see Table\,\ref{table:sed}) and the calculated SED from the best dusty model in red. The right panels show $\chi^2$ as a function of the temperature at the dust condensation radius, $T_c$, and the logarithm of the optical depth at 10 $\mu$m, $\log \tau_{10}$. The white contours correspond to 1, 2, and 3\,$\sigma$ levels.} \label{fig:sed}
\end{figure*}

\setcounter{figure}{1}
\begin{figure*}
\centering
\includegraphics[angle=0,width=0.88\textwidth]{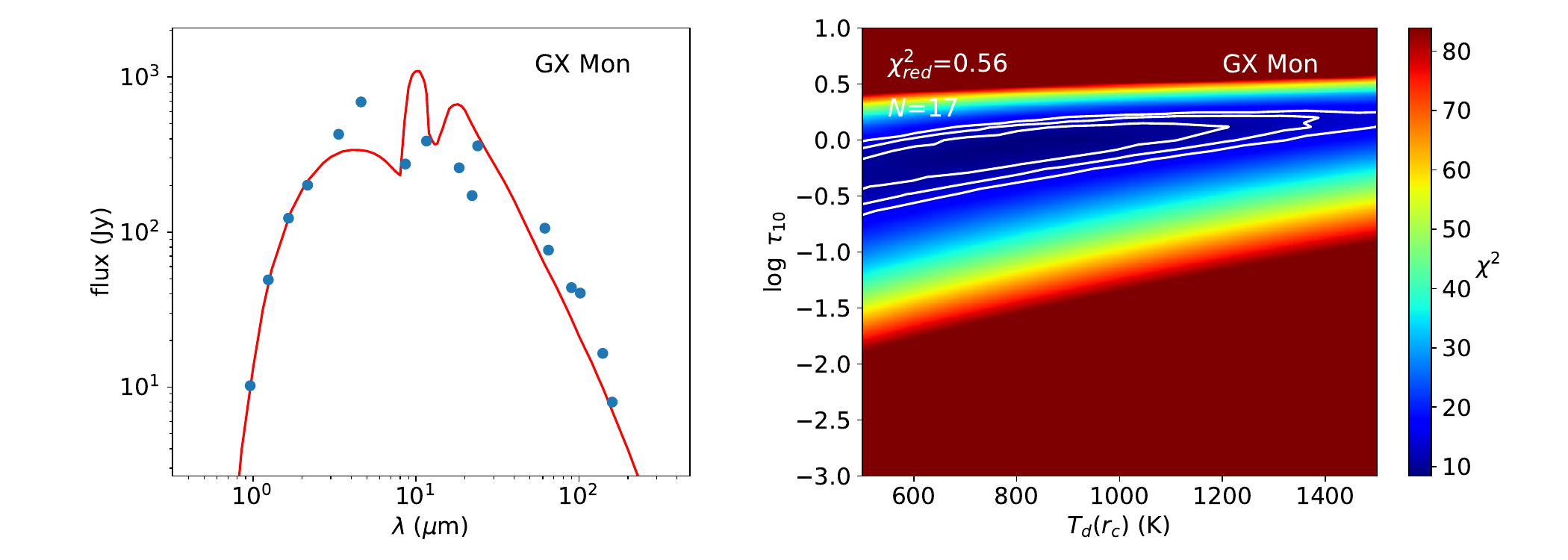} \includegraphics[angle=0,width=0.88\textwidth]{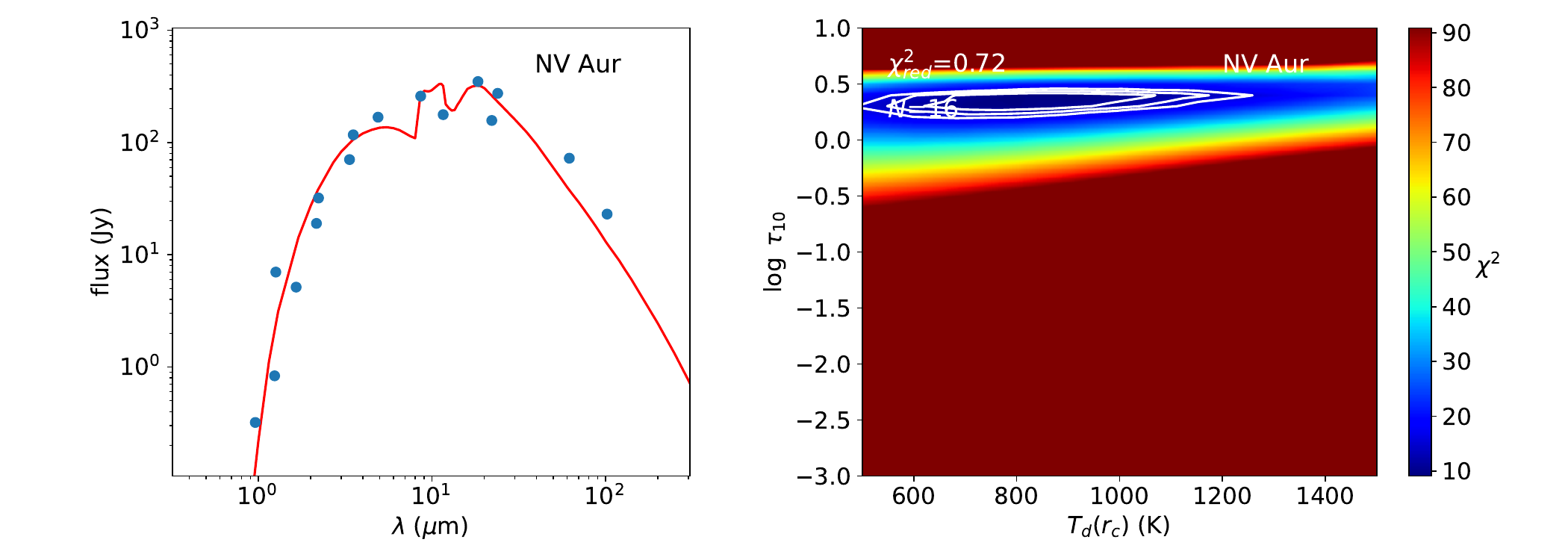} \includegraphics[angle=0,width=0.88\textwidth]{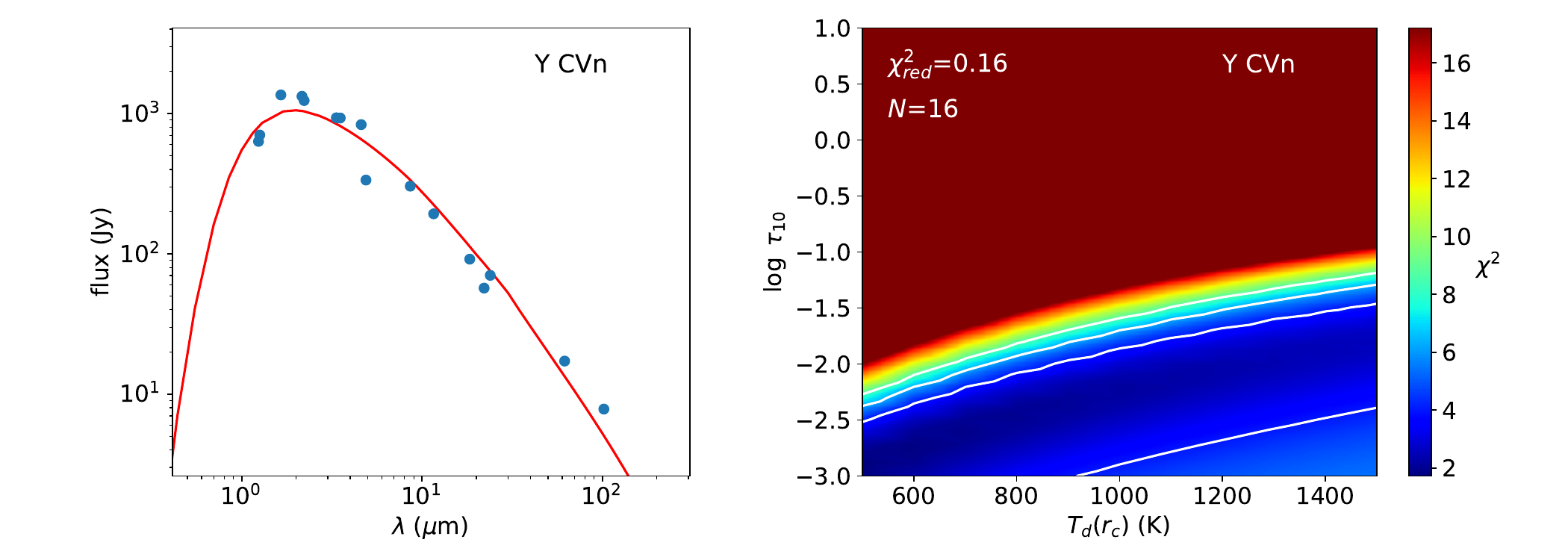} \includegraphics[angle=0,width=0.88\textwidth]{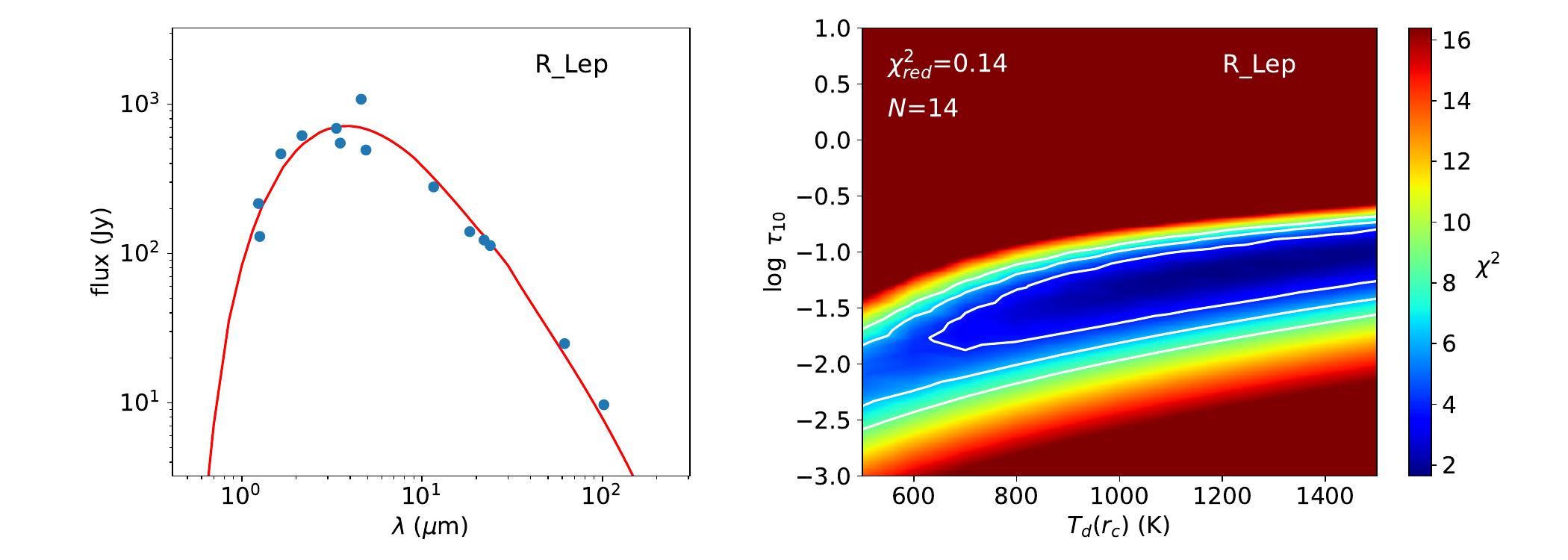} 
\caption{continued.}
\end{figure*}

\setcounter{figure}{1}
\begin{figure*}
\centering
\includegraphics[angle=0,width=0.88\textwidth]{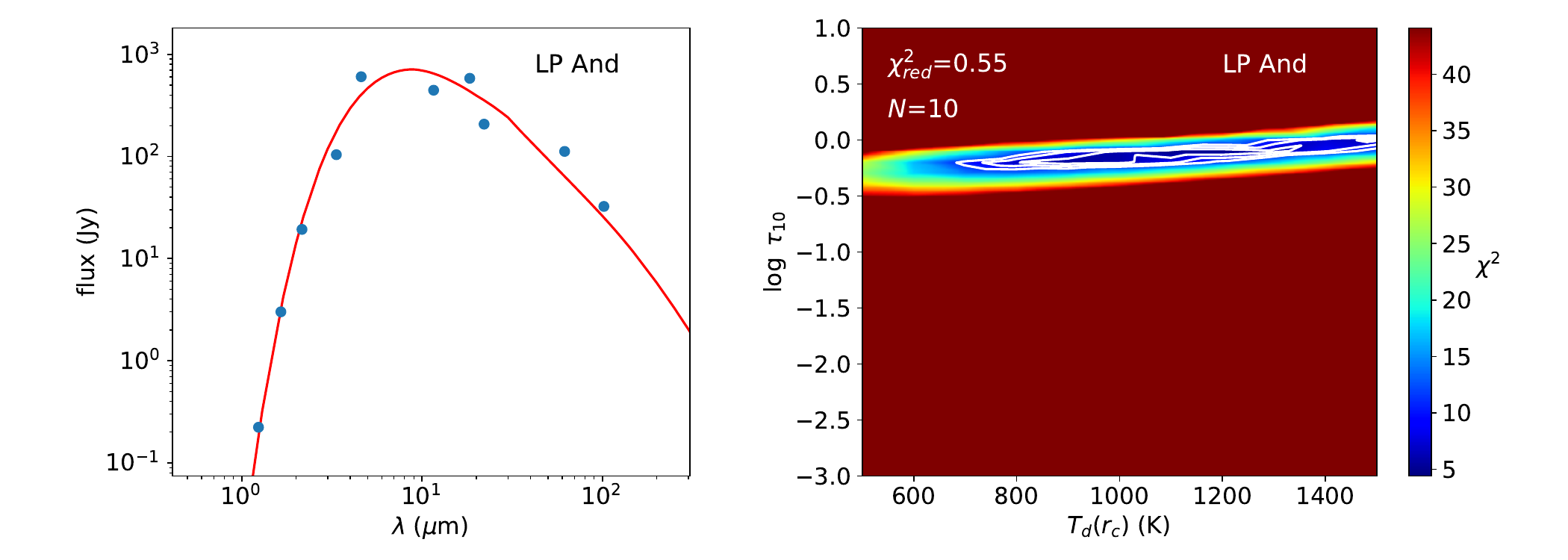} \includegraphics[angle=0,width=0.88\textwidth]{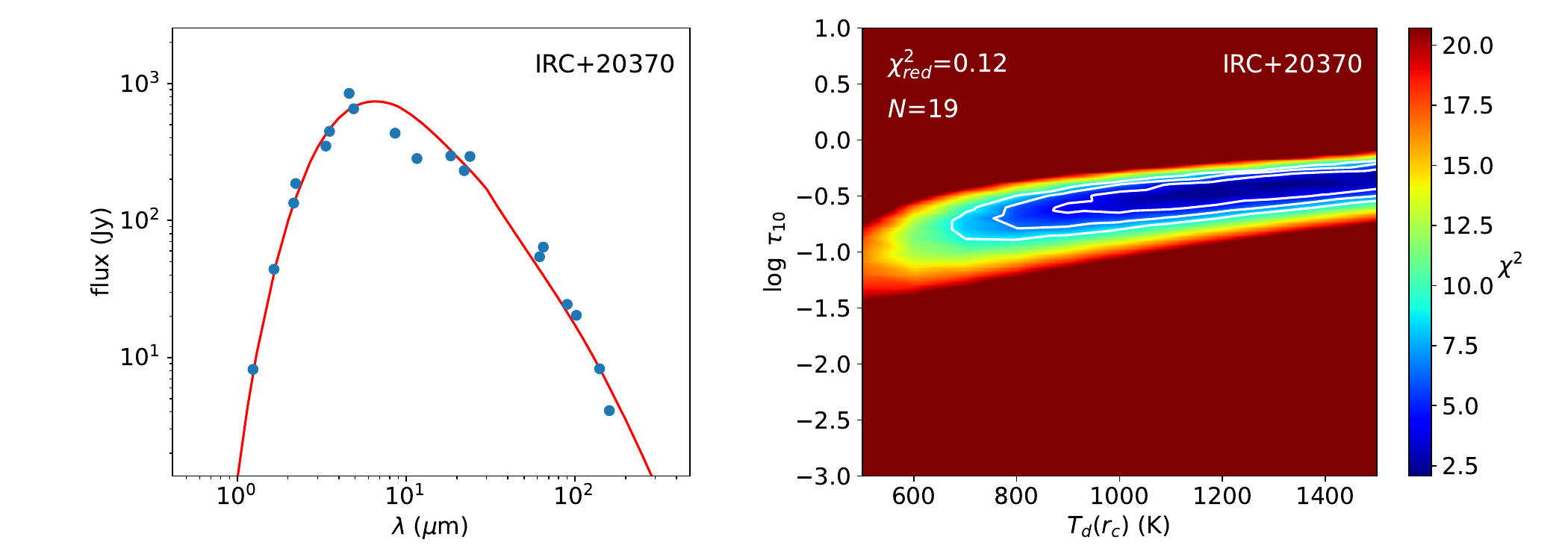} \includegraphics[angle=0,width=0.88\textwidth]{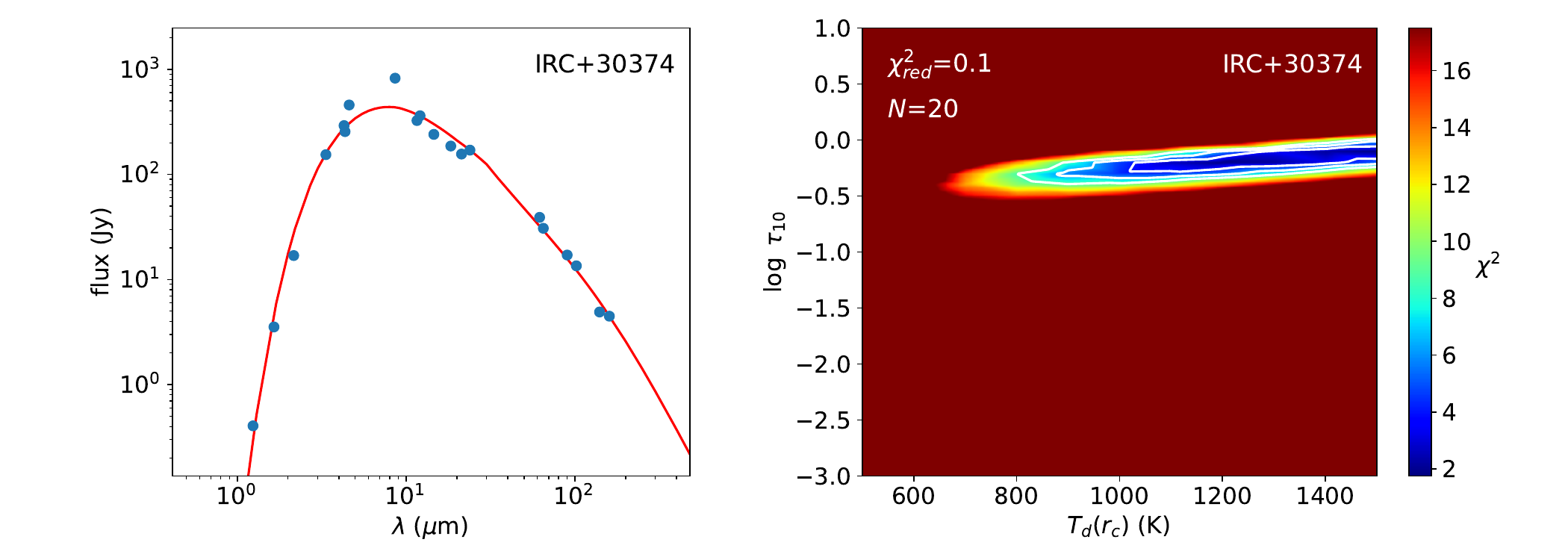} \includegraphics[angle=0,width=0.88\textwidth]{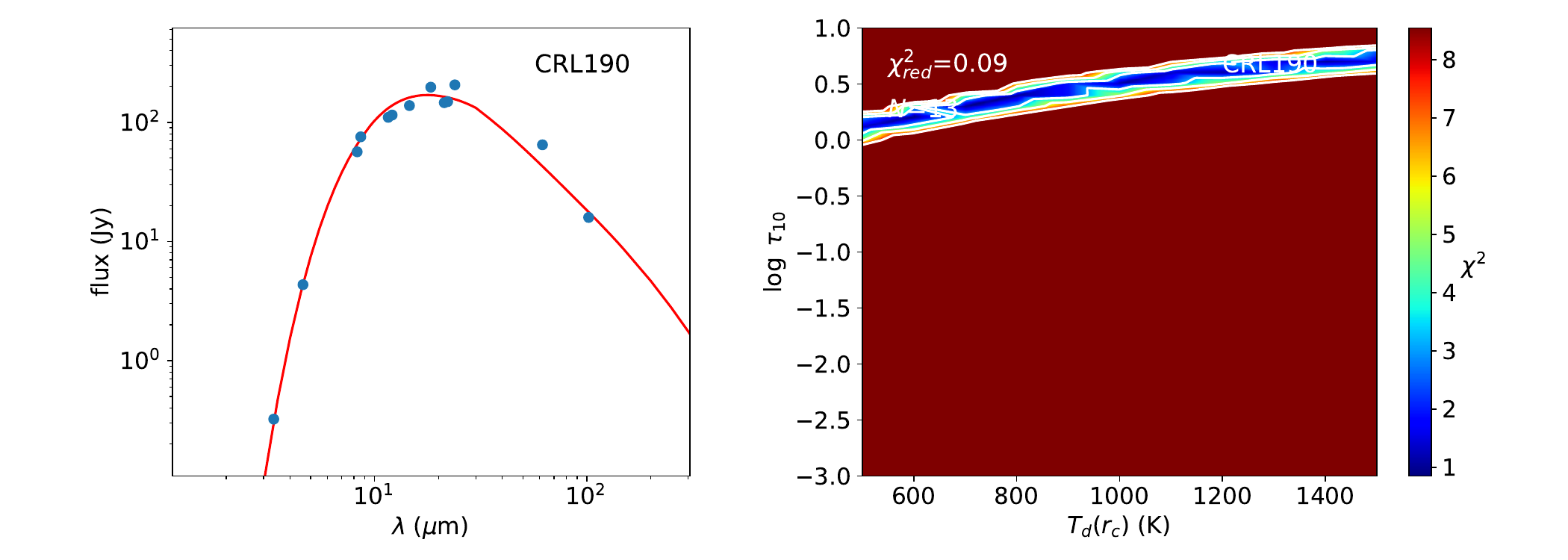} 
\caption{continued.}
\end{figure*}

\clearpage

\small
\begin{longtable}{rclrcrclrcrclr}
\caption{\label{table:co} Summary of CO observations included in the analysis.} \\
\hline\hline
\multicolumn{1}{c}{Line} & \multicolumn{1}{c}{$\int T_{\rm mb} dv$} & \multicolumn{2}{l}{Telescope} & & \multicolumn{1}{c}{Line} & \multicolumn{1}{c}{$\int T_{\rm mb} dv$} & \multicolumn{2}{l}{Telescope} & & \multicolumn{1}{c}{Line} & \multicolumn{1}{c}{$\int T_{\rm mb} dv$} & \multicolumn{2}{l}{Telescope} \\
 & \multicolumn{1}{c}{(K km s$^{-1}$)} & \multicolumn{2}{r}{Reference} & & & \multicolumn{1}{c}{(K km s$^{-1}$)} & \multicolumn{2}{r}{Reference} & & & \multicolumn{1}{c}{(K km s$^{-1}$)} & \multicolumn{2}{r}{Reference} \\
\hline
\endfirsthead
\caption{continued.}\\
\hline\hline
\multicolumn{1}{c}{Line} & \multicolumn{1}{c}{$\int T_{\rm mb} dv$} & \multicolumn{2}{l}{Telescope} & & \multicolumn{1}{c}{Line} & \multicolumn{1}{c}{$\int T_{\rm mb} dv$} & \multicolumn{2}{l}{Telescope} & & \multicolumn{1}{c}{Line} & \multicolumn{1}{c}{$\int T_{\rm mb} dv$} & \multicolumn{2}{l}{Telescope} \\
 & \multicolumn{1}{c}{(K km s$^{-1}$)} & \multicolumn{2}{r}{Reference} & & & \multicolumn{1}{c}{(K km s$^{-1}$)} & \multicolumn{2}{r}{Reference} & & & \multicolumn{1}{c}{(K km s$^{-1}$)} & \multicolumn{2}{r}{Reference} \\
\hline
\endhead
\hline
\endfoot
\multicolumn{4}{c}{R\,Leo} & & \multicolumn{4}{c}{R\,Cas} & & \multicolumn{4}{c}{TX\,Cam} \\
\hline
$J$\,=\,1-0          & 4.1                  & IRAM                 & (1)                  & & $J$\,=\,1-0          & 29.1                 & IRAM                 & (3)                  & & $J$\,=\,1-0          & 20.0                 & OSO                  & (5)                  \\
$J$\,=\,2-1          & 28.5                 & IRAM                 & (1)                  & & $J$\,=\,2-1          & 56.5                 & JCMT                 & (2)                  & & $J$\,=\,2-1          & 61.0                 & JCMT                 & (5)                  \\
$J$\,=\,3-2          & 33.5                 & APEX                 & (2)                  & & $J$\,=\,3-2          & 99.0                 & JCMT                 & (2)                  & & $J$\,=\,3-2          & 71.0                 & JCMT                 & (5)                  \\
$J$\,=\,4-3          & 28.1                 & CSO                  & (1)                  & & $J$\,=\,4-3          & 110.0                & JCMT                 & (2)                  & & $J$\,=\,4-3          & 149.0                & JCMT                 & (5)                  \\
$J$\,=\,5-4          & 5.86                 & HIFI                 & (1)                  & & $J$\,=\,6-5          & 14.7                 & HIFI                 & (4)                  & & $J$\,=\,6-5          & 14.2                 & HIFI                 & (4)                  \\
$J$\,=\,6-5          & 38.1                 & CSO                  & (2)                  & & $J$\,=\,10-9         & 11.8                 & HIFI                 & (4)                  & & $J$\,=\,10-9         & 9.7                  & HIFI                 & (4)                  \\
$J$\,=\,9-8          & 8.35                 & HIFI                 & (1)                  & & $J$\,=\,16-15        & 5.5                  & HIFI                 & (4)                  & &                      &                      &                      &                      \\
                     &                      &                      &                      & &                      &                      &                      &                      & &                      &                      &                      &                      \\
\multicolumn{14}{c}{} \\
\multicolumn{4}{c}{IK\,Tau} & & \multicolumn{4}{c}{V1111\,Oph} & & \multicolumn{4}{c}{GX\,Mon} \\
\hline
$J$\,=\,1-0          & 59.42                & IRAM                 & (6)                  & & $J$\,=\,1-0          & 47.2                 & IRAM                 & (1)                  & & $J$\,=\,1-0          & 64.2                 & IRAM                 & (1)                  \\
$J$\,=\,2-1          & 194.01               & IRAM                 & (6)                  & & $J$\,=\,2-1          & 82.1                 & IRAM                 & (1)                  & & $J$\,=\,2-1          & 128.1                & IRAM                 & (1)                  \\
$J$\,=\,3-2          & 246.13               & IRAM                 & (6)                  & & $J$\,=\,3-2          & 59.1                 & JCMT                 & (7)                  & & $J$\,=\,3-2          & 80.0                 & JCMT                 & (5)                  \\
$J$\,=\,4-3          & 127.27               & JCMT                 & (2)                  & & $J$\,=\,4-3          & 46.1                 & APEX                 & (1)                  & & $J$\,=\,4-3          & 79.0                 & JCMT                 & (5)                  \\
$J$\,=\,6-5          & 11.6                 & HIFI                 & (4)                  & & $J$\,=\,5-4          & 3.13                 & HIFI                 & (1)                  & & $J$\,=\,5-4          & 5.83                 & HIFI                 & (1)                  \\
$J$\,=\,7-6          & 122.61               & APEX                 & (2)                  & & $J$\,=\,9-8          & 2.50                 & HIFI                 & (1)                  & & $J$\,=\,9-8          & 4.07                 & HIFI                 & (1)                  \\
$J$\,=\,10-9         & 10.7                 & HIFI                 & (4)                  & &                      &                      &                      &                      & &                      &                      &                      &                      \\
$J$\,=\,16-15        & 12.9                 & HIFI                 & (4)                  & &                      &                      &                      &                      & &                      &                      &                      &                      \\
\multicolumn{14}{c}{} \\
\multicolumn{4}{c}{NV\,Aur} & & \multicolumn{4}{c}{Y\,CVn} & & \multicolumn{4}{c}{R\,Lep} \\
\hline
$J$\,=\,1-0          & 43.3                 & IRAM                 & (1)                  & & $J$\,=\,1-0          & 9.1                  & IRAM                 & (2)                  & & $J$\,=\,1-0          & 32.5                 & IRAM                 & (1)                  \\
$J$\,=\,2-1          & 59.1                 & IRAM                 & (1)                  & & $J$\,=\,2-1          & 24.6                 & IRAM                 & (2)                  & & $J$\,=\,2-1          & 100.3                & IRAM                 & (1)                  \\
$J$\,=\,3-2          & 38.8                 & JCMT                 & (2)                  & & $J$\,=\,3-2          & 20.9                 & JCMT                 & (8)                  & & $J$\,=\,3-2          & 32.3                 & APEX                 & (1)                  \\
$J$\,=\,4-3          & 35.4                 & JCMT                 & (2)                  & & $J$\,=\,4-3          & 17.0                 & CSO                  & (1)                  & & $J$\,=\,4-3          & 36.6                 & APEX                 & (1)                  \\
$J$\,=\,5-4          & 2.54                 & HIFI                 & (1)                  & & $J$\,=\,6-5          & 16.9                 & CSO                  & (1)                  & & $J$\,=\,5-4          & 4.23                 & HIFI                 & (1)                  \\
$J$\,=\,9-8          & 1.84                 & HIFI                 & (1)                  & &                      &                      &                      &                      & & $J$\,=\,9-8          & 4.57                 & HIFI                 & (1)                  \\
                     &                      &                      &                      & &                      &                      &                      &                      & &                      &                      &                      &                      \\
\multicolumn{14}{c}{} \\
\multicolumn{4}{c}{LP\,And} & & \multicolumn{4}{c}{IRC\,+20370} & & \multicolumn{4}{c}{IRC\,+30374} \\
\hline
$J$\,=\,1-0          & 83.9                 & IRAM                 & (2)                  & & $J$\,=\,1-0          & 82.8                 & IRAM                 & (1)                  & & $J$\,=\,1-0          & 54.4                 & IRAM                 & (9)                 \\
$J$\,=\,2-1          & 153.6                & IRAM                 & (2)                  & & $J$\,=\,2-1          & 136.5                & IRAM                 & (1)                  & & $J$\,=\,2-1          & 110.5                & IRAM                 & (9)                 \\
$J$\,=\,3-2          & 67.6                 & JCMT                 & (2)                  & & $J$\,=\,3-2          & 60.3                 & APEX                 & (2)                  & & $J$\,=\,3-2          & 38.3                 & JCMT                 & (10)                 \\
$J$\,=\,4-3          & 74.2                 & JCMT                 & (2)                  & & $J$\,=\,4-3          & 66.1                 & APEX                 & (2)                  & &                      &                      &                      &                      \\
$J$\,=\,6-5          & 54.9                 & CSO                  & (2)                  & & $J$\,=\,5-4          & 6.32                 & HIFI                 & (1)                  & &                      &                      &                      &                      \\
                     &                      &                      &                      & & $J$\,=\,7-6          & 49.8                 & APEX                 & (2)                  & &                      &                      &                      &                      \\
                     &                      &                      &                      & & $J$\,=\,9-8          & 7.46                 & HIFI                 & (1)                  & &                      &                      &                      &                      \\
\multicolumn{14}{c}{} \\
\multicolumn{4}{c}{IRC\,+10216} & & \multicolumn{4}{c}{CRL\,190} & & \multicolumn{4}{c}{} \\
\hline
$J$\,=\,1-0          & 446.2                & IRAM                 & (11)                  & & $J$\,=\,1-0          & 31.0                 & IRAM                 & (9)                 & &                      &                      &                      &                      \\
$J$\,=\,2-1          & 919.4                & IRAM                 & (11)                  & & $J$\,=\,2-1          & 42.7                 & IRAM                 & (9)                 & &                      &                      &                      &                      \\
$J$\,=\,3-2          & 1573.                & IRAM                 & (11)                  & &                      &                      &                      &                      & &                      &                      &                      &                      \\
$J$\,=\,4-3          & 639.17               & CSO                  & (2)                  & &                      &                      &                      &                      & &                      &                      &                      &                      \\
$J$\,=\,5-4          & 246.7                & HIFI                 & (11)                  & &                      &                      &                      &                      & &                      &                      &                      &                      \\
$J$\,=\,6-5          & 271.6                & HIFI                 & (11)                  & &                      &                      &                      &                      & &                      &                      &                      &                      \\
$J$\,=\,7-6          & 283.6                & HIFI                 & (11)                  & &                      &                      &                      &                      & &                      &                      &                      &                      \\
$J$\,=\,8-7          & 314.2                & HIFI                 & (11)                  & &                      &                      &                      &                      & &                      &                      &                      &                      \\
$J$\,=\,9-8          & 324.3                & HIFI                 & (11)                  & &                      &                      &                      &                      & &                      &                      &                      &                      \\
$J$\,=\,10-9         & 340.2                & HIFI                 & (11)                  & &                      &                      &                      &                      & &                      &                      &                      &                      \\
$J$\,=\,11-10        & 322.3                & HIFI                 & (11)                  & &                      &                      &                      &                      & &                      &                      &                      &                      \\
$J$\,=\,13-12        & 325.0                & HIFI                 & (11)                  & &                      &                      &                      &                      & &                      &                      &                      &                      \\
$J$\,=\,14-13        & 364.6                & HIFI                 & (11)                  & &                      &                      &                      &                      & &                      &                      &                      &                      \\
$J$\,=\,15-14        & 335.8                & HIFI                 & (11)                  & &                      &                      &                      &                      & &                      &                      &                      &                      \\
$J$\,=\,16-15        & 345.0                & HIFI                 & (11)                  & &                      &                      &                      &                      & &                      &                      &                      &                      \\
\end{longtable}
\hspace{-0.2cm} \tablenotee{References: (1)\,\cite{Danilovich2015}. (2)\,\cite{DeBeck2010}. (3)\,\cite{Neri1998}. (4)\,\cite{Justtanont2012}. (5)\,\cite{Ramstedt2008}. (6)\,\cite{Velilla-Prieto2017}. (7)\,\cite{Ramstedt2014}. (8)\,\cite{Schoier2001}. (9)\,This work. (10)\,JCMT archive. (11)\,J. Cernicharo (priv.~comm.).} \\

\clearpage

\begin{figure*}
\centering
\includegraphics[angle=0,width=0.88\textwidth]{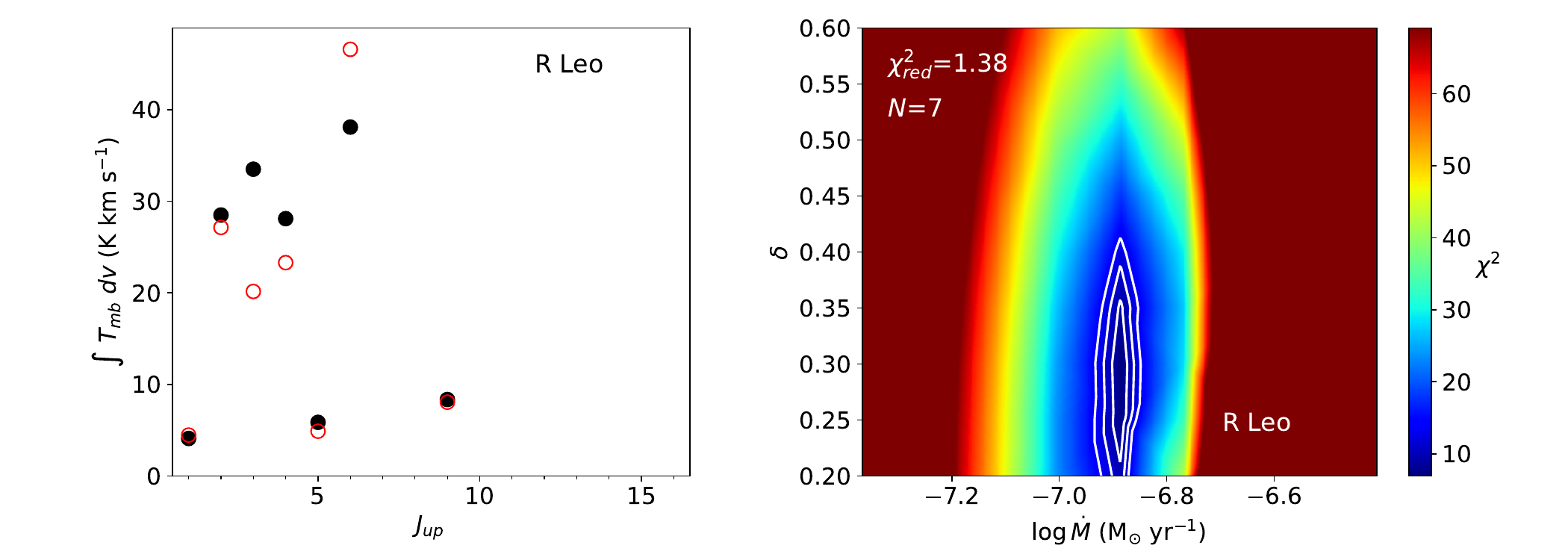} \includegraphics[angle=0,width=0.88\textwidth]{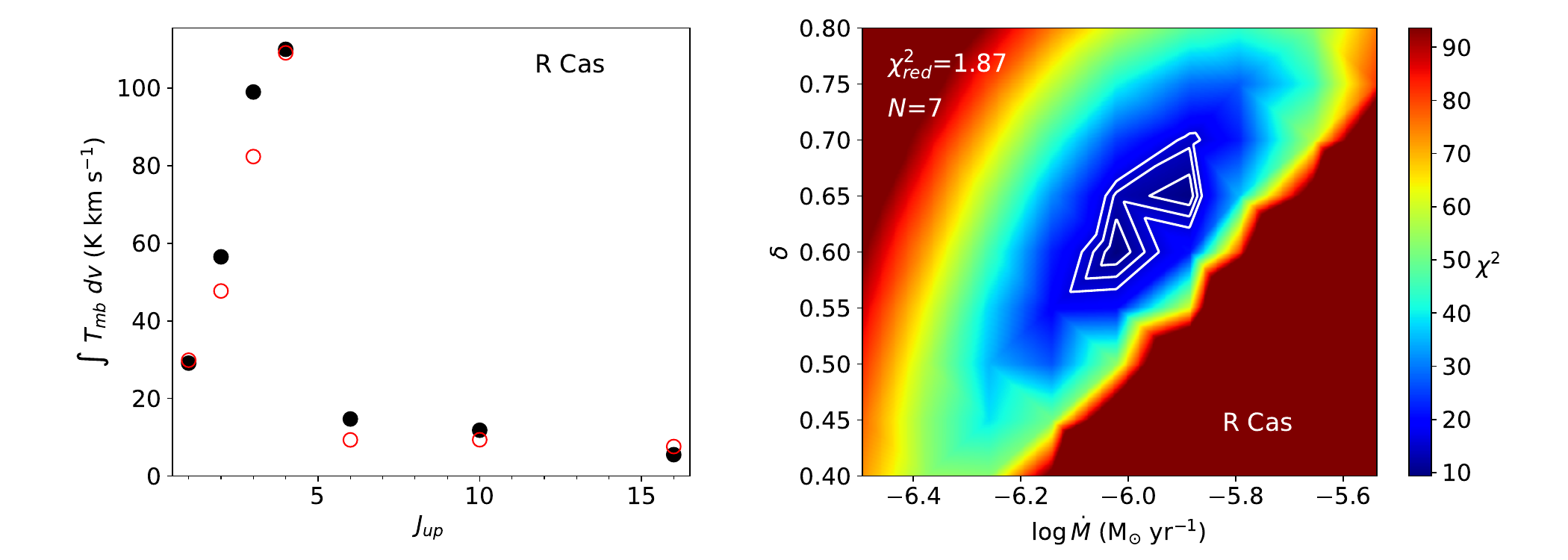} \includegraphics[angle=0,width=0.88\textwidth]{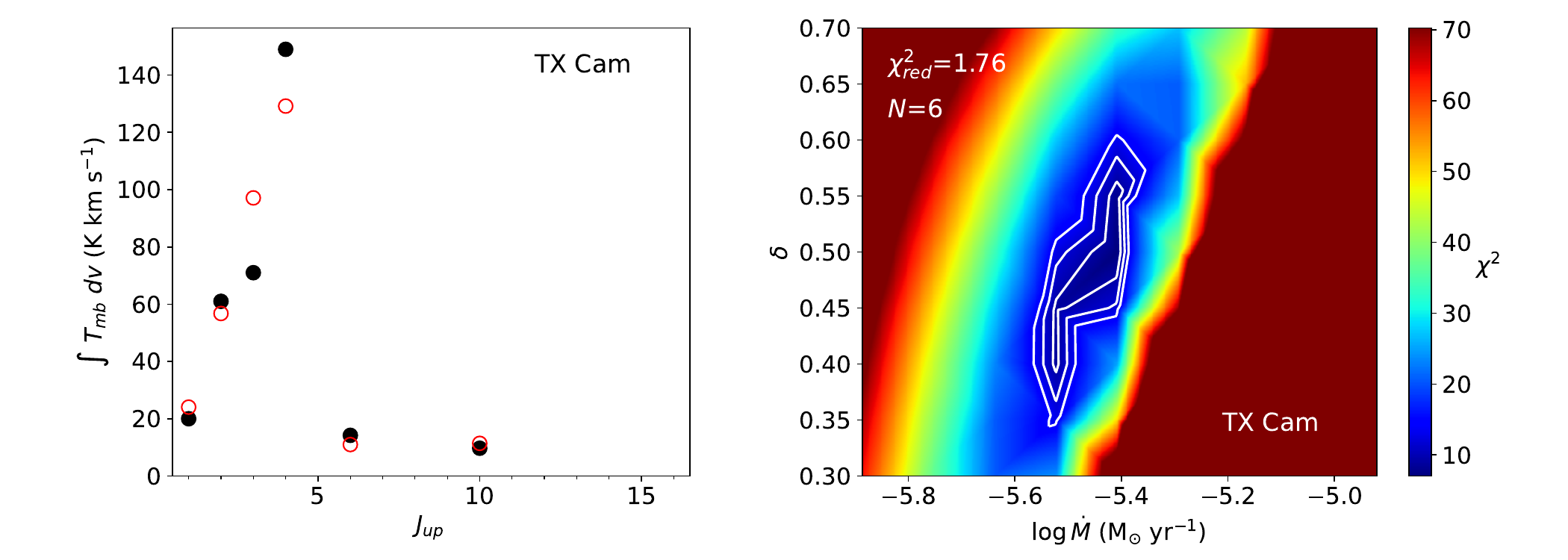} \includegraphics[angle=0,width=0.88\textwidth]{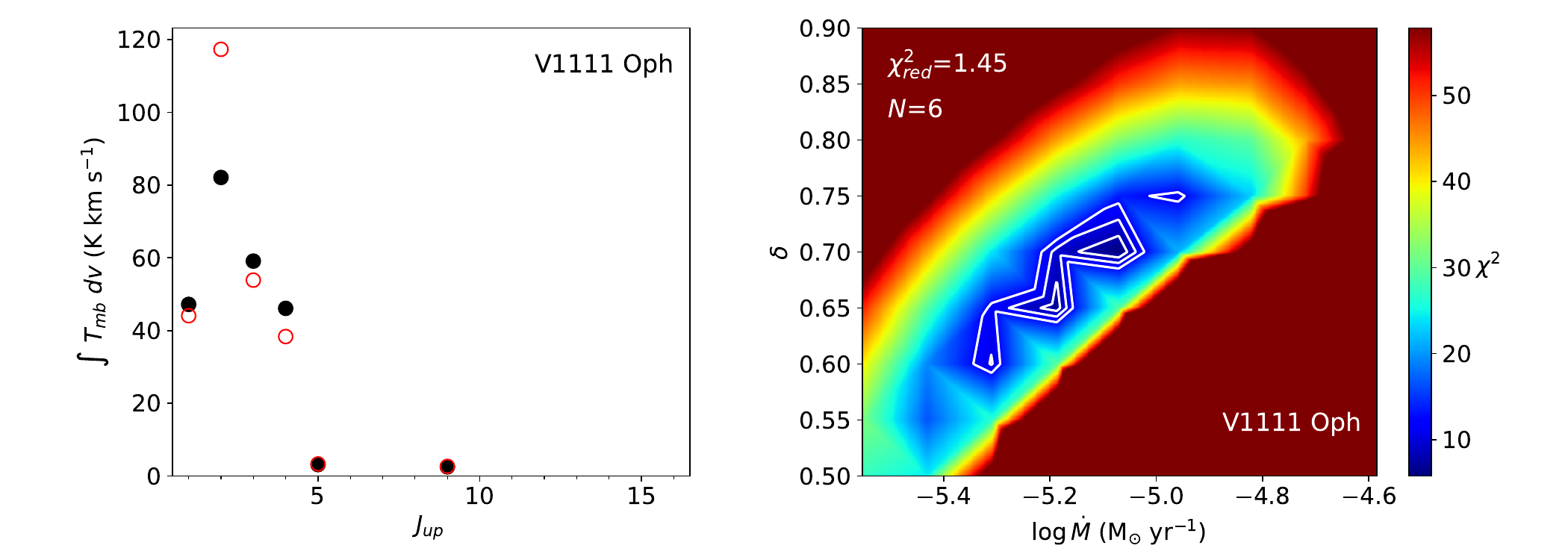}
\caption{Results from CO analysis for all envelopes except IK\,Tau and IRC\,+10216 (shown in Fig.\,\ref{fig:co_iktau} and Fig.\,\ref{fig:co_irc10216}). The left panels show the observed velocity-integrated intensities as black filled circles (see Table\,\ref{table:co}) and the calculated ones as red empty circles. The right panels show $\chi^2$ as a function of the logarithm of the mass loss rate, $\log \dot{M}$, and the exponent of the gas kinetic temperature radial profile, $\delta$. The white contours correspond to 1, 2, and 3\,$\sigma$ levels.} \label{fig:co}
\end{figure*}

\setcounter{figure}{2}
\begin{figure*}
\centering
\includegraphics[angle=0,width=0.88\textwidth]{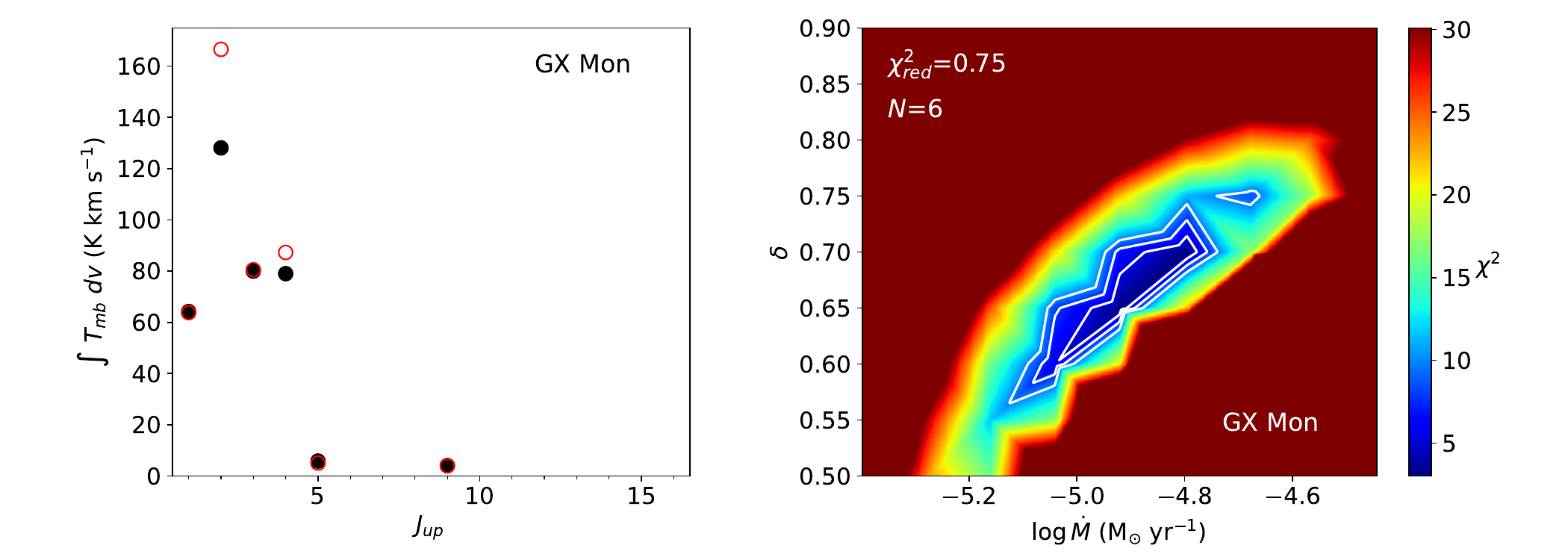} \includegraphics[angle=0,width=0.88\textwidth]{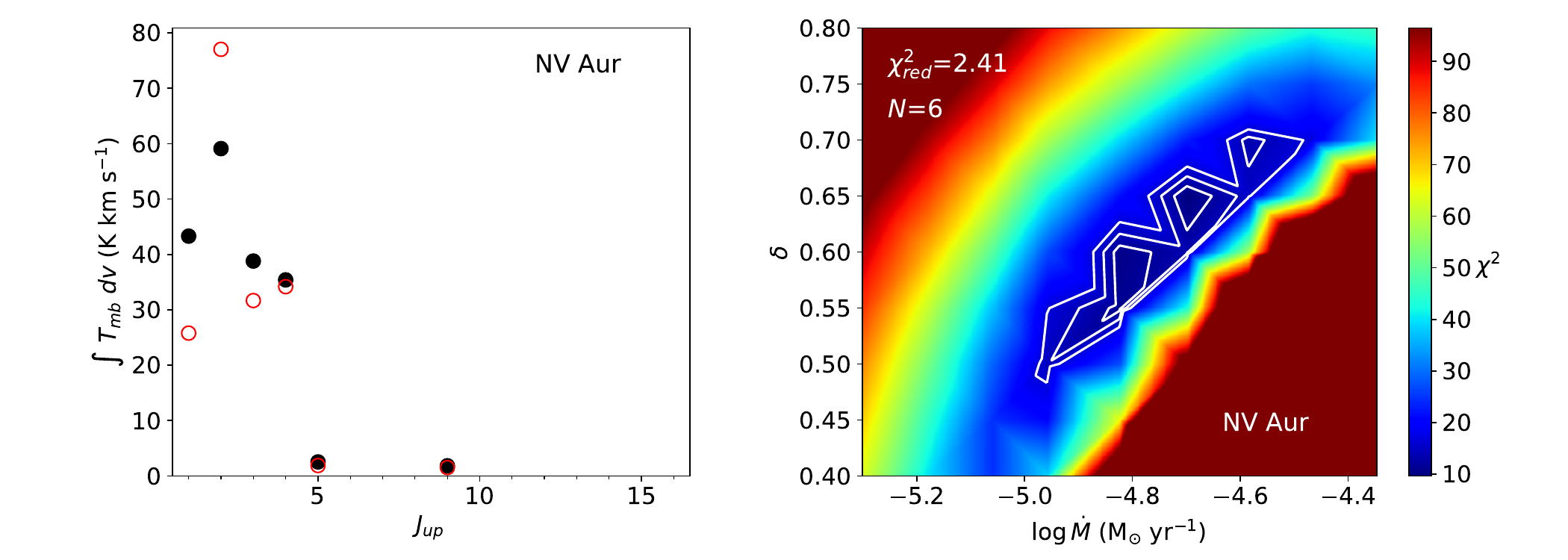} \includegraphics[angle=0,width=0.88\textwidth]{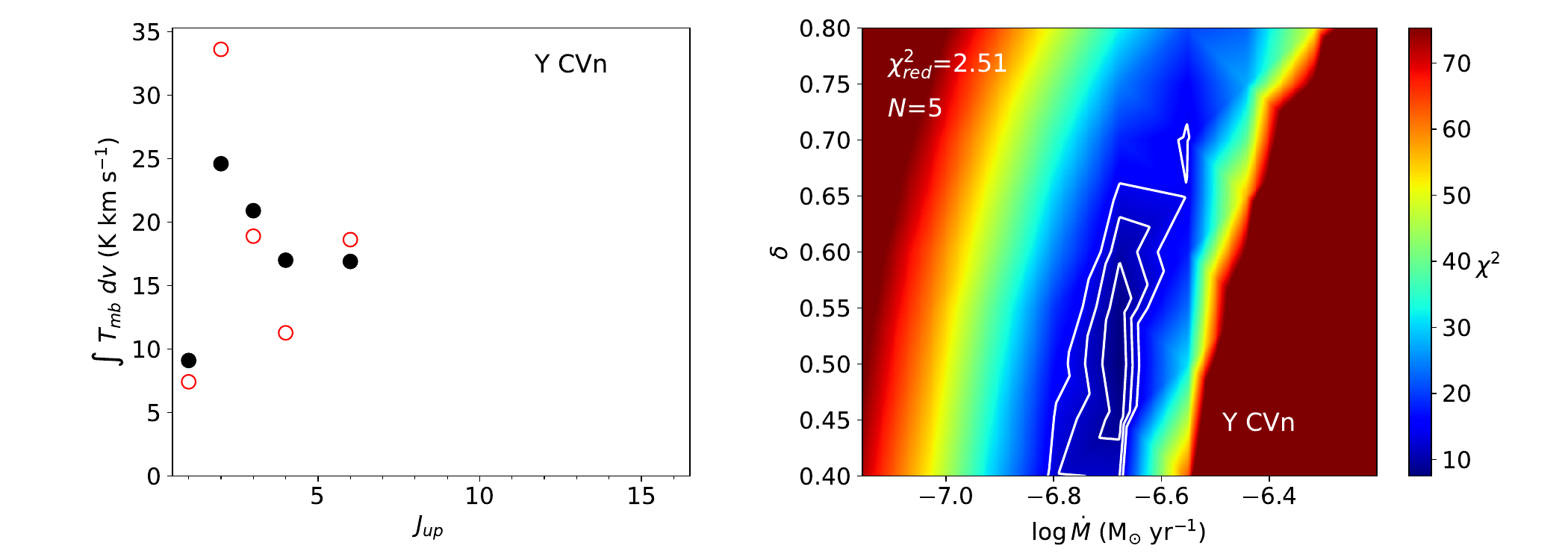} \includegraphics[angle=0,width=0.88\textwidth]{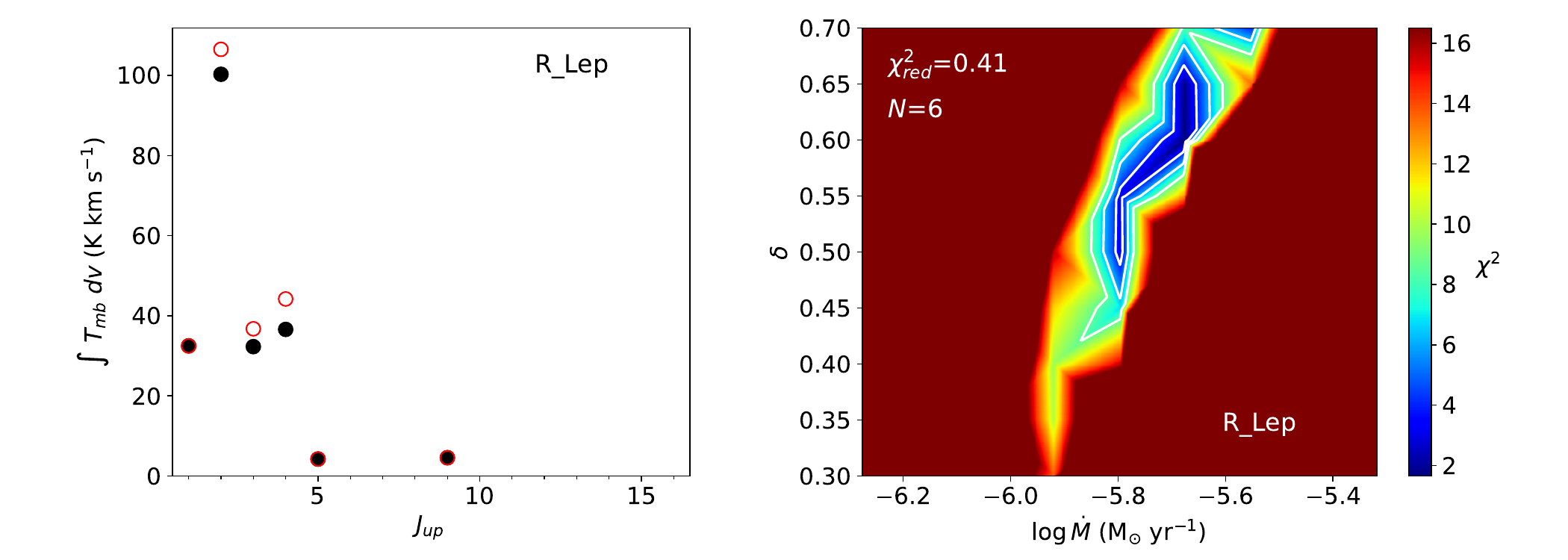} 
\caption{continued.}
\end{figure*}

\setcounter{figure}{2}
\begin{figure*}
\centering
\includegraphics[angle=0,width=0.88\textwidth]{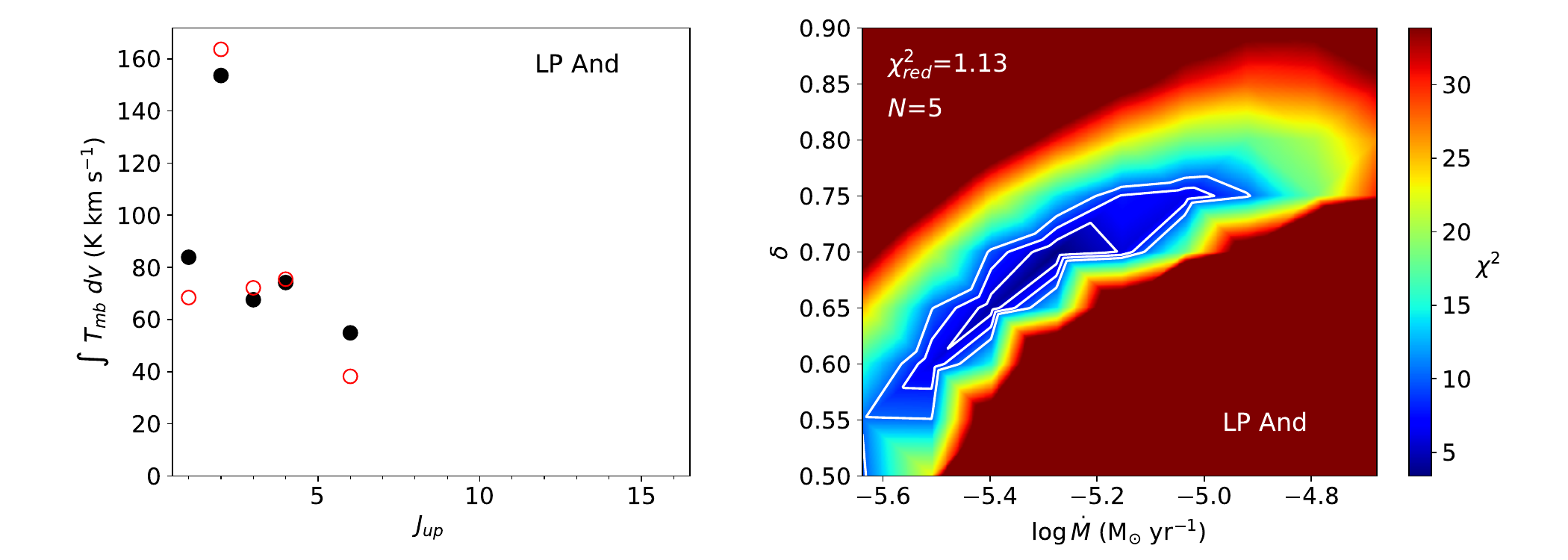} \includegraphics[angle=0,width=0.88\textwidth]{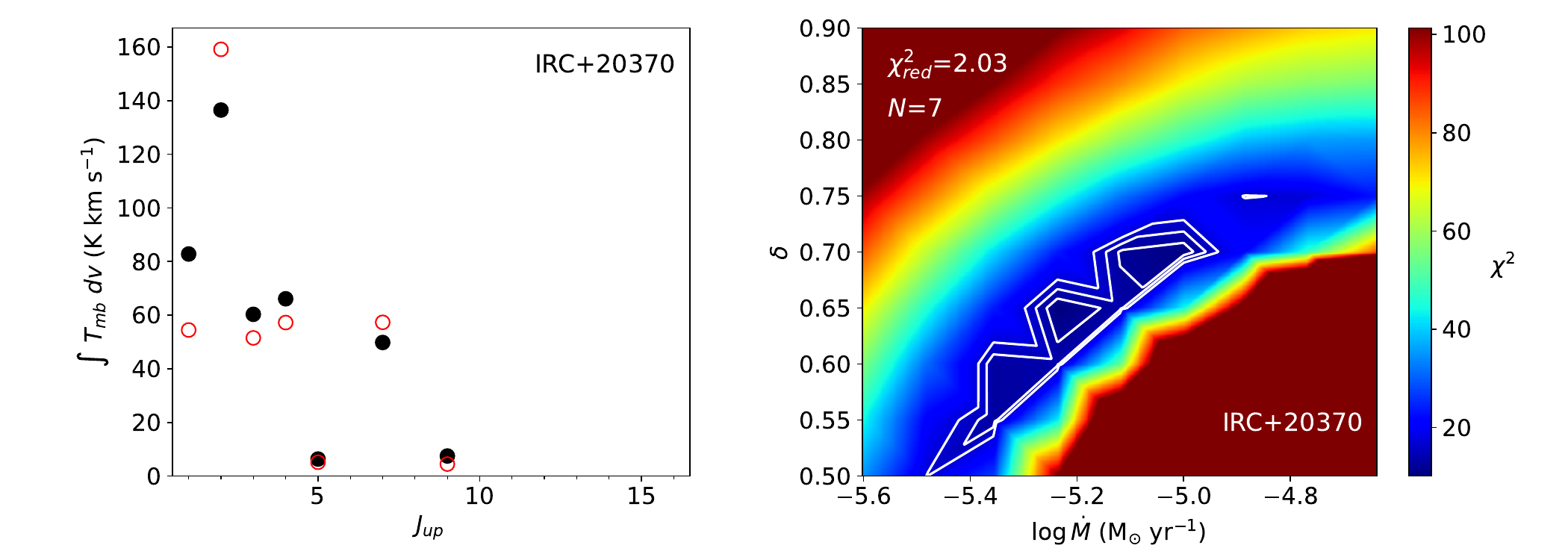} \includegraphics[angle=0,width=0.88\textwidth]{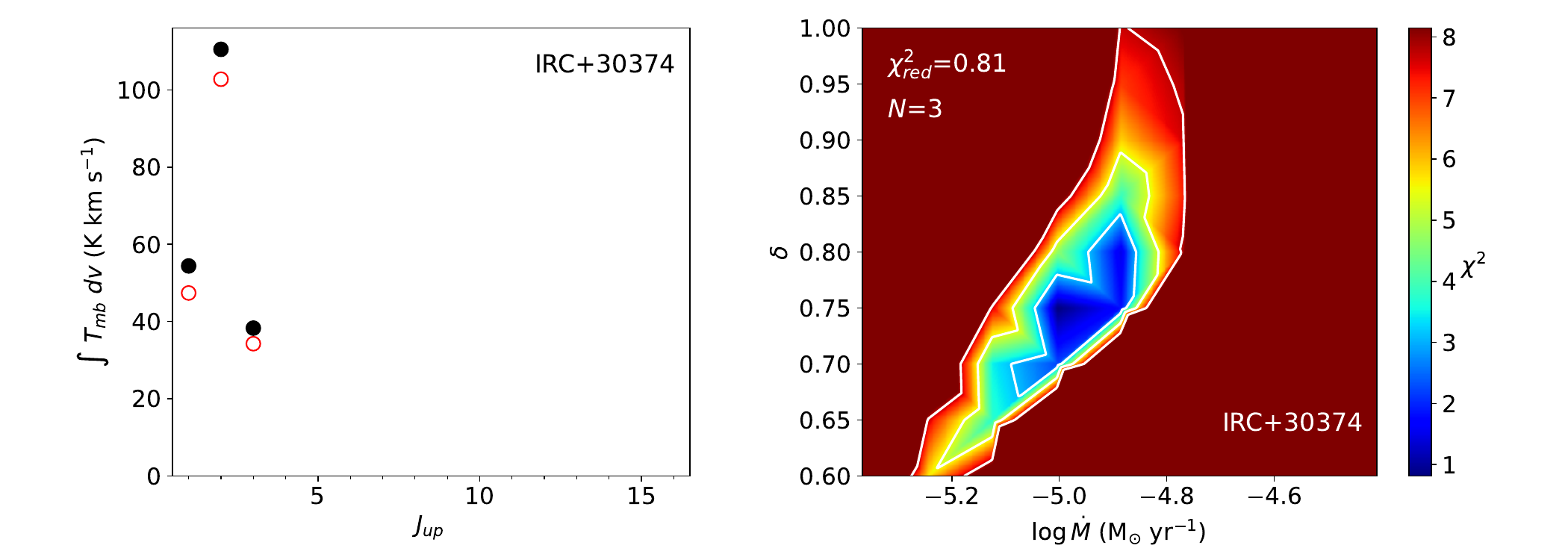} \includegraphics[angle=0,width=0.88\textwidth]{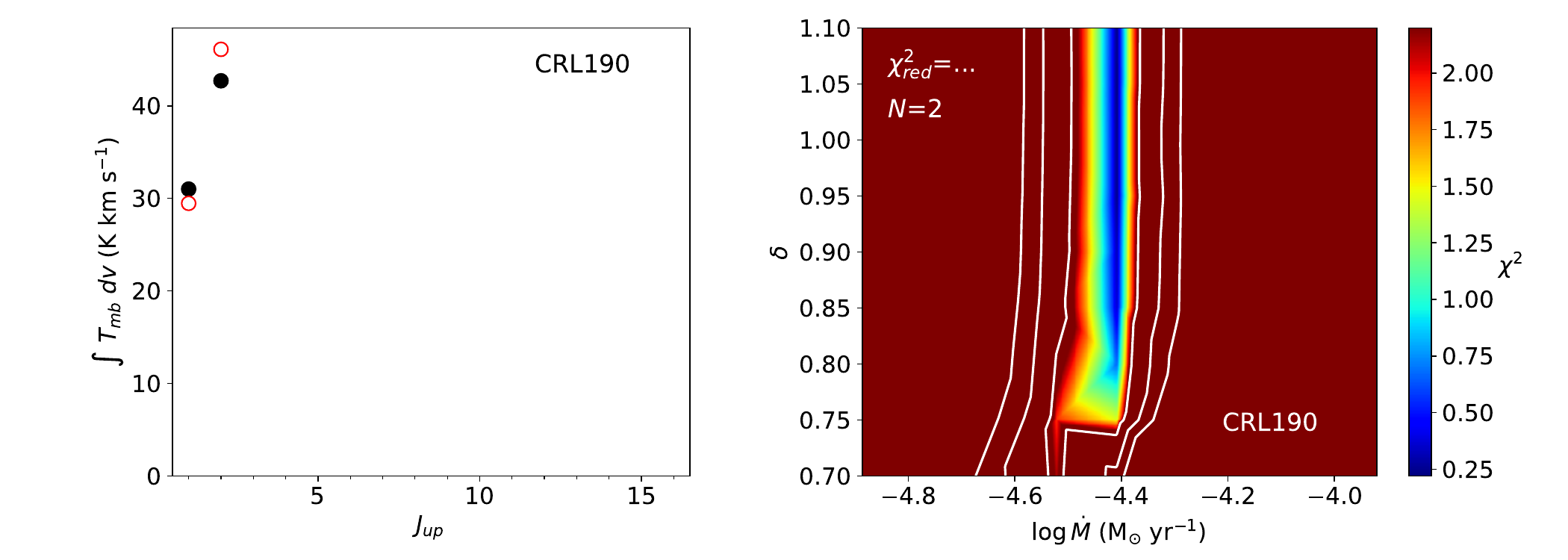} 
\caption{continued.}
\end{figure*}

\clearpage

\small
\begin{longtable}{rcll c rcll c rcll}
\caption{\label{table:lines_all} Summary of SiO, CS, and SiS observations included in the analysis.} \\
\hline\hline
\multicolumn{1}{c}{Line} & \multicolumn{1}{c}{$\int T_{\rm mb} dv$} & \multicolumn{2}{l}{Telescope} & & \multicolumn{1}{c}{Line} & \multicolumn{1}{c}{$\int T_{\rm mb} dv$} & \multicolumn{2}{l}{Telescope} & & \multicolumn{1}{c}{Line} & \multicolumn{1}{c}{$\int T_{\rm mb} dv$} & \multicolumn{2}{l}{Telescope} \\
 & \multicolumn{1}{c}{(K km s$^{-1}$)} & \multicolumn{2}{r}{Reference} & & & \multicolumn{1}{c}{(K km s$^{-1}$)} & \multicolumn{2}{r}{Reference} & & & \multicolumn{1}{c}{(K km s$^{-1}$)} & \multicolumn{2}{r}{Reference} \\
\hline
\endfirsthead
\caption{continued.}\\
\hline\hline
\multicolumn{1}{c}{Line} & \multicolumn{1}{c}{$\int T_{\rm mb} dv$} & \multicolumn{2}{l}{Telescope} & & \multicolumn{1}{c}{Line} & \multicolumn{1}{c}{$\int T_{\rm mb} dv$} & \multicolumn{2}{l}{Telescope} & & \multicolumn{1}{c}{Line} & \multicolumn{1}{c}{$\int T_{\rm mb} dv$} & \multicolumn{2}{l}{Telescope} \\
 & \multicolumn{1}{c}{(K km s$^{-1}$)} & \multicolumn{2}{r}{Reference} & & & \multicolumn{1}{c}{(K km s$^{-1}$)} & \multicolumn{2}{r}{Reference} & & & \multicolumn{1}{c}{(K km s$^{-1}$)} & \multicolumn{2}{r}{Reference} \\
\hline
\endhead
\hline
\endfoot
\multicolumn{14}{c}{} \\
\multicolumn{14}{l}{R\,Leo} \\
\multicolumn{4}{c}{SiO} & & \multicolumn{4}{c}{CS} & & \multicolumn{4}{c}{SiS} \\
\hline
$J$\,=\,1-0          &   3.9                & Yebes                & (1)                  & & $J$\,=\,3-2          &   0.10               & IRAM                 & (3)                  & &                      &                      &                      &                      \\
$J$\,=\,2-1          &  16.1                & IRAM                 & (1)                  & & $J$\,=\,5-4          &   0.41               & IRAM                 & (1)                  & &                      &                      &                      &                      \\
$J$\,=\,2-1          &   5.10               & OSO                  & (6)                  & & $J$\,=\,6-5          &   0.59               & IRAM                 & (1)                  & &                      &                      &                      &                      \\
$J$\,=\,2-1          &   4.90               & SEST                 & (6)                  & &                      &                      &                      &                      & &                      &                      &                      &                      \\
$J$\,=\,3-2          &  25.6                & IRAM                 & (3)                  & &                      &                      &                      &                      & &                      &                      &                      &                      \\
$J$\,=\,5-4          &  10.8                & SEST                 & (6)                  & &                      &                      &                      &                      & &                      &                      &                      &                      \\
$J$\,=\,6-5          &  43                  & IRAM                 & (1)                  & &                      &                      &                      &                      & &                      &                      &                      &                      \\
$J$\,=\,7-6          &  49                  & IRAM                 & (1)                  & &                      &                      &                      &                      & &                      &                      &                      &                      \\
\multicolumn{14}{c}{} \\
\multicolumn{14}{l}{R\,Cas} \\
\multicolumn{4}{c}{SiO} & & \multicolumn{4}{c}{CS} & & \multicolumn{4}{c}{SiS} \\
\hline
$J$\,=\,1-0          &  10.2                & Yebes                & (1)                  & & $J$\,=\,2-1          &   0.10               & IRAM                 & (1)                  & & $J$\,=\,8-7          &   0.18               & IRAM                 & (3)                  \\
$J$\,=\,2-1          &  28.3                & IRAM                 & (1)                  & & $J$\,=\,3-2          &   0.32               & IRAM                 & (3)                  & & $J$\,=\,14-13        &   0.92               & IRAM                 & (1)                  \\
$J$\,=\,2-1          &   8.60               & OSO                  & (6)                  & & $J$\,=\,5-4          &   0.84               & IRAM                 & (1)                  & & $J$\,=\,19-18        &   0.80               & JCMT                 & (11)                 \\
$J$\,=\,3-2          &  31.9                & IRAM                 & (3)                  & &                      &                      &                      &                      & &                      &                      &                      &                      \\
$J$\,=\,5-4          &   7.2                & SMT                  & (8)                  & &                      &                      &                      &                      & &                      &                      &                      &                      \\
$J$\,=\,6-5          &  53.7                & IRAM                 & (1)                  & &                      &                      &                      &                      & &                      &                      &                      &                      \\
$J$\,=\,8-7          &  15.0                & SMT                  & (8)                  & &                      &                      &                      &                      & &                      &                      &                      &                      \\
\multicolumn{14}{c}{} \\
\multicolumn{14}{l}{TX\,Cam} \\
\multicolumn{4}{c}{SiO} & & \multicolumn{4}{c}{CS} & & \multicolumn{4}{c}{SiS} \\
\hline
$J$\,=\,1-0          &  10.4                & Yebes                & (1)                  & & $J$\,=\,2-1          &   1.69               & IRAM                 & (1)                  & & $J$\,=\,5-4          &   1.80               & IRAM                 & (9)                  \\
$J$\,=\,2-1          &  30.4                & IRAM                 & (1)                  & & $J$\,=\,2-1          &   1.00               & OSO                  & (10)                 & & $J$\,=\,5-4          &   0.60               & OSO                  & (10)                 \\
$J$\,=\,2-1          &  13.2                & OSO                  & (6)                  & & $J$\,=\,3-2          &   3.06               & IRAM                 & (3)                  & & $J$\,=\,6-5          &   0.80               & OSO                  & (11)                 \\
$J$\,=\,3-2          &  36.4                & IRAM                 & (3)                  & & $J$\,=\,5-4          &   3.8                & IRAM                 & (9)                  & & $J$\,=\,8-7          &   2.87               & IRAM                 & (3)                  \\
$J$\,=\,5-4          &   9.6                & SMT                  & (8)                  & &                      &                      &                      &                      & & $J$\,=\,12-11        &   1.50               & JCMT                 & (11)                 \\
$J$\,=\,8-7          &  11.6                & SMT                  & (8)                  & &                      &                      &                      &                      & &                      &                      &                      &                      \\
\multicolumn{14}{c}{} \\
\multicolumn{14}{l}{IK\,Tau} \\
\multicolumn{4}{c}{SiO} & & \multicolumn{4}{c}{CS} & & \multicolumn{4}{c}{SiS} \\
\hline
$J$\,=\,1-0          &  24.8                & Yebes                & (1)                  & & $J$\,=\,2-1          &   0.93               & IRAM                 & (4)                  & & $J$\,=\,5-4          &   1.19               & IRAM                 & (4)                  \\
$J$\,=\,2-1          &  36.0                & IRAM                 & (4)                  & & $J$\,=\,2-1          &   0.50               & OSO                  & (10)                 & & $J$\,=\,6-5          &   2.44               & IRAM                 & (4)                  \\
$J$\,=\,2-1          &  16.1                & OSO                  & (6)                  & & $J$\,=\,3-2          &   3.11               & IRAM                 & (3)                  & & $J$\,=\,8-7          &   6.1                & IRAM                 & (3)                  \\
$J$\,=\,3-2          &  56                  & IRAM                 & (3)                  & & $J$\,=\,4-3          &   1.46               & APEX                 & (13)                 & & $J$\,=\,9-8          &   7.7                & IRAM                 & (4)                  \\
$J$\,=\,4-3          &  66                  & IRAM                 & (4)                  & & $J$\,=\,5-4          &   9.7                & IRAM                 & (4)                  & & $J$\,=\,12-11        &  14.3                & IRAM                 & (4)                  \\
$J$\,=\,5-4          &  91                  & IRAM                 & (4)                  & & $J$\,=\,6-5          &  10.8                & IRAM                 & (4)                  & & $J$\,=\,12-11        &   4.7                & JCMT                 & (11)                 \\
$J$\,=\,5-4          &  18.9                & SEST                 & (6)                  & & $J$\,=\,7-6          &  12.4                & IRAM                 & (4)                  & & $J$\,=\,12-11        &   3.4                & APEX                 & (13)                 \\
$J$\,=\,5-4          &  16.5                & SMT                  & (8)                  & & $J$\,=\,7-6          &   2.57               & APEX                 & (13)                 & & $J$\,=\,13-12        &  18.1                & IRAM                 & (4)                  \\
$J$\,=\,6-5          & 101                  & IRAM                 & (4)                  & &                      &                      &                      &                      & & $J$\,=\,14-13        &  20.4                & IRAM                 & (4)                  \\
$J$\,=\,6-5          &  13.7                & SEST                 & (6)                  & &                      &                      &                      &                      & & $J$\,=\,14-13        &   4.3                & APEX                 & (13)                 \\
$J$\,=\,7-6          & 110                  & IRAM                 & (4)                  & &                      &                      &                      &                      & & $J$\,=\,16-15        &  23.9                & IRAM                 & (4)                  \\
$J$\,=\,8-7          &  98                  & IRAM                 & (4)                  & &                      &                      &                      &                      & & $J$\,=\,17-16        &  20.3                & IRAM                 & (4)                  \\
                     &                      &                      &                      & &                      &                      &                      &                      & & $J$\,=\,19-18        &  28.1                & IRAM                 & (4)                  \\
                     &                      &                      &                      & &                      &                      &                      &                      & & $J$\,=\,19-18        &   9.2                & JCMT                 & (11)                 \\
                     &                      &                      &                      & &                      &                      &                      &                      & & $J$\,=\,19-18        &   5.6                & APEX                 & (13)                 \\
\multicolumn{14}{c}{} \\
\multicolumn{14}{l}{V1111\,Oph} \\
\multicolumn{4}{c}{SiO} & & \multicolumn{4}{c}{CS} & & \multicolumn{4}{c}{SiS} \\
\hline
$J$\,=\,1-0          &   6.4                & Yebes                & (1)                  & & $J$\,=\,2-1          &   0.32               & IRAM                 & (1)                  & & $J$\,=\,8-7          &   2.17               & IRAM                 & (3)                  \\
$J$\,=\,2-1          &  12.8                & IRAM                 & (1)                  & & $J$\,=\,3-2          &   1.18               & IRAM                 & (3)                  & & $J$\,=\,14-13        &   3.00               & IRAM                 & (1)                  \\
$J$\,=\,2-1          &   6.30               & OSO                  & (6)                  & & $J$\,=\,5-4          &   1.93               & IRAM                 & (1)                  & & $J$\,=\,16-15        &   4.7                & IRAM                 & (1)                  \\
$J$\,=\,3-2          &  21.0                & IRAM                 & (3)                  & & $J$\,=\,6-5          &   2.88               & IRAM                 & (1)                  & & $J$\,=\,17-16        &   4.3                & IRAM                 & (1)                  \\
$J$\,=\,5-4          &   5.3                & SMT                  & (8)                  & &                      &                      &                      &                      & &                      &                      &                      &                      \\
$J$\,=\,6-5          &  19.4                & IRAM                 & (1)                  & &                      &                      &                      &                      & &                      &                      &                      &                      \\
$J$\,=\,7-6          &  24.5                & IRAM                 & (1)                  & &                      &                      &                      &                      & &                      &                      &                      &                      \\
$J$\,=\,8-7          &   3.8                & SMT                  & (8)                  & &                      &                      &                      &                      & &                      &                      &                      &                      \\
\multicolumn{14}{c}{} \\
\multicolumn{14}{l}{GX\,Mon} \\
\multicolumn{4}{c}{SiO} & & \multicolumn{4}{c}{CS} & & \multicolumn{4}{c}{SiS} \\
\hline
$J$\,=\,1-0          &   9.9                & Yebes                & (1)                  & & $J$\,=\,2-1          &   0.75               & IRAM                 & (1)                  & & $J$\,=\,6-5          &   1.04               & IRAM                 & (14)                 \\
$J$\,=\,2-1          &  23.1                & IRAM                 & (1)                  & & $J$\,=\,3-2          &   1.51               & IRAM                 & (3)                  & & $J$\,=\,8-7          &   2.10               & IRAM                 & (3)                  \\
$J$\,=\,2-1          &   9.40               & OSO                  & (6)                  & & $J$\,=\,4-3          &   0.94               & APEX                 & (13)                 & & $J$\,=\,12-11        &   1.40               & JCMT                 & (11)                 \\
$J$\,=\,2-1          &   5.50               & SEST                 & (6)                  & & $J$\,=\,5-4          &   3.4                & IRAM                 & (1)                  & & $J$\,=\,12-11        &   1.03               & APEX                 & (13)                 \\
$J$\,=\,3-2          &  25.0                & IRAM                 & (3)                  & & $J$\,=\,6-5          &   3.4                & IRAM                 & (1)                  & & $J$\,=\,14-13        &   4.2                & IRAM                 & (1)                  \\
$J$\,=\,5-4          &   8.9                & SEST                 & (6)                  & & $J$\,=\,7-6          &   0.71               & APEX                 & (13)                 & & $J$\,=\,14-13        &   0.85               & APEX                 & (13)                 \\
$J$\,=\,5-4          &   3.8                & SMT                  & (8)                  & &                      &                      &                      &                      & & $J$\,=\,16-15        &   4.4                & IRAM                 & (1)                  \\
$J$\,=\,6-5          &  29.1                & IRAM                 & (1)                  & &                      &                      &                      &                      & & $J$\,=\,17-16        &   4.0                & IRAM                 & (1)                  \\
$J$\,=\,7-6          &  31.7                & IRAM                 & (1)                  & &                      &                      &                      &                      & & $J$\,=\,19-18        &   1.02               & APEX                 & (13)                 \\
$J$\,=\,8-7          &   3.4                & SMT                  & (8)                  & &                      &                      &                      &                      & &                      &                      &                      &                      \\
\multicolumn{14}{c}{} \\
\multicolumn{14}{l}{NV\,Aur} \\
\multicolumn{4}{c}{SiO} & & \multicolumn{4}{c}{CS} & & \multicolumn{4}{c}{SiS} \\
\hline
$J$\,=\,1-0          &   5.1                & Yebes                & (1)                  & & $J$\,=\,2-1          &   0.31               & IRAM                 & (1)                  & & $J$\,=\,6-5          &   0.65               & OSO                  & (11)                 \\
$J$\,=\,2-1          &   7.61               & IRAM                 & (1)                  & & $J$\,=\,3-2          &   0.77               & IRAM                 & (3)                  & & $J$\,=\,6-5          &   1.25               & IRAM                 & (14)                 \\
$J$\,=\,3-2          &  13.7                & IRAM                 & (3)                  & & $J$\,=\,5-4          &   1.76               & IRAM                 & (1)                  & & $J$\,=\,8-7          &   2.98               & IRAM                 & (3)                  \\
$J$\,=\,5-4          &   1.52               & SMT                  & (8)                  & &                      &                      &                      &                      & & $J$\,=\,12-11        &   1.50               & JCMT                 & (11)                 \\
$J$\,=\,6-5          &  12.7                & IRAM                 & (1)                  & &                      &                      &                      &                      & & $J$\,=\,14-13        &   5.3                & IRAM                 & (1)                  \\
$J$\,=\,8-7          &   3.04               & SMT                  & (8)                  & &                      &                      &                      &                      & &                      &                      &                      &                      \\
\multicolumn{14}{c}{} \\
\multicolumn{14}{l}{Y\,CVn} \\
\multicolumn{4}{c}{SiO} & & \multicolumn{4}{c}{CS} & & \multicolumn{4}{c}{SiS} \\
\hline
$J$\,=\,2-1          &   0.16               & IRAM                 & (1)                  & & $J$\,=\,2-1          &   1.80               & IRAM                 & (1)                  & &                      &                      &                      &                      \\
$J$\,=\,3-2          &   0.38               & IRAM                 & (2)                  & & $J$\,=\,2-1          &   0.70               & OSO                  & (10)                 & &                      &                      &                      &                      \\
$J$\,=\,7-6          &   1.08               & IRAM                 & (1)                  & & $J$\,=\,3-2          &   5.5                & IRAM                 & (2)                  & &                      &                      &                      &                      \\
                     &                      &                      &                      & & $J$\,=\,6-5          &  10.3                & IRAM                 & (1)                  & &                      &                      &                      &                      \\
\multicolumn{14}{c}{} \\
\multicolumn{14}{l}{R\,Lep} \\
\multicolumn{4}{c}{SiO} & & \multicolumn{4}{c}{CS} & & \multicolumn{4}{c}{SiS} \\
\hline
$J$\,=\,1-0          &   0.26               & Yebes                & (1)                  & & $J$\,=\,2-1          &   1.46               & IRAM                 & (1)                  & &                      &                      &                      &                      \\
$J$\,=\,2-1          &   1.22               & IRAM                 & (1)                  & & $J$\,=\,3-2          &   5.8                & IRAM                 & (2)                  & &                      &                      &                      &                      \\
$J$\,=\,2-1          &   0.39               & SEST                 & (7)                  & & $J$\,=\,4-3          &   2.30               & APEX                 & (13)                 & &                      &                      &                      &                      \\
$J$\,=\,3-2          &   3.45               & IRAM                 & (2)                  & & $J$\,=\,5-4          &  14.0                & IRAM                 & (1)                  & &                      &                      &                      &                      \\
$J$\,=\,3-2          &   0.99               & SEST                 & (7)                  & & $J$\,=\,6-5          &   4.1                & APEX                 & (13)                 & &                      &                      &                      &                      \\
$J$\,=\,5-4          &   3.26               & SEST                 & (7)                  & &                      &                      &                      &                      & &                      &                      &                      &                      \\
$J$\,=\,6-5          &   8.2                & IRAM                 & (1)                  & &                      &                      &                      &                      & &                      &                      &                      &                      \\
$J$\,=\,8-7          &   2.73               & APEX                 & (7)                  & &                      &                      &                      &                      & &                      &                      &                      &                      \\
\multicolumn{14}{c}{} \\
\multicolumn{14}{l}{LP\,And} \\
\multicolumn{4}{c}{SiO} & & \multicolumn{4}{c}{CS} & & \multicolumn{4}{c}{SiS} \\
\hline
$J$\,=\,1-0          &   0.89               & Yebes                & (1)                  & & $J$\,=\,2-1          &  14.2                & IRAM                 & (1)                  & & $J$\,=\,2-1          &   0.30               & Yebes                & (1)                  \\
$J$\,=\,2-1          &   3.09               & IRAM                 & (1)                  & & $J$\,=\,2-1          &   9.56               & OSO                  & (12)                 & & $J$\,=\,5-4          &   4.20               & IRAM                 & (9)                  \\
$J$\,=\,2-1          &   0.94               & OSO                  & (12)                 & & $J$\,=\,3-2          &  35.6                & IRAM                 & (2)                  & & $J$\,=\,5-4          &   1.80               & OSO                  & (11)                 \\
$J$\,=\,3-2          &   8.6                & IRAM                 & (2)                  & & $J$\,=\,5-4          &  31.2                & IRAM                 & (1)                  & & $J$\,=\,6-5          &   1.91               & OSO                  & (12)                 \\
$J$\,=\,3-2          &   0.62               & ARO                  & (7)                  & &                      &                      &                      &                      & & $J$\,=\,7-6          &   7.0                & IRAM                 & (2)                  \\
$J$\,=\,5-4          &   2.76               & ARO                  & (7)                  & &                      &                      &                      &                      & & $J$\,=\,8-7          &  12.3                & IRAM                 & (2)                  \\
$J$\,=\,6-5          &   8.5                & IRAM                 & (1)                  & &                      &                      &                      &                      & & $J$\,=\,14-13        &  15.7                & IRAM                 & (1)                  \\
\multicolumn{14}{c}{} \\
\multicolumn{14}{l}{IRC\,+20370} \\
\multicolumn{4}{c}{SiO} & & \multicolumn{4}{c}{CS} & & \multicolumn{4}{c}{SiS} \\
\hline
$J$\,=\,1-0          &   0.57               & Yebes                & (1)                  & & $J$\,=\,2-1          &   7.74               & IRAM                 & (1)                  & & $J$\,=\,6-5          &   1.33               & IRAM                 & (14)                 \\
$J$\,=\,2-1          &   3.08               & IRAM                 & (1)                  & & $J$\,=\,3-2          &  19.9                & IRAM                 & (2)                  & & $J$\,=\,7-6          &   3.10               & IRAM                 & (2)                  \\
$J$\,=\,3-2          &   7.2                & IRAM                 & (2)                  & & $J$\,=\,4-3          &   6.0                & APEX                 & (13)                 & & $J$\,=\,8-7          &   5.6                & IRAM                 & (2)                  \\
$J$\,=\,6-5          &  17.3                & IRAM                 & (1)                  & & $J$\,=\,5-4          &  45                  & IRAM                 & (1)                  & & $J$\,=\,9-8          &   1.66               & APEX                 & (13)                 \\
$J$\,=\,7-6          &  19.4                & IRAM                 & (1)                  & & $J$\,=\,6-5          &  41                  & IRAM                 & (1)                  & & $J$\,=\,12-11        &   2.39               & APEX                 & (13)                 \\
                     &                      &                      &                      & & $J$\,=\,6-5          &  10.8                & APEX                 & (13)                 & & $J$\,=\,14-13        &  16.8                & IRAM                 & (1)                  \\
                     &                      &                      &                      & & $J$\,=\,7-6          &   9.6                & APEX                 & (13)                 & & $J$\,=\,16-15        &  14.3                & IRAM                 & (1)                  \\
                     &                      &                      &                      & &                      &                      &                      &                      & & $J$\,=\,16-15        &   3.6                & APEX                 & (13)                 \\
                     &                      &                      &                      & &                      &                      &                      &                      & & $J$\,=\,17-16        &  13.0                & IRAM                 & (1)                  \\
                     &                      &                      &                      & &                      &                      &                      &                      & & $J$\,=\,19-18        &   3.27               & APEX                 & (13)                 \\
\multicolumn{14}{c}{} \\
\multicolumn{14}{l}{IRC\,+30374} \\
\multicolumn{4}{c}{SiO} & & \multicolumn{4}{c}{CS} & & \multicolumn{4}{c}{SiS} \\
\hline
$J$\,=\,2-1          &   3.69               & IRAM                 & (1)                  & & $J$\,=\,2-1          &  15.6                & IRAM                 & (1)                  & & $J$\,=\,7-6          &   1.83               & IRAM                 & (2)                  \\
$J$\,=\,3-2          &   7.1                & IRAM                 & (2)                  & & $J$\,=\,3-2          &  26.7                & IRAM                 & (2)                  & & $J$\,=\,8-7          &   2.99               & IRAM                 & (2)                  \\
$J$\,=\,6-5          &  14.2                & IRAM                 & (1)                  & & $J$\,=\,5-4          &  42                  & IRAM                 & (1)                  & & $J$\,=\,14-13        &   5.4                & IRAM                 & (1)                  \\
$J$\,=\,7-6          &  18.6                & IRAM                 & (1)                  & & $J$\,=\,6-5          &  48                  & IRAM                 & (1)                  & & $J$\,=\,16-15        &   7.8                & IRAM                 & (1)                  \\
                     &                      &                      &                      & &                      &                      &                      &                      & & $J$\,=\,17-16        &   5.2                & IRAM                 & (1)                  \\
\multicolumn{14}{c}{} \\
\multicolumn{14}{l}{IRC\,+10216} \\
\multicolumn{4}{c}{SiO} & & \multicolumn{4}{c}{CS} & & \multicolumn{4}{c}{SiS} \\
\hline
$J$\,=\,1-0          &  15.4                & Yebes                & (1)                  & & $J$\,=\,1-0          &  45.0                & Yebes                & (1)                  & & $J$\,=\,2-1          &   9.08               & Yebes                & (1)                  \\
$J$\,=\,2-1          &  60.1                & IRAM                 & (5)                  & & $J$\,=\,2-1          & 137                  & IRAM                 & (5)                  & & $J$\,=\,5-4          & 114                  & IRAM                 & (5)                  \\
$J$\,=\,2-1          &  21.7                & SEST                 & (7)                  & & $J$\,=\,2-1          & 130                  & OSO                  & (10)                 & & $J$\,=\,5-4          &  60.0                & OSO                  & (11)                 \\
$J$\,=\,2-1          &  29.9                & OSO                  & (10)                 & & $J$\,=\,3-2          & 379                  & IRAM                 & (2)                  & & $J$\,=\,6-5          & 106                  & IRAM                 & (5)                  \\
$J$\,=\,3-2          & 131                  & IRAM                 & (2)                  & & $J$\,=\,4-3          & 407                  & IRAM                 & (5)                  & & $J$\,=\,7-6          & 163                  & IRAM                 & (2)                  \\
$J$\,=\,3-2          &  27.9                & ARO                  & (7)                  & & $J$\,=\,5-4          & 518                  & IRAM                 & (5)                  & & $J$\,=\,8-7          & 271                  & IRAM                 & (2)                  \\
$J$\,=\,3-2          &  48.4                & SEST                 & (7)                  & & $J$\,=\,6-5          & 778                  & IRAM                 & (5)                  & & $J$\,=\,9-8          & 335                  & IRAM                 & (5)                  \\
$J$\,=\,4-3          & 206                  & IRAM                 & (5)                  & & $J$\,=\,7-6          & 607                  & IRAM                 & (5)                  & & $J$\,=\,10-9         & 245                  & IRAM                 & (5)                  \\
$J$\,=\,5-4          & 221                  & IRAM                 & (5)                  & &                      &                      &                      &                      & & $J$\,=\,11-10        & 343                  & IRAM                 & (5)                  \\
$J$\,=\,5-4          &  84                  & SEST                 & (7)                  & &                      &                      &                      &                      & & $J$\,=\,12-11        & 272                  & IRAM                 & (5)                  \\
$J$\,=\,5-4          &  68                  & SMT                  & (8)                  & &                      &                      &                      &                      & & $J$\,=\,13-12        & 350                  & IRAM                 & (5)                  \\
$J$\,=\,6-5          & 288                  & IRAM                 & (5)                  & &                      &                      &                      &                      & & $J$\,=\,14-13        & 525                  & IRAM                 & (5)                  \\
$J$\,=\,7-6          & 334                  & IRAM                 & (5)                  & &                      &                      &                      &                      & & $J$\,=\,15-14        & 621                  & IRAM                 & (5)                  \\
$J$\,=\,8-7          & 367                  & IRAM                 & (5)                  & &                      &                      &                      &                      & & $J$\,=\,16-15        & 590                  & IRAM                 & (5)                  \\
$J$\,=\,8-7          & 101                  & SMT                  & (8)                  & &                      &                      &                      &                      & & $J$\,=\,17-16        & 365                  & IRAM                 & (5)                  \\
                     &                      &                      &                      & &                      &                      &                      &                      & & $J$\,=\,18-17        & 500                  & IRAM                 & (5)                  \\
                     &                      &                      &                      & &                      &                      &                      &                      & & $J$\,=\,19-18        & 459                  & IRAM                 & (5)                  \\
                     &                      &                      &                      & &                      &                      &                      &                      & & $J$\,=\,20-19        & 160                  & APEX                 & (11)                 \\
\multicolumn{14}{c}{} \\
\multicolumn{14}{l}{CRL190} \\
\multicolumn{4}{c}{SiO} & & \multicolumn{4}{c}{CS} & & \multicolumn{4}{c}{SiS} \\
\hline
$J$\,=\,2-1          &   0.19               & IRAM                 & (1)                  & & $J$\,=\,1-0          &   0.83               & Yebes                & (1)                  & & $J$\,=\,7-6          &   1.45               & IRAM                 & (2)                  \\
$J$\,=\,3-2          &   0.39               & IRAM                 & (2)                  & & $J$\,=\,2-1          &   3.90               & IRAM                 & (1)                  & & $J$\,=\,8-7          &   1.93               & IRAM                 & (2)                  \\
                     &                      &                      &                      & & $J$\,=\,3-2          &   6.8                & IRAM                 & (2)                  & & $J$\,=\,14-13        &   1.67               & IRAM                 & (1)                  \\
                     &                      &                      &                      & & $J$\,=\,5-4          &   8.7                & IRAM                 & (1)                  & & $J$\,=\,16-15        &   2.06               & IRAM                 & (1)                  \\
                     &                      &                      &                      & & $J$\,=\,6-5          &   7.1                & IRAM                 & (1)                  & &                      &                      &                      &                      \\
\end{longtable}
\hspace{0.1cm} \tablenotec{References: (1)\,This work. (2)\,\cite{Massalkhi2019}. (3)\,\cite{Massalkhi2020}. (4)\,\cite{Velilla-Prieto2017}. (5)\,\cite{Agundez2012}. (6)\,\cite{Gonzalez-Delgado2003}. (7)\,\cite{Schoier2006a}. (8)\,\cite{Bieging2000}. (9)\,\cite{Bujarrabal1994}. (10)\,\cite{Olofsson1998}. (11)\,\cite{Schoier2007}. (12)\,\cite{Woods2003}. (13)\,\cite{Danilovich2018}. (14)\,\cite{Danilovich2015}. $^a$\,Values have been revised with respect to those given in the original reference.} \\

\clearpage

\begin{figure*}
\centering
\includegraphics[angle=0,width=0.88\textwidth]{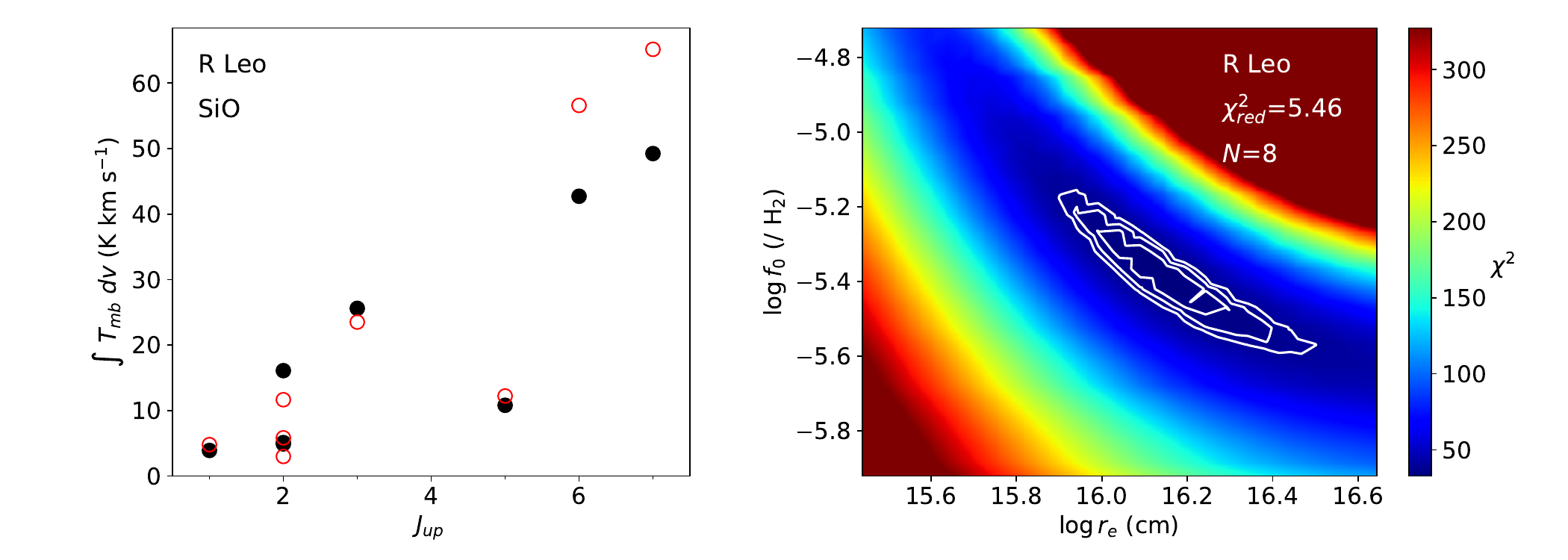} \includegraphics[angle=0,width=0.88\textwidth]{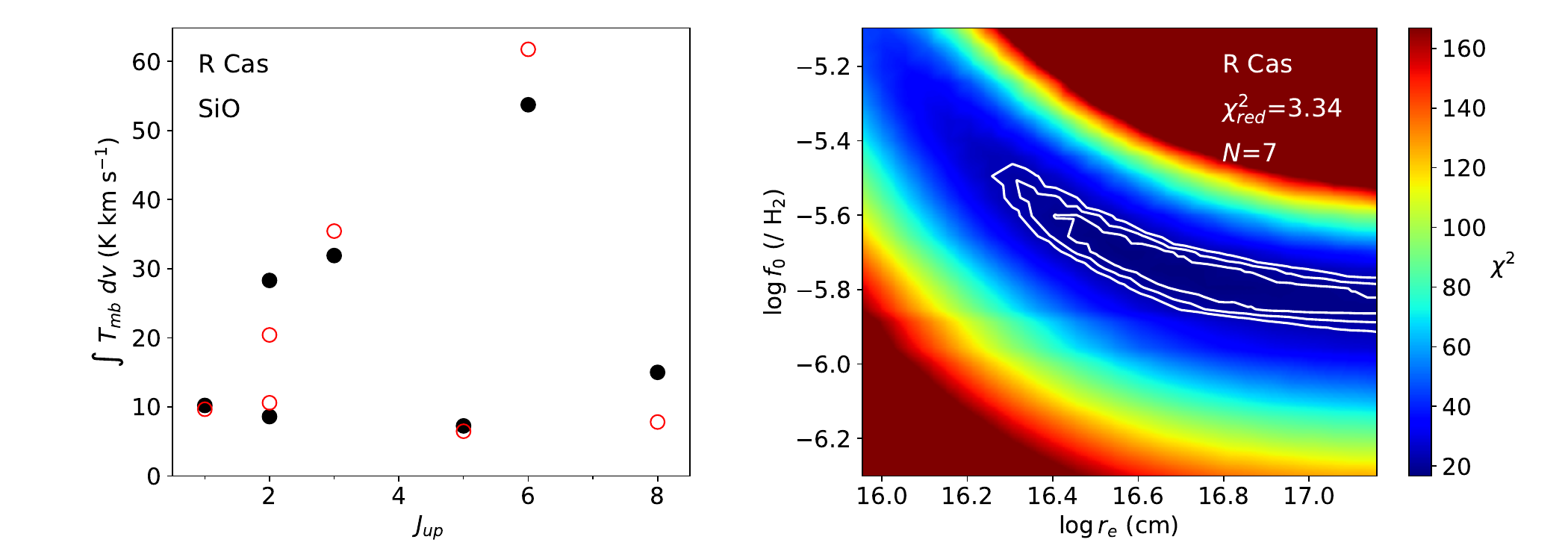} \includegraphics[angle=0,width=0.88\textwidth]{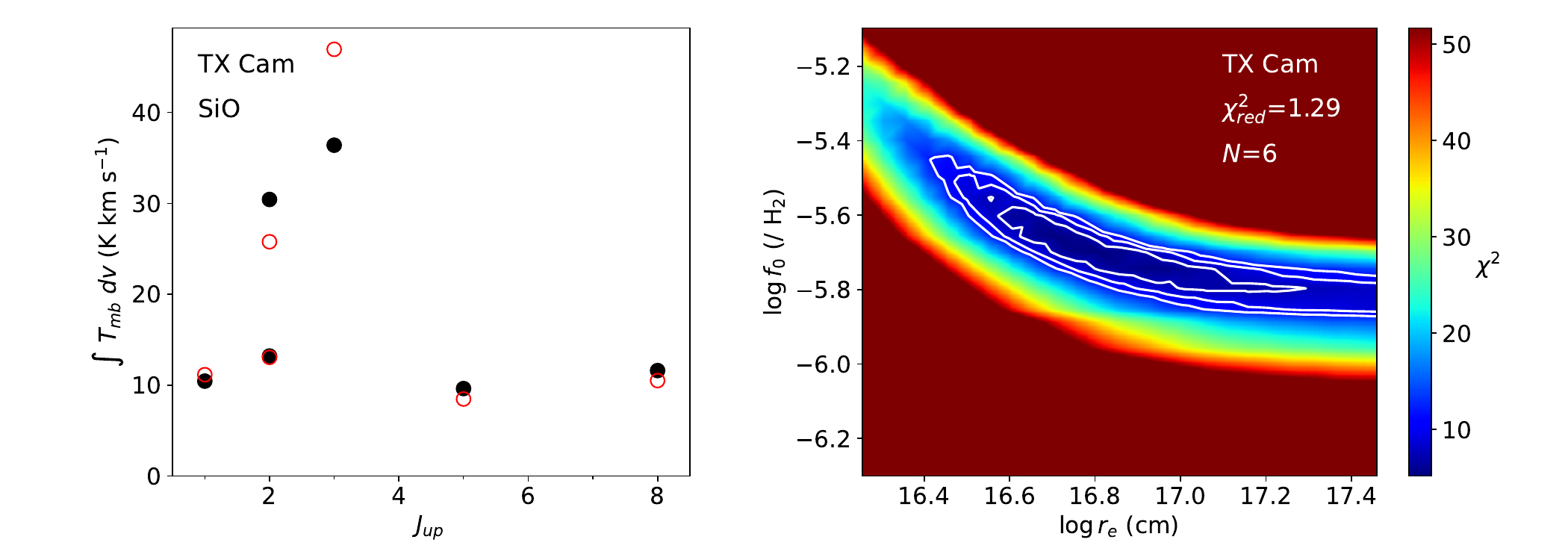} \includegraphics[angle=0,width=0.88\textwidth]{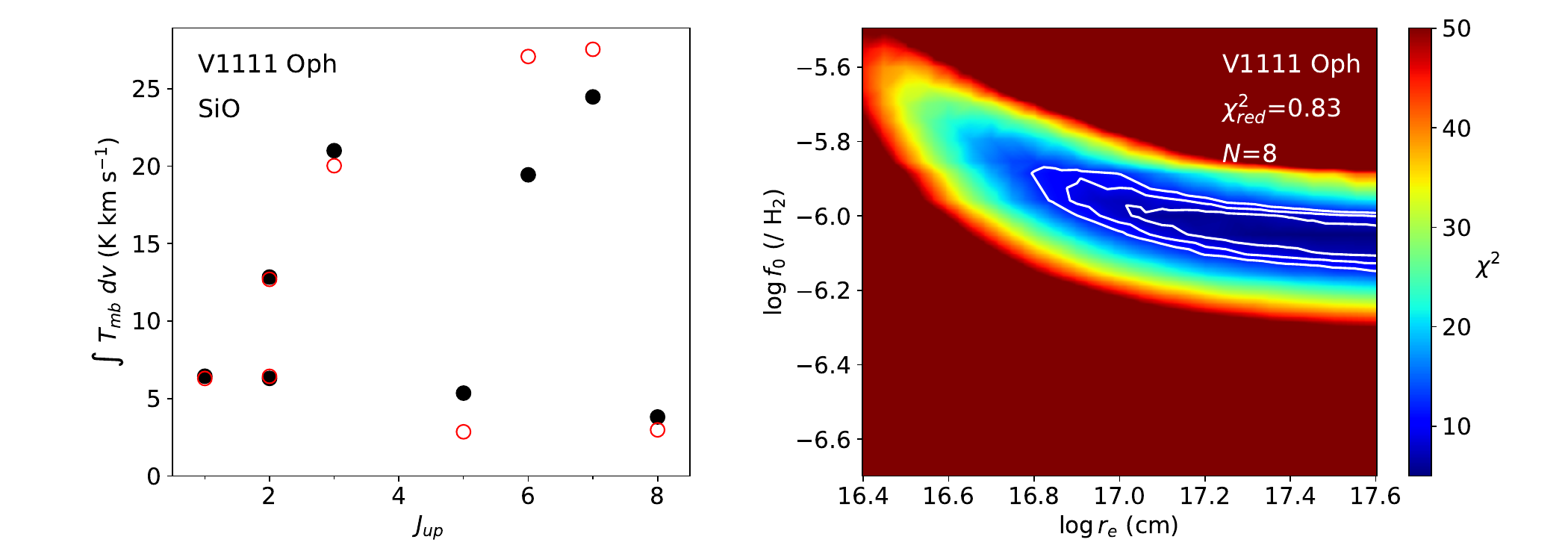}
\caption{Results from SiO analysis for all envelopes except IK\,Tau and IRC\,+10216 (shown in Fig.\,\ref{fig:iktau} and Fig.\,\ref{fig:irc10216}). The left panels show the observed velocity-integrated intensities as black filled circles (see Table\,\ref{table:lines_all}) and the calculated ones as red empty circles. The right panels show $\chi^2$ as a function of the logarithm of the fractional abundance of SiO relative to H$_2$, $\log f_0$, and the logarithm of the $e$-folding radius, $\log r_e$. The white contours correspond to 1, 2, and 3\,$\sigma$ levels.} \label{fig:sio}
\end{figure*}

\setcounter{figure}{3}
\begin{figure*}
\centering
\includegraphics[angle=0,width=0.88\textwidth]{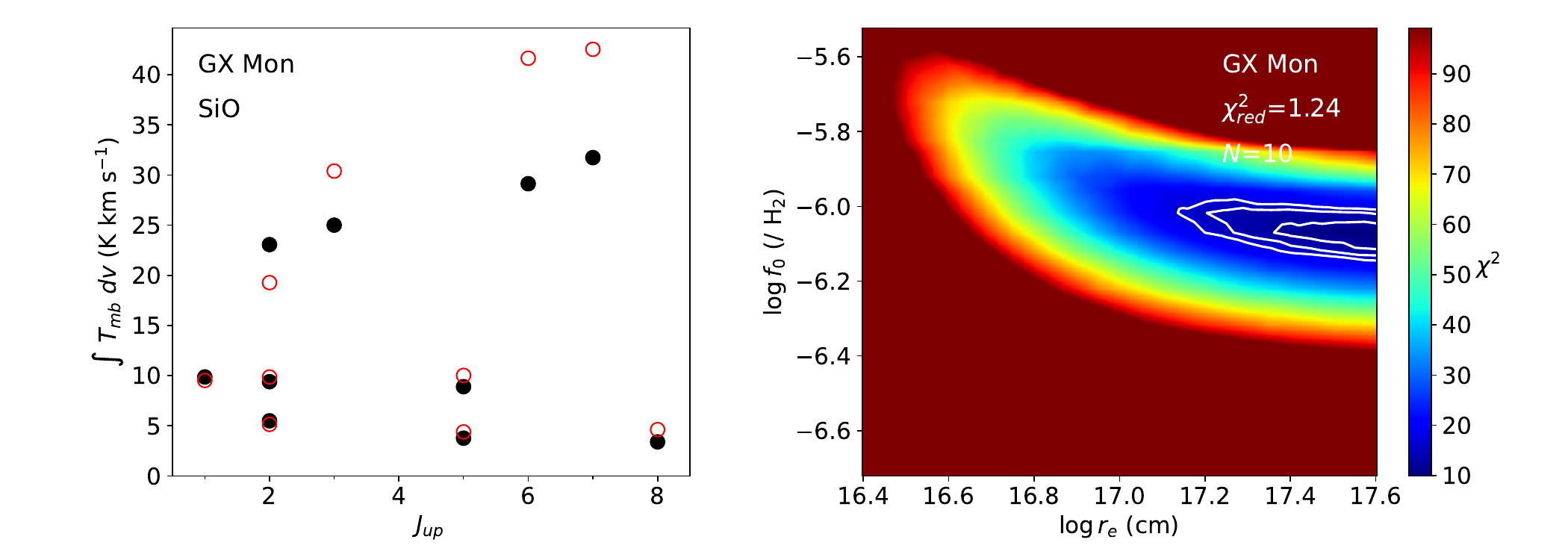} \includegraphics[angle=0,width=0.88\textwidth]{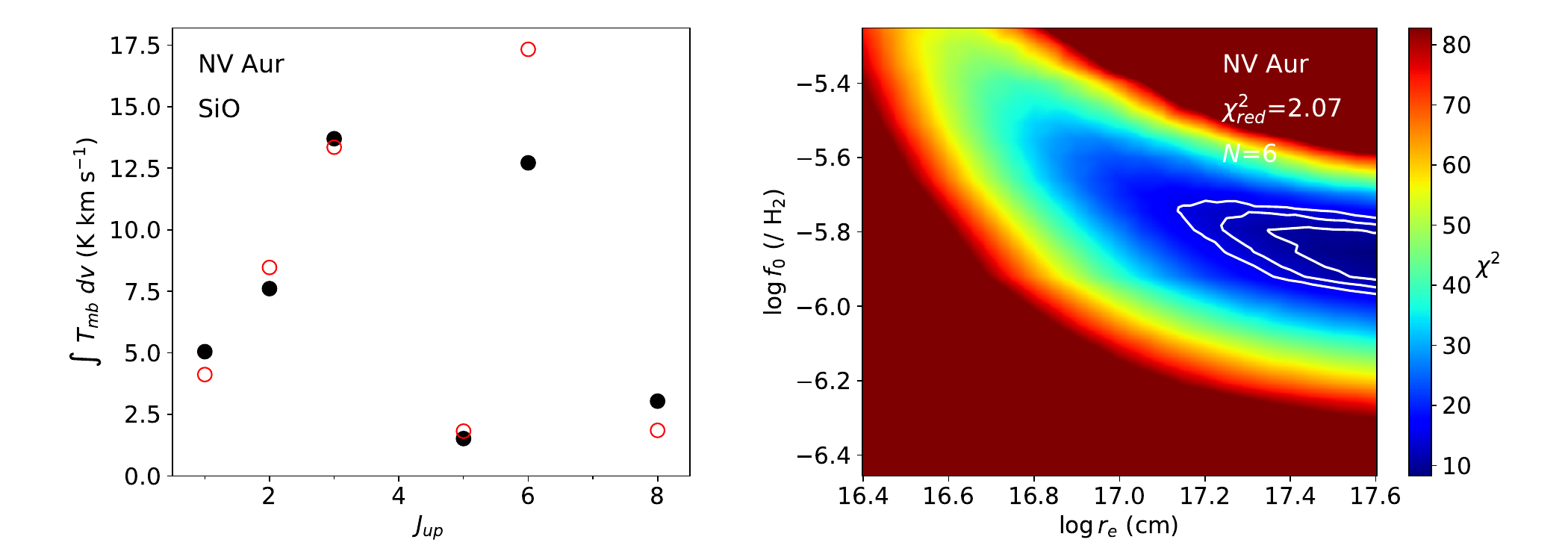} \includegraphics[angle=0,width=0.88\textwidth]{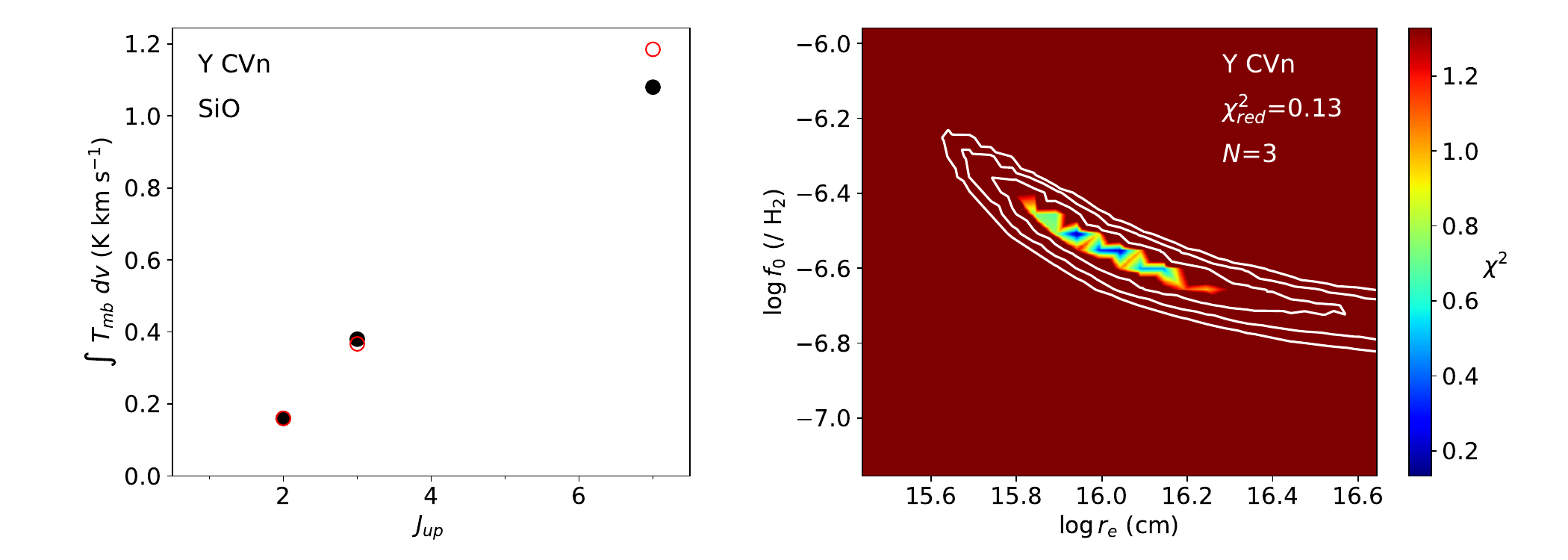} \includegraphics[angle=0,width=0.88\textwidth]{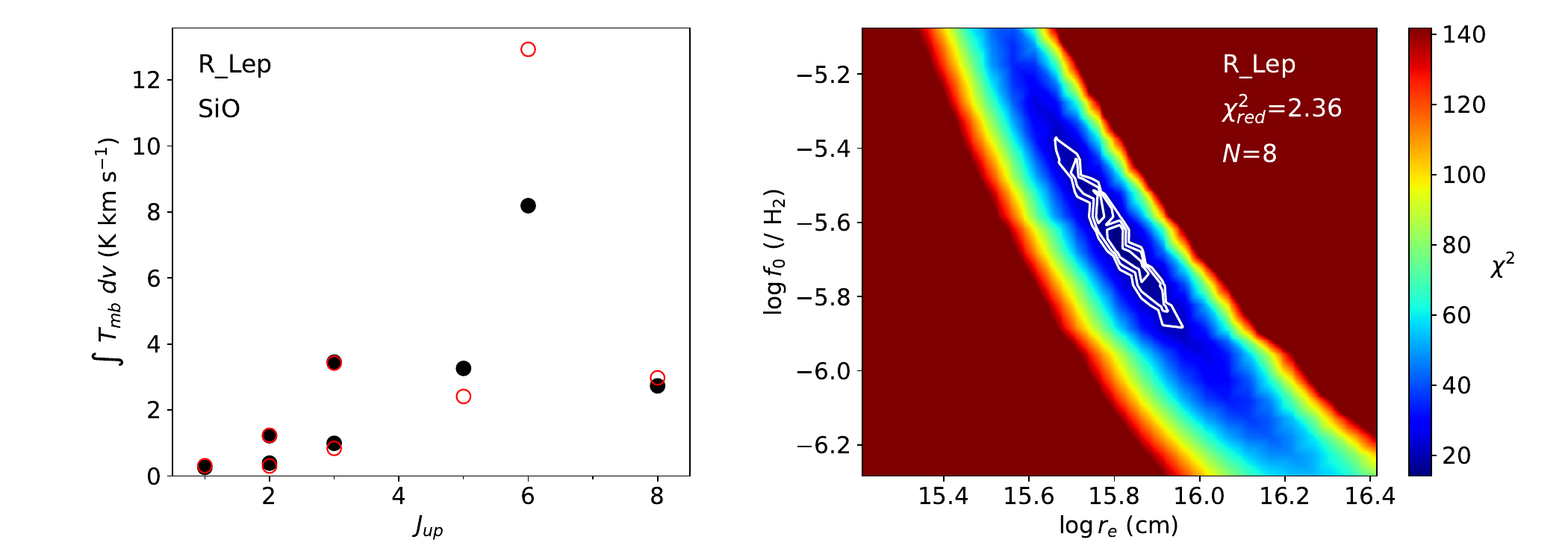} 
\caption{continued.}
\end{figure*}

\setcounter{figure}{3}
\begin{figure*}
\centering
\includegraphics[angle=0,width=0.88\textwidth]{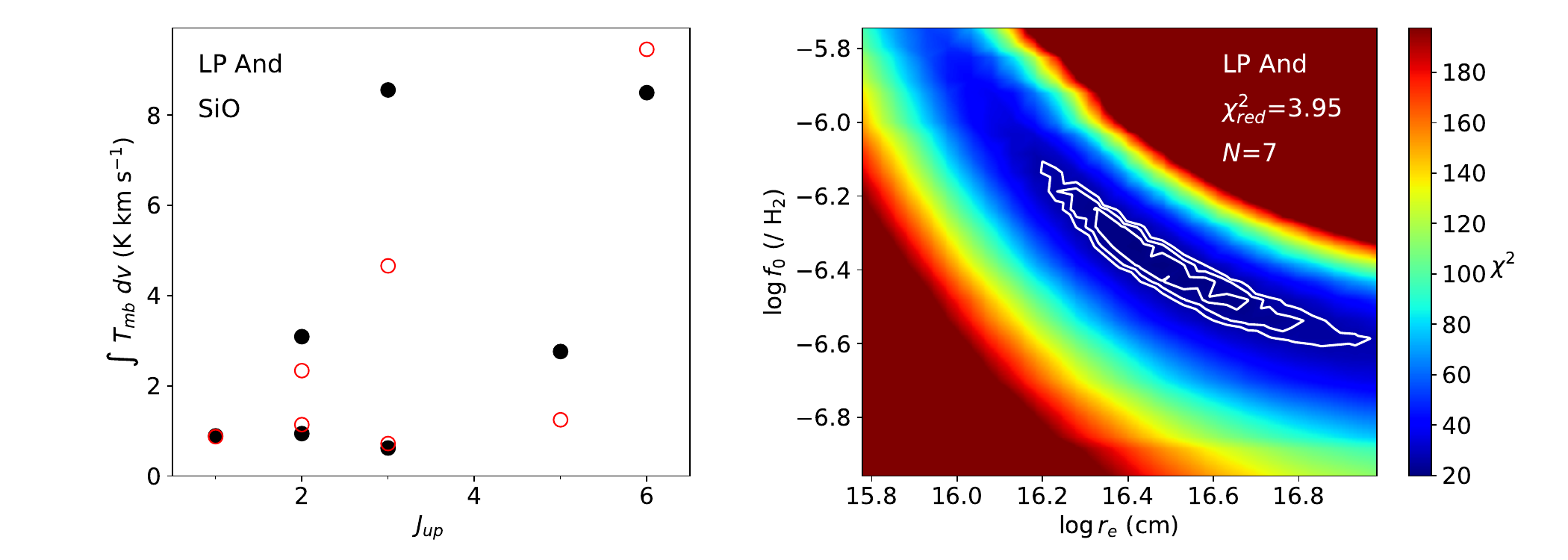} \includegraphics[angle=0,width=0.88\textwidth]{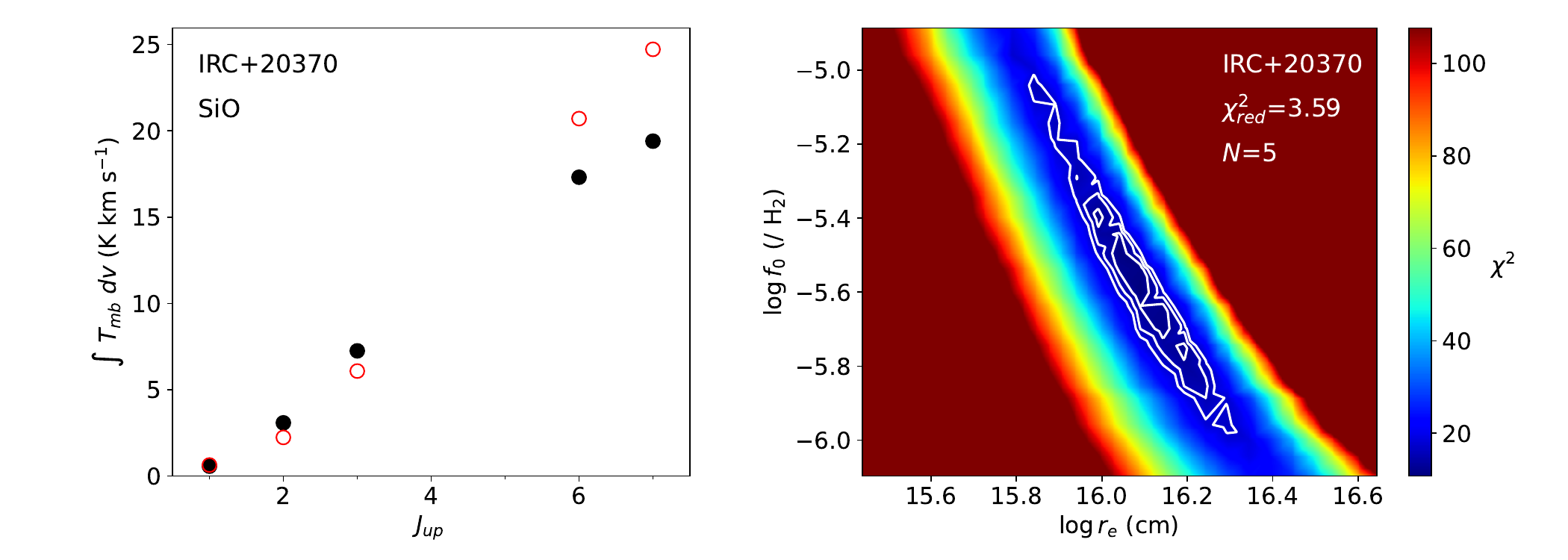} \includegraphics[angle=0,width=0.88\textwidth]{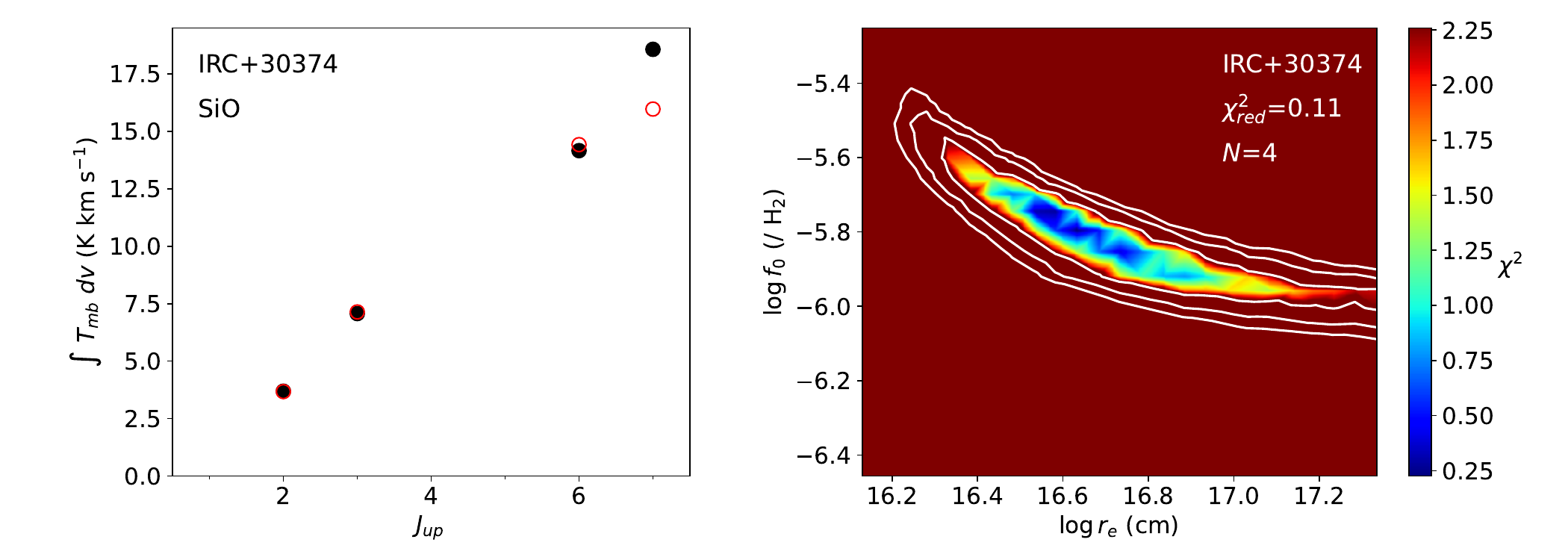} \includegraphics[angle=0,width=0.88\textwidth]{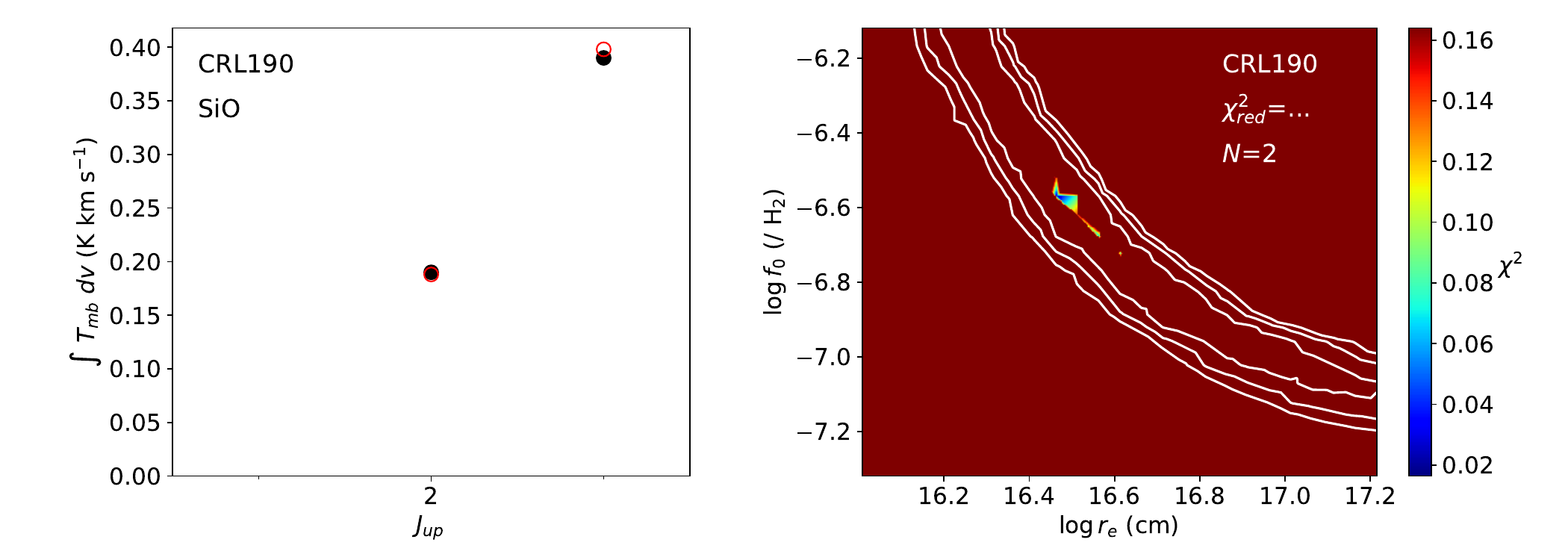} 
\caption{continued.}
\end{figure*}

\clearpage

\begin{figure*}
\centering
\includegraphics[angle=0,width=0.88\textwidth]{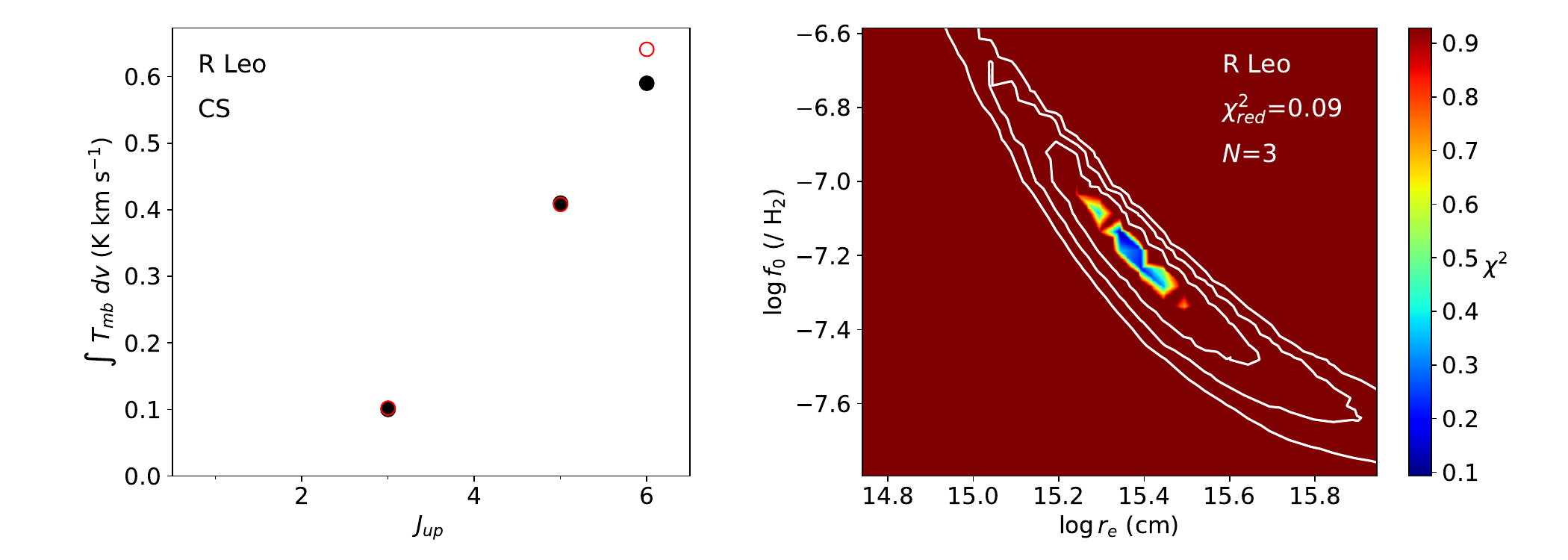} \includegraphics[angle=0,width=0.88\textwidth]{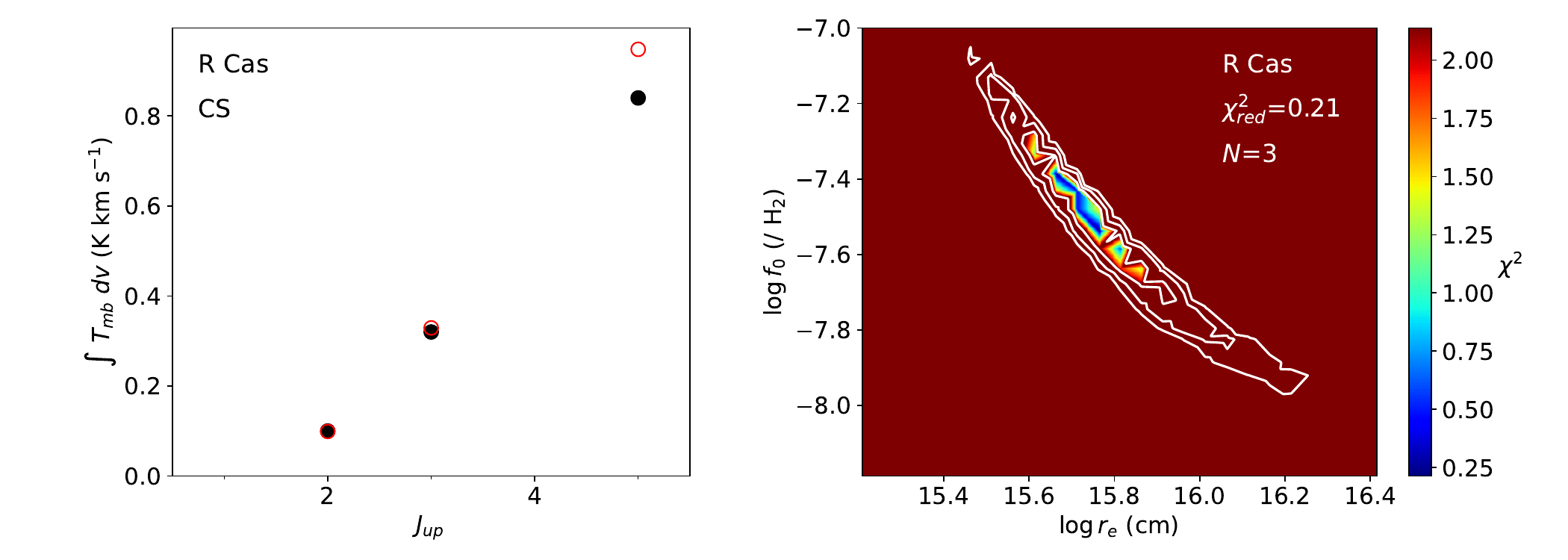} \includegraphics[angle=0,width=0.88\textwidth]{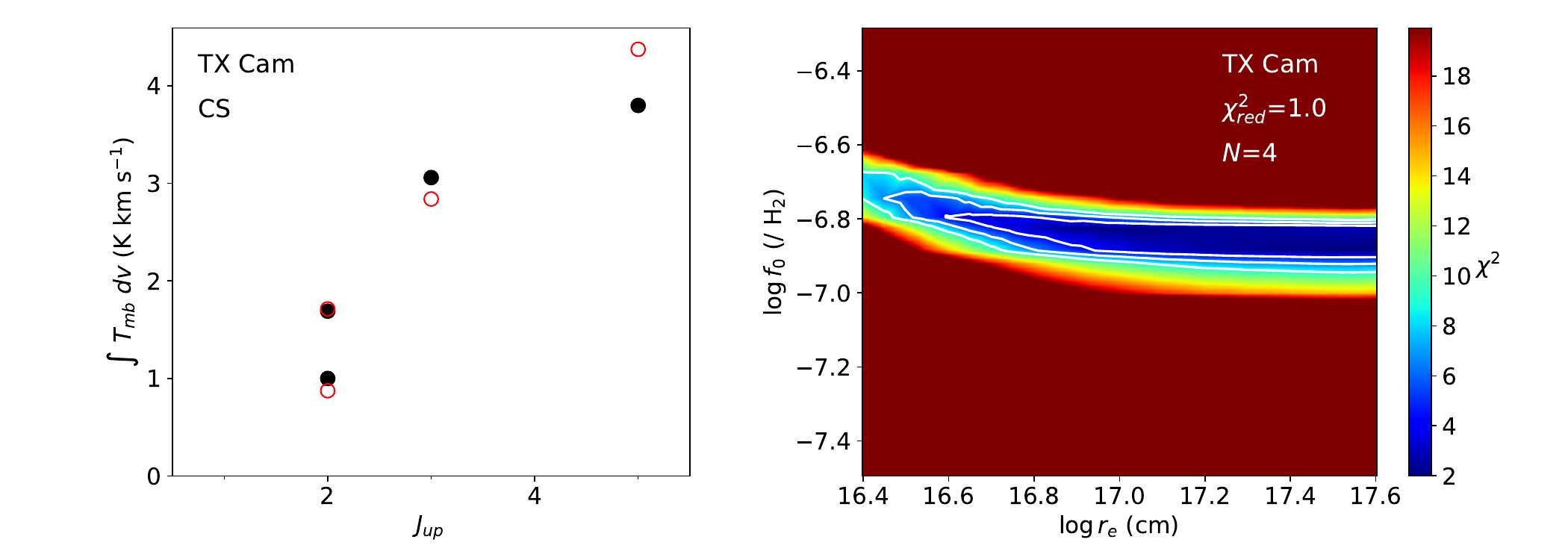} \includegraphics[angle=0,width=0.88\textwidth]{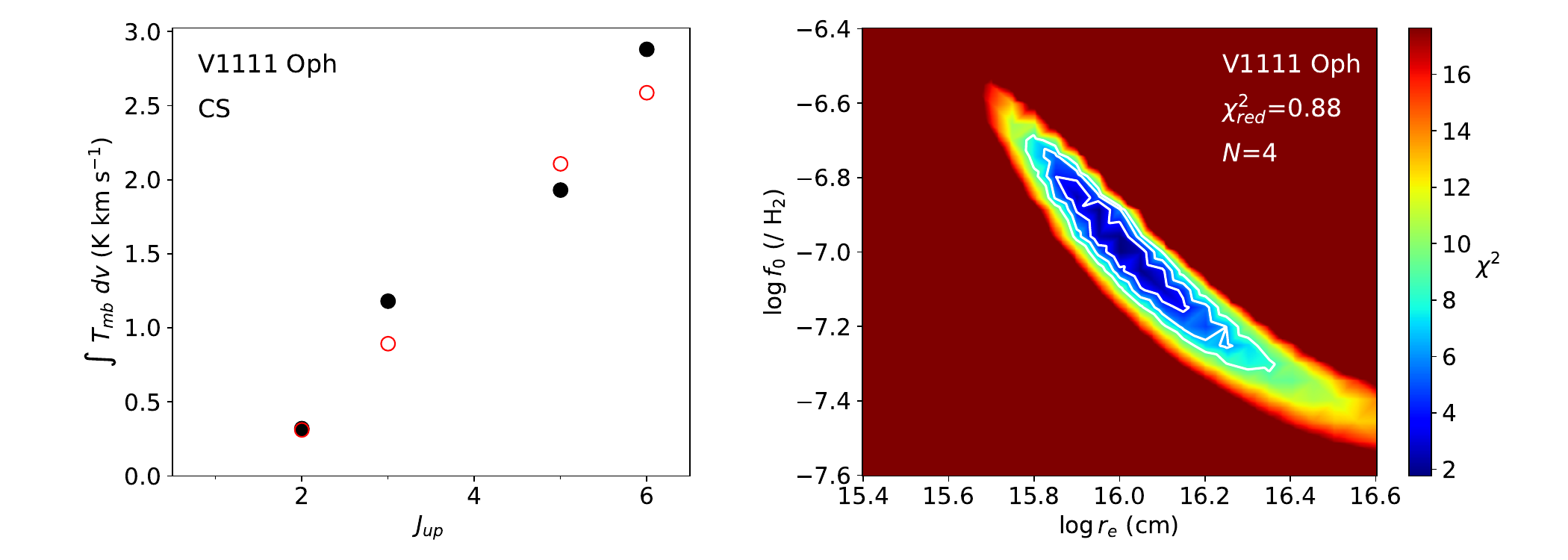}
\caption{Results from CS analysis for all envelopes except IK\,Tau and IRC\,+10216 (shown in Fig.\,\ref{fig:iktau} and Fig.\,\ref{fig:irc10216}). The left panels show the observed velocity-integrated intensities as black filled circles (see Table\,\ref{table:lines_all}) and the calculated ones as red empty circles. The right panels show $\chi^2$ as a function of the logarithm of the fractional abundance of CS relative to H$_2$, $\log f_0$, and the logarithm of the $e$-folding radius, $\log r_e$. The white contours correspond to 1, 2, and 3\,$\sigma$ levels.} \label{fig:cs}
\end{figure*}

\setcounter{figure}{4}
\begin{figure*}
\centering
\includegraphics[angle=0,width=0.88\textwidth]{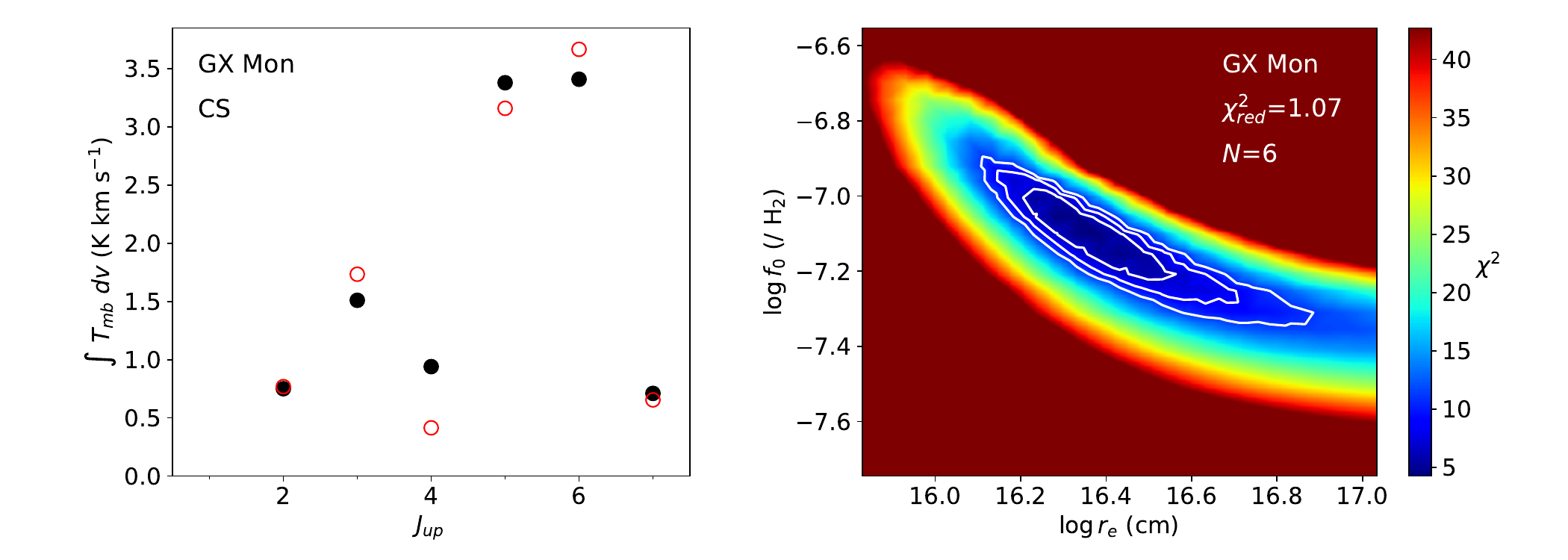} \includegraphics[angle=0,width=0.88\textwidth]{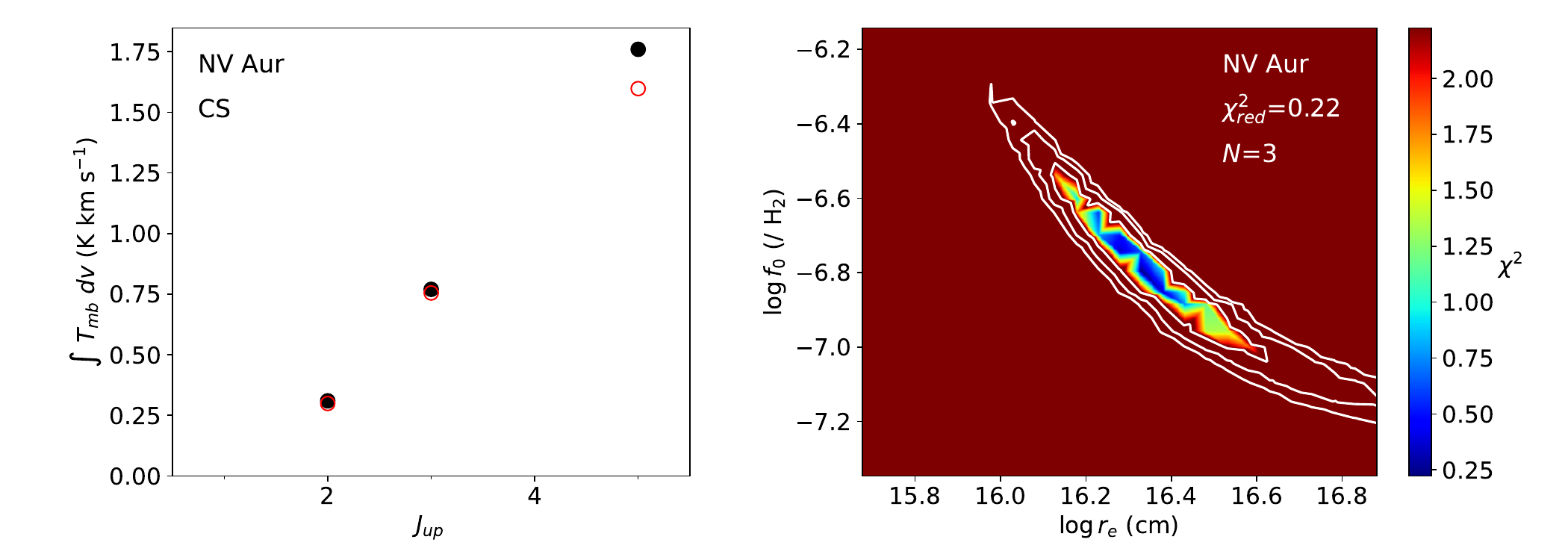} \includegraphics[angle=0,width=0.88\textwidth]{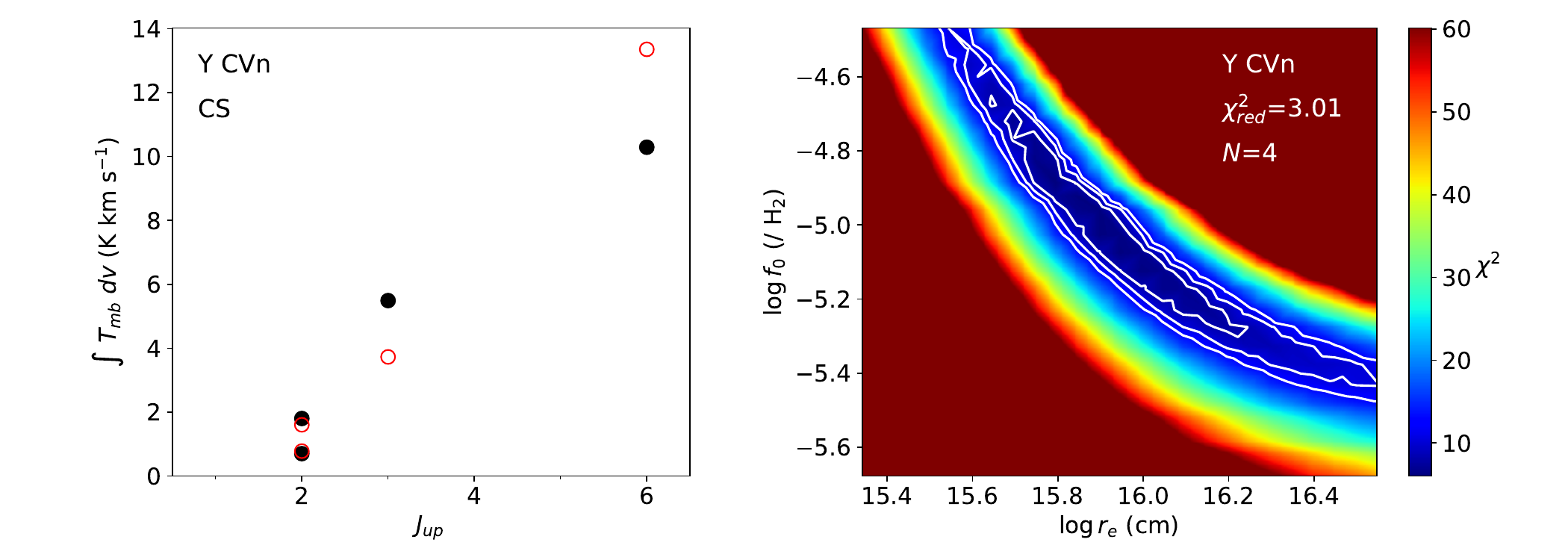} \includegraphics[angle=0,width=0.88\textwidth]{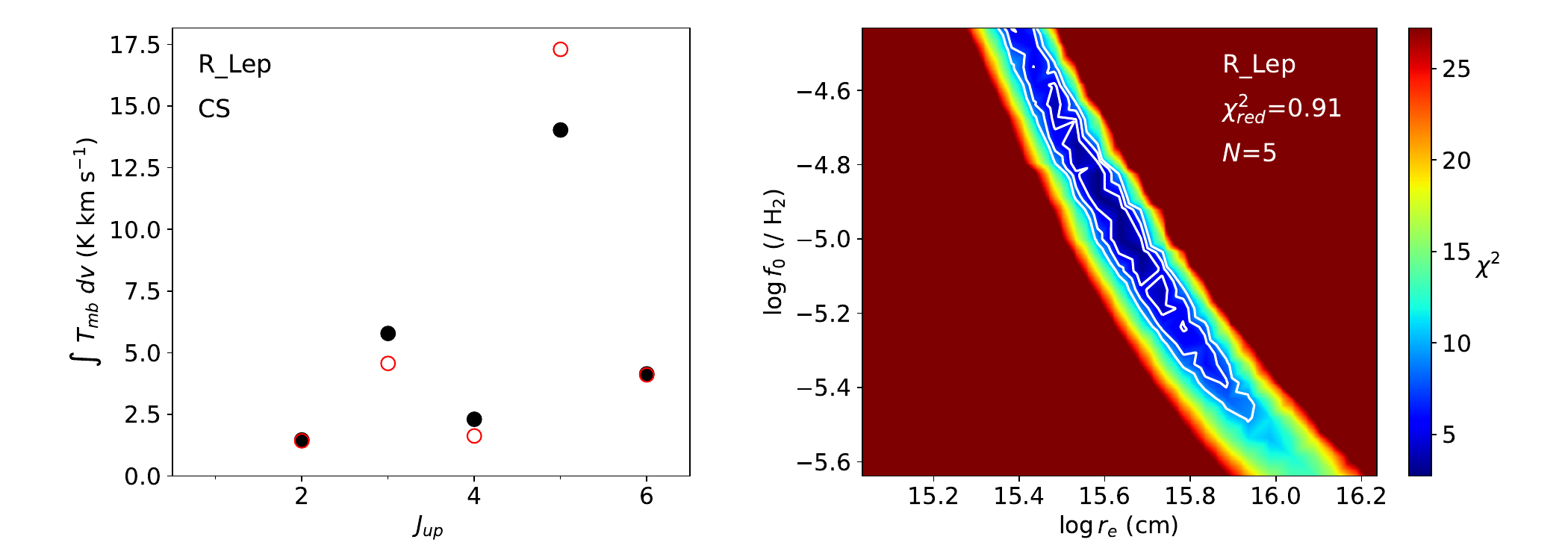} 
\caption{continued.}
\end{figure*}

\setcounter{figure}{4}
\begin{figure*}
\centering
\includegraphics[angle=0,width=0.88\textwidth]{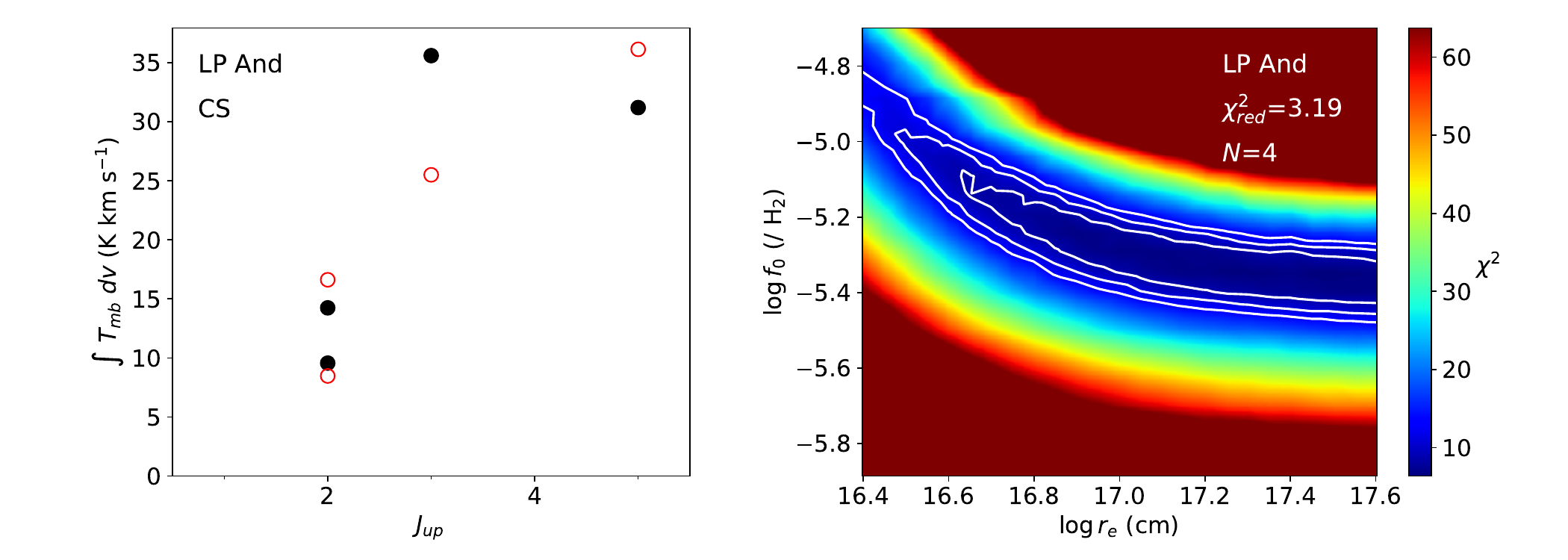} \includegraphics[angle=0,width=0.88\textwidth]{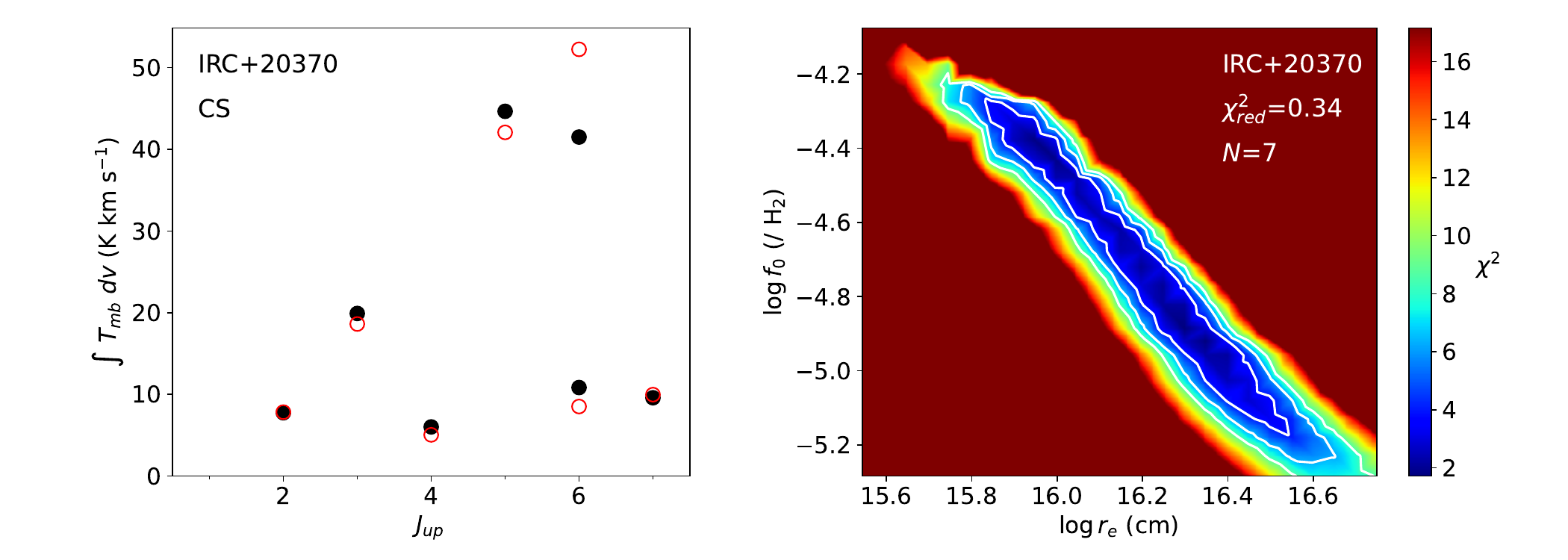} \includegraphics[angle=0,width=0.88\textwidth]{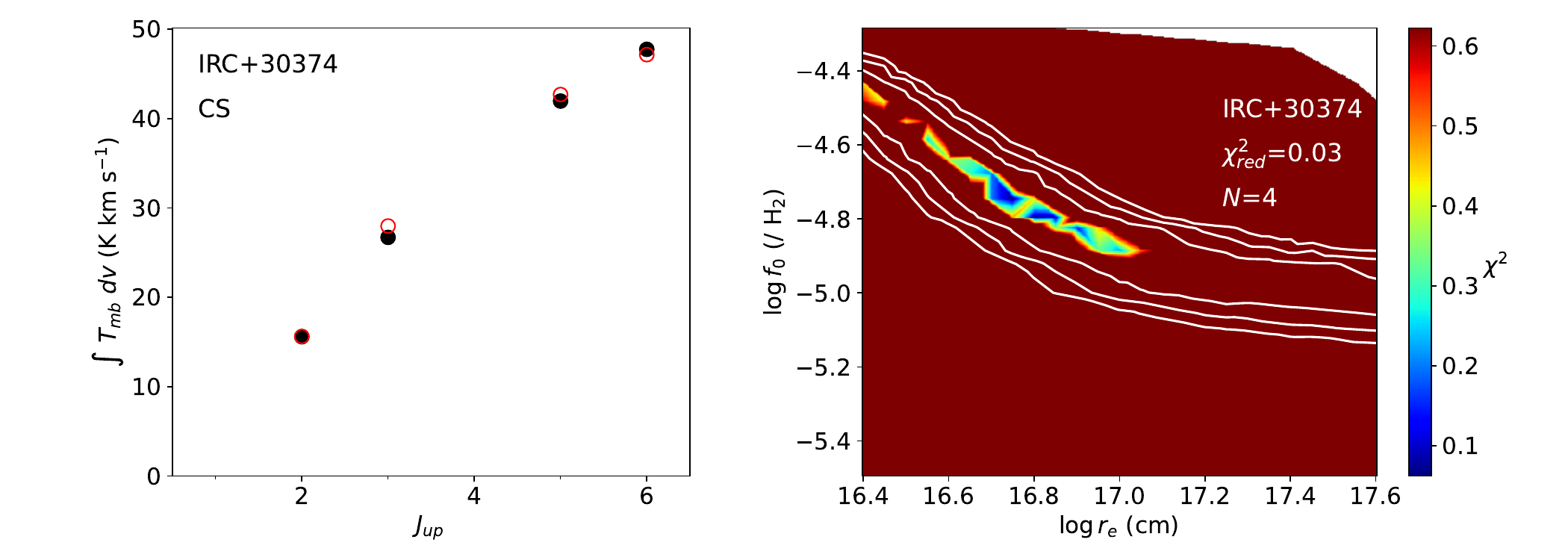} 
\caption{continued.}
\end{figure*}

\clearpage

\begin{figure*}
\centering
\includegraphics[angle=0,width=0.88\textwidth]{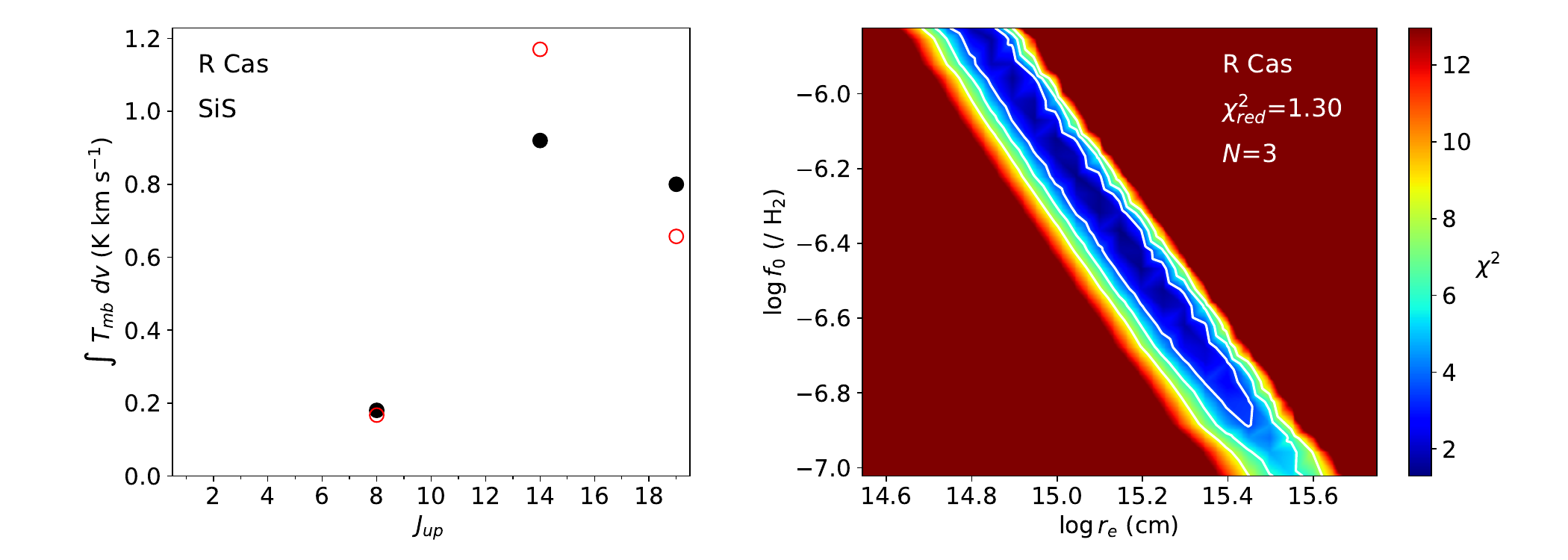} \includegraphics[angle=0,width=0.88\textwidth]{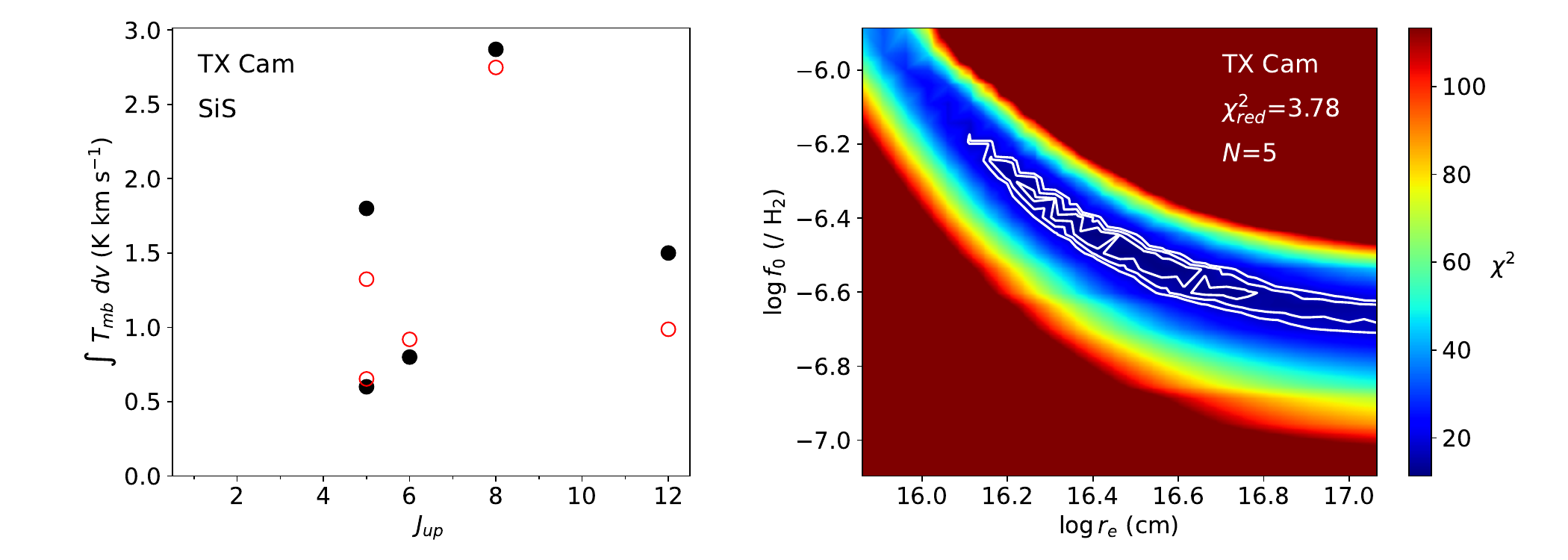} \includegraphics[angle=0,width=0.88\textwidth]{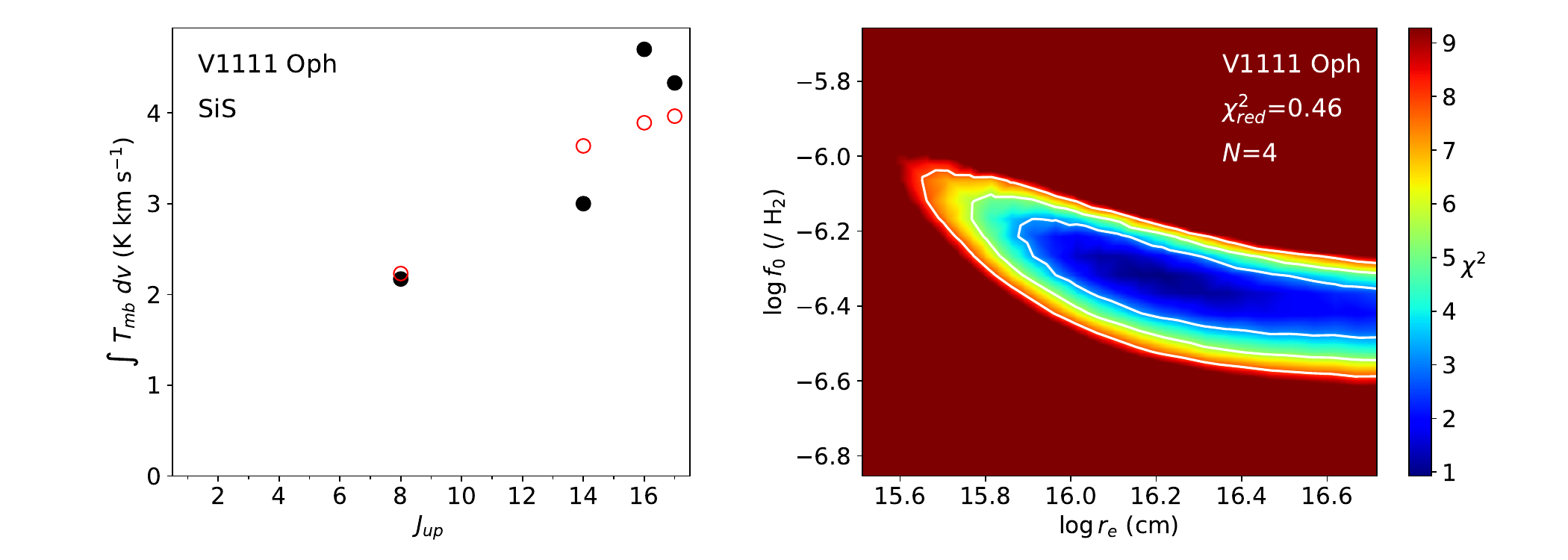} \includegraphics[angle=0,width=0.88\textwidth]{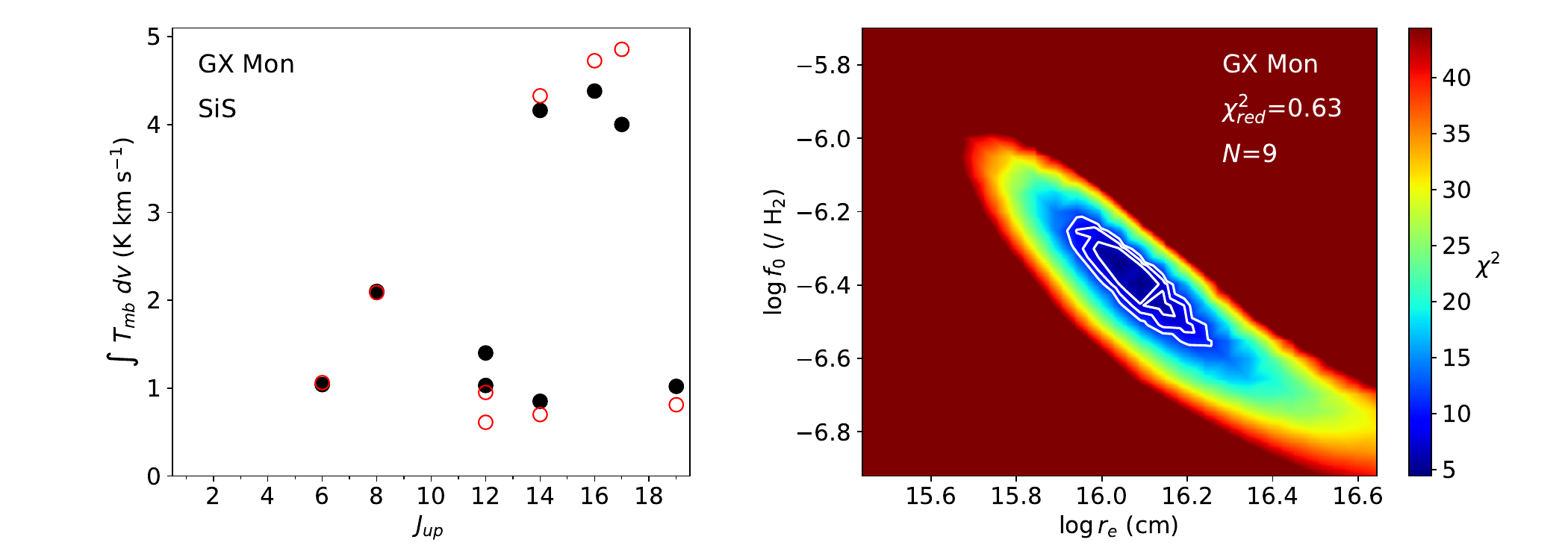}
\caption{Results from SiS analysis for all envelopes except IK\,Tau and IRC\,+10216 (shown in Fig.\,\ref{fig:iktau} and Fig.\,\ref{fig:irc10216}). The left panels show the observed velocity-integrated intensities as black filled circles (see Table\,\ref{table:lines_all}) and the calculated ones as red empty circles. The right panels show $\chi^2$ as a function of the logarithm of the fractional abundance of SiS relative to H$_2$, $\log f_0$, and the logarithm of the $e$-folding radius, $\log r_e$. The white contours correspond to 1, 2, and 3\,$\sigma$ levels.} \label{fig:sis}
\end{figure*}

\setcounter{figure}{5}
\begin{figure*}
\centering
\includegraphics[angle=0,width=0.88\textwidth]{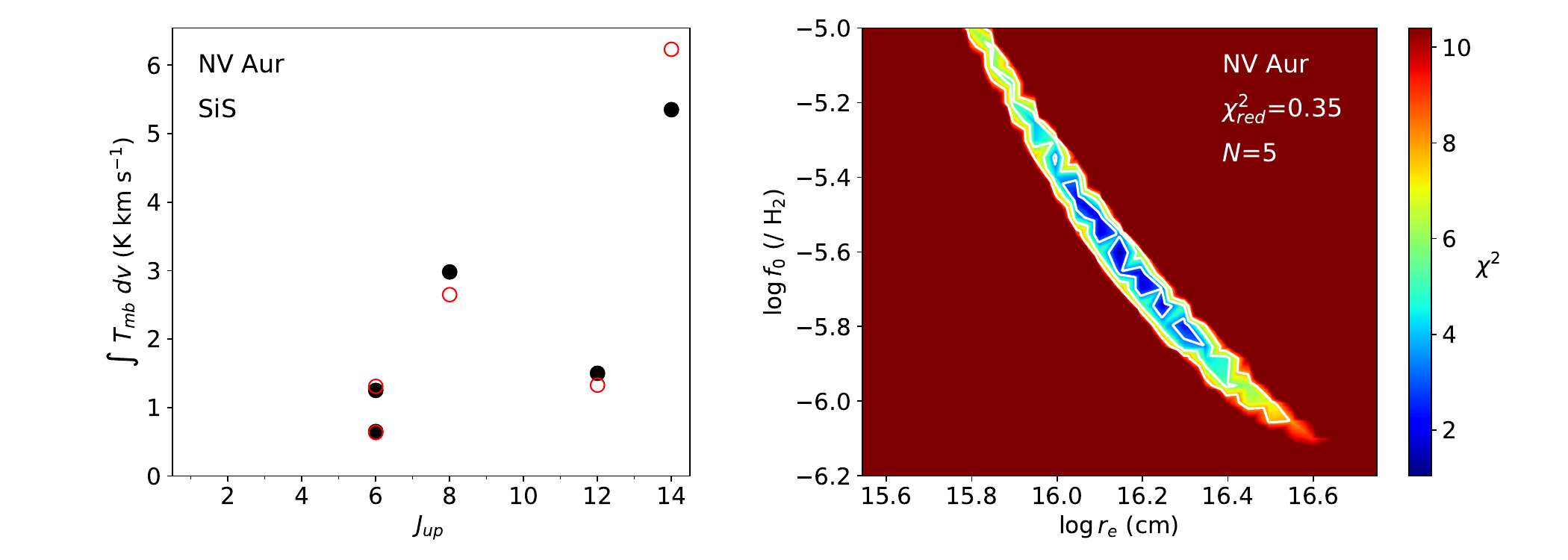} \includegraphics[angle=0,width=0.88\textwidth]{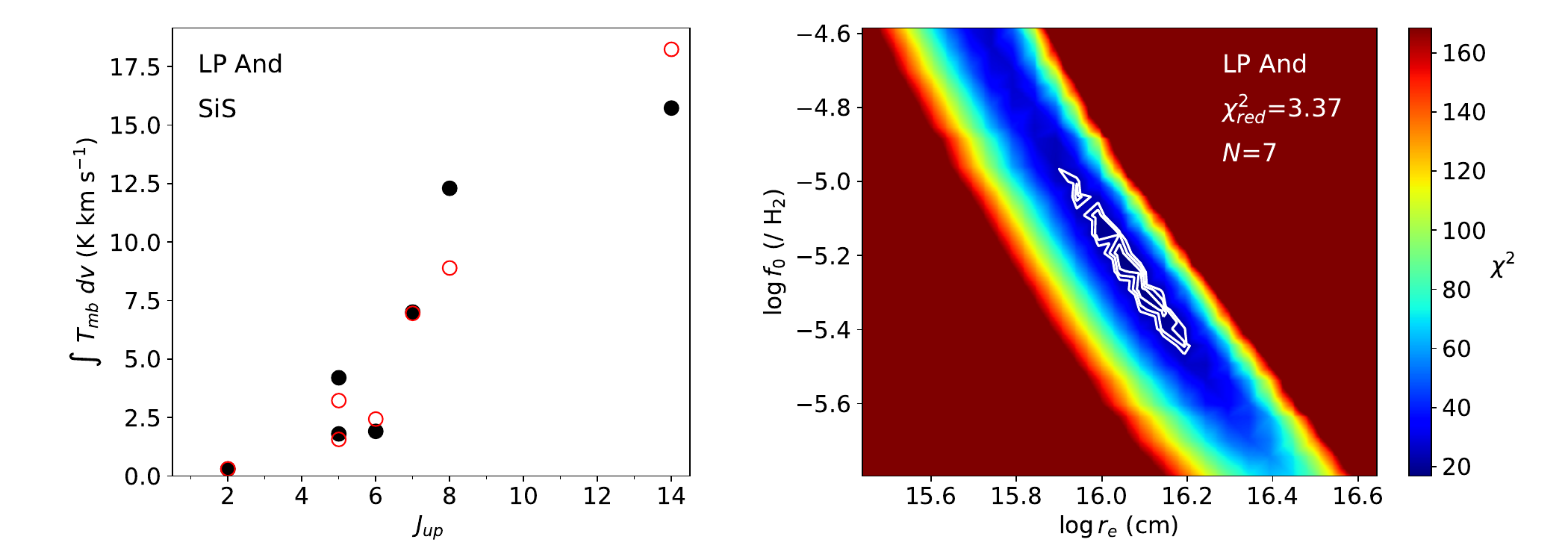} \includegraphics[angle=0,width=0.88\textwidth]{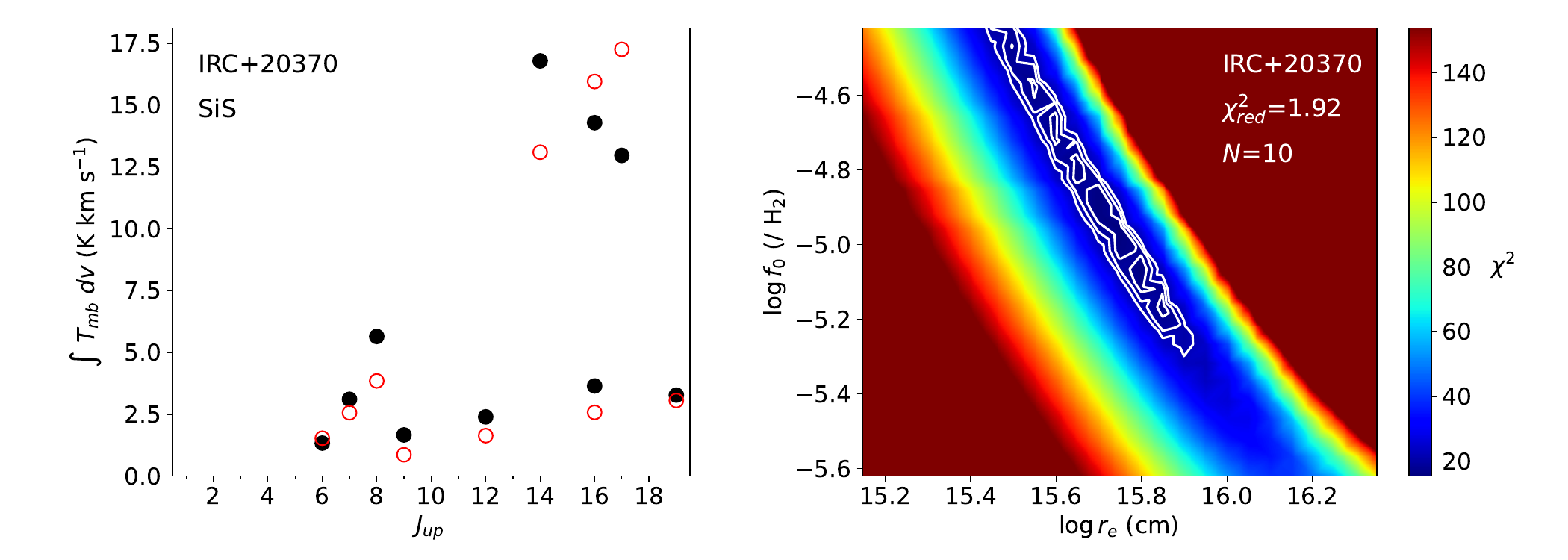} \includegraphics[angle=0,width=0.88\textwidth]{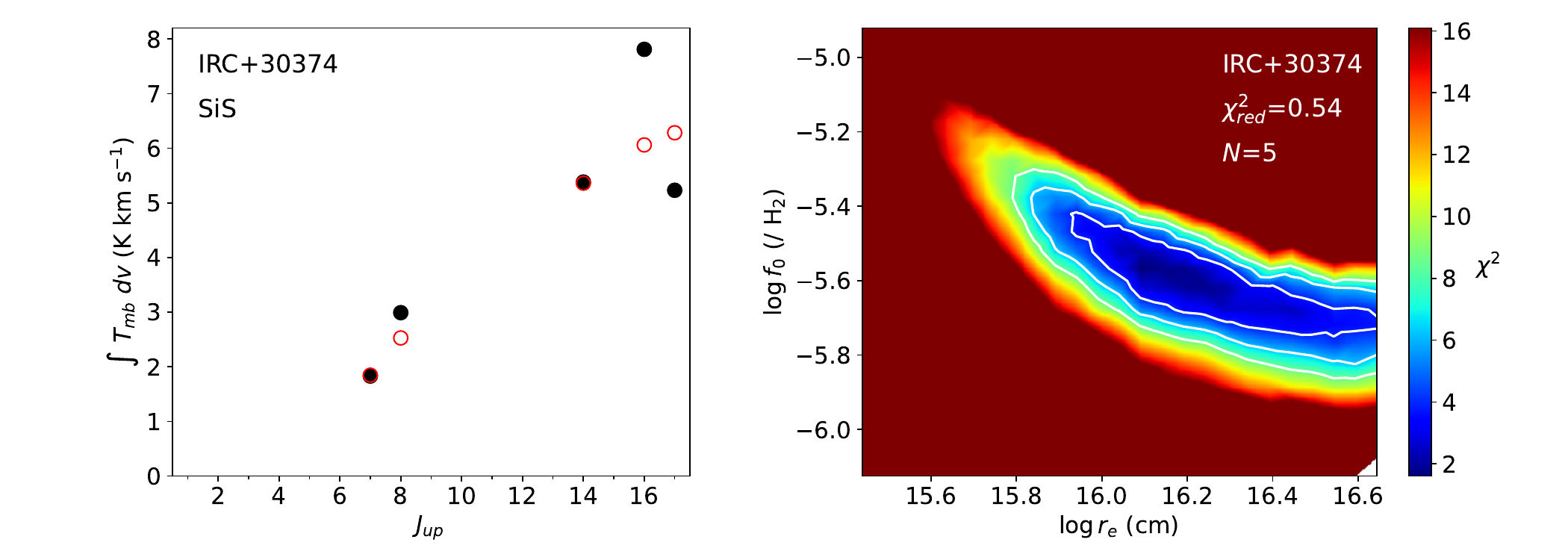} 
\caption{continued.}
\end{figure*}

\setcounter{figure}{5}
\begin{figure*}
\centering
\includegraphics[angle=0,width=0.88\textwidth]{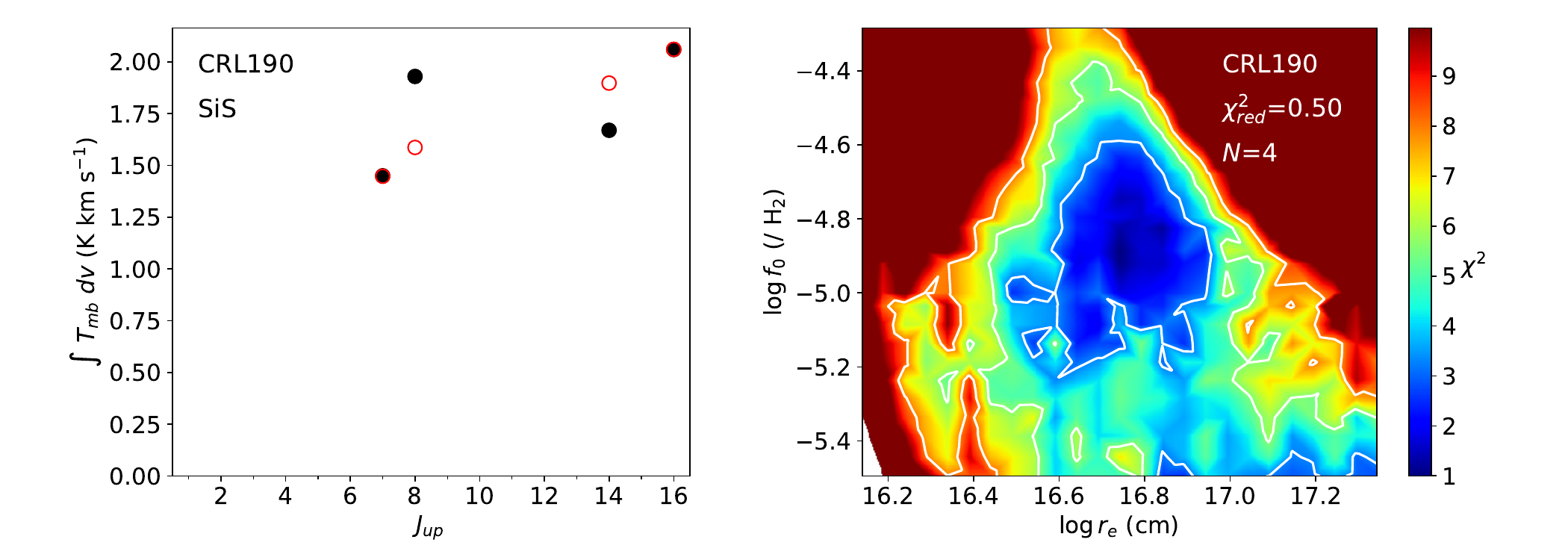}
\caption{continued.}
\end{figure*}


\begin{thebibliography}{}

\bibitem[Adam \& Ohnaka(2019)]{Adam2019} Adam, C. \& Ohnaka, K. 2019, \aap, 628, A132
\bibitem[Ag\'undez et al.(2012)]{Agundez2012} Ag\'undez, M., Fonfr\'ia, J. P., Cernicharo, J., et al. 2012, \aap, 543, A48
\bibitem[Ag\'undez et al.(2017)]{Agundez2017} Ag\'undez, M., Cernicharo, J., Quintana-Lacaci, G., et al. 2017, \aap, 601, A4
\bibitem[Ag\'undez et al.(2018)]{Agundez2018} Ag\'undez, M., Roueff, E., Le Petit, F., \& Le Bourlot, J. 2018, \aap, 616, A19
\bibitem[Ag\'undez et al.(2020)]{Agundez2020} Ag\'undez, M., Mart\'inez, J. I., de Andres, P. L., et al. 2020, \aap, 637, A59
\bibitem[Andriantsaralaza et al.(2022)]{Andriantsaralaza2022} Andriantsaralaza, M., Ramstedt, S., Vlemmings, W. H. T., \& De Beck, E. 2022, \aap, 667, A74
\bibitem[Balan{\c{c}}a \& Dayou(2017)]{Balanca2017} Balan{\c{c}}a, C. \& Dayou, F. 2017, \mnras, 469, 1673
\bibitem[Balan{\c{c}}a et al.(2018)]{Balanca2018} Balan{\c{c}}a, C., Dayou, F., Faure, A., et al. 2018, \mnras, 479, 2692
\bibitem[Beichman et al.(1988)]{Beichman1988} Beichman, C. A., Neugebauer, G., Habing, H. J., Clegg, P. E. \& Chester, T. J. 1988, Infrared Astronomical Satellite (IRAS) Catalogs and Atlases, vol. 1, Explanatory Supplement
\bibitem[Bergeat et al.(2001)]{Bergeat2001} Bergeat, J., Knapik, A., \& Rutily, B. 2001, \aap, 369, 178
\bibitem[Bieging et al.(2000)]{Bieging2000} Bieging, J. H., Shaked, S., \& Gensheimer, P. D. 2000, \apj, 543, 897
\bibitem[Bohlin et al.(1978)]{Bohlin1978} Bohlin, R. C., Savage, B. D., \& Drake, J. F. 1978, \apjs, 224, 132
\bibitem[Bruna et al.(1975)]{Bruna1975} Bruna, P. J., Kammer, W. E., \& Vasudevan, K. 1975, \cp, 9, 91
\bibitem[Bujarrabal et al.(1994)]{Bujarrabal1994} Bujarrabal, V., Fuente, A., \& Omont, A. 1994, \aap, 285, 247
\bibitem[Castro-Carrizo et al.(2010)]{Castro-Carrizo2010} Castro-Carrizo, A., Quintana-Lacaci, G., Neri, R., et al. 2010, \aap, 523, A59
\bibitem[Cecchi-Pestellini et al.(2002)]{Cecchi-Pestellini2002} Cecchi-Pestellini, C., Bodo, E., Balakrishnan, N., \& Dalgarno, A. 2002, \apj, 571, 1015
\bibitem[Cernicharo et al.(2000)]{Cernicharo2000} Cernicharo, J., Gu\'elin, M., \& Kahane, C. 2000, \aaps, 142, 181
\bibitem[Cernicharo et al.(2015)]{Cernicharo2015} Cernicharo, J., Marcelino, N., Ag\'undez, M., \& Gu\'elin, M. 2015, \aap, 575, A91
\bibitem[Chambers et al.(2019)]{Chambers2019} Chambers, K. C., Magnier, E. A., Metcalfe, N., et al. 2019, arXiv:1612.05560
\bibitem[Chandra et al.(1995)]{Chandra1995} Chandra, S., Kegel, W. H., Le Roy, R. J., \& Hertenstein, T. 1995, \aaps, 114, 175
\bibitem[Daniel et al.(2012)]{Daniel2012} Daniel, F., Ag\'undez, M., Cernicharo, J., et al. 2012, \aap, 542, A37
\bibitem[Danilovich et al.(2015)]{Danilovich2015} Danilovich, T., Teyssier, D., Justtanont, K., et al. 2015, \aap, 581, A60
\bibitem[Danilovich et al.(2018)]{Danilovich2018} Danilovich, T., Ramstedt, S., Gobrecht, D., et al. 2018, \aap, 617, A132
\bibitem[Danilovich et al.(2019)]{Danilovich2019} Danilovich, T., Richards, A. M. S., Karakas, A. I., et al. 2019, \mnras, 484, 494
\bibitem[Dayou \& Balan{\c{c}}a(2006)]{Dayou2006} Dayou, F. \& Balan{\c{c}}a, C. 2006, \aap, 459, 297
\bibitem[De Beck et al.(2010)]{DeBeck2010} De Beck, E., Decin, L., de Koter, A., et al. 2010, \aap, 523, A18
\bibitem[Decin et al.(2010)]{Decin2010} Decin, L., Justtanont, K., De Beck, E., et al. 2010, \aap, 521, L4
\bibitem[Denis-Alpizar et al.(2018)]{Denis-Alpizar2018} Denis-Alpizar, O., Stoecklin, T., Guilloteau, S., \& Dutrey, A. 2018, \mnras, 478, 1811
\bibitem[Drira et al.(1997)]{Drira1997} Drira, I., Hur\'e, J. M., Spielfiedel, A., et al. 1997, \aap, 319, 720
\bibitem[Egan et al.(2003)]{Egan2003} Egan, M. P., Price, S. D., Kraemer, K. E., et al. 2003, Air Force Research Laboratory Technical Report, AFRL-VS-TR-2003-1589
\bibitem[Flewelling et al.(2019)]{Flewelling2019} Flewelling, H. A., Magnier, E. A., Chambers, K. C., et al. 2019, arXiv:1612.05243
\bibitem[Fonfr\'ia Exp\'osito et al.(2006)]{Fonfria2006} Fonfr\'ia Exp\'osito, J. P., Ag\'undez, M., Tercero, B., et al. 2006, \apj, 646, L127
\bibitem[Fonfr\'ia et al.(2022)]{Fonfria2022} Fonfr\'ia, J. P., DeWitt, C. N., Montiel, E. J., et al. 2022, \apj, 927, L33
\bibitem[Gaia Collaboration(2023)]{Gaia2023} Gaia Collaboration (Vallenari, A., et al.) 2023, \aap, 674, A1
\bibitem[Gendriesch et al.(2009)]{Gendriesch2009} Gendriesch, R., Lewen, F., Klapper, G., et al. 2009, \aap, 497, 927
\bibitem[Gonz\'alez Delgado et al.(2003)]{Gonzalez-Delgado2003} Gonz\'alez Delgado, D., Olofsson, H., Kerschbaum, F., et al. 2003, \aap, 411, 123
\bibitem[Goorvitch(1994)]{Goorvitch1994} Goorvitch, D. 1994, \apjs, 95, 535
\bibitem[Groenewegen \& Whitelock(1996)]{Groenewegen1996} Groenewegen, M. A. T. \& Whitelock, P. A. 1996, \mnras, 281, 1347
\bibitem[Groenewegen et al.(1998)]{Groenewegen1998} Groenewegen, M. A. T., Whitelock, P. A., Smith, C. H., \& Kerschbaum, F. 1998, \mnras, 293, 18
\bibitem[Groenewegen et al.(2002)]{Groenewegen2002} Groenewegen, M. A. T., Sevenster, M., Spoon, H. W. W., \& P\'erez, I. 2002, \aap, 390, 511
\bibitem[Groenewegen et al.(2012)]{Groenewegen2012} Groenewegen, M. A. T., Barlow, M. J., Blommaert, J. A. D. L., et al. 2012, \aap, 543, L8
\bibitem[Groenewegen et al.(2017)]{Groenewegen2017} Groenewegen, M. A. T. 2017, \aap, 606, A67
\bibitem[Groenewegen \& Saberi(2021)]{Groenewegen2021} Groenewegen, M. A. T. \& Saberi, M. 2021, \aap, 649, A172
\bibitem[Gu\'elin et al.(2018)]{Guelin2018} Gu\'elin, M., Patel, N. A., Bremer, M., et al. 2018, \aap, 610, A4
\bibitem[Heays et al.(2017)]{Heays2017} Heays, A. N., Bosman, A. D., \& van Dishoeck, E. F. 2017, \aap, 602, A105
\bibitem[H\"ofner \& Olofsson(2018)]{Hofner2018} H\"ofner, S. \& Olofsson, H. 2018, \aapr, 26, 1
\bibitem[Hogerheijde \& van der Tak(2000)]{Hogerheijde2000} Hogerheijde, M. R. \& van der Tak, F. F. S. 2000, \aap, 362, 697
\bibitem[Ishihara et al.(2010)]{Ishihara2010} Ishihara, D., Onaka, T., Kataza, H. et al. 2010, \aap, 514, A1
\bibitem[Ivezi\'c \& Elitzur(1997)]{Ivezic1997} Ivezi\'c, $\hat{\rm Z}$. \& Elitzur, M. 1997, \mnras, 287, 799
\bibitem[Justtanont et al.(2012)]{Justtanont2012} Justtanont, K., Khouri, T., Maercker, M., et al. 2012, \aap, 537, A144
\bibitem[Karovicova et al.(2013)]{Karovicova2013} Karovicova, I., Wittkowski, M., Ohnaka, K., et al. 2013, \aap, 560, A75
\bibitem[K{\l}os \& Lique(2008)]{Klos2008} K{\l}os, J. \& Lique, F. 2008, \mnras, 390, 239
\bibitem[Lang(2104)]{Lang2014} Lang, D., 2014, \aj, 147, 108
\bibitem[Lique et al.(2006)]{Lique2006} Lique, F., Spielfiedel, A., \& Cernicharo, J. 2006, \aap, 451, 1125
\bibitem[Lique \& Spielfiedel(2007)]{Lique2007} Lique, F. \& Spielfiedel, A. 2007, \aap, 462, 1179
\bibitem[Lucas et al.(1992)]{Lucas1992} Lucas, R., Bujarrabal, V., Guilloteau, S., et al. 1992, \aap, 262, 491
\bibitem[Mainzer et al.(2011)]{Mainzer2011} Mainzer, A., Bauer, J., Grav, T., et al. 2011, \apj, 731, 53
\bibitem[Massalkhi et al.(2018)]{Massalkhi2018} Massalkhi, S., Ag\'undez, M., Cernicharo, J., et al. 2018, \aap, 611, A29
\bibitem[Massalkhi et al.(2019)]{Massalkhi2019} Massalkhi, S., Ag\'undez, M., \& Cernicharo, J. 2019, \aap, 628, A62
\bibitem[Massalkhi et al.(2020)]{Massalkhi2020} Massalkhi, S., Ag\'undez, M., Cernicharo, J., \& Velilla-Prieto, L. 2020, \aap, 641, A57
\bibitem[Meisner et al.(2016)]{Meisner2016} Meisner, A. M., Lang, D., \& Schlegel, D. J. 2016, arXiv:1603.05664
\bibitem[Meisner et al.(2017)]{Meisner2017} Meisner, A. M., Lang, D., \& Schlegel, D. J. 2017, arXiv:1705.06746
\bibitem[Morris \& Jura(1983)]{Morris1983} Morris, M. \& Jura, M. 1983, \apj, 264, 546
\bibitem[M\"uller et al.(2005)]{Muller2005} M\"uller, H. S. P., Schl\"oder, F., Stutzki, J., \& Winnewiser, G. 2005, \jmst, 742, 215
\bibitem[M\"uller et al.(2007)]{Muller2007} M\"uller, H. S. P., McCarthy, M. C., Bizzocchi, L., et al. 2007, \pccp, 9, 1579
\bibitem[Neri et al.(1998)]{Neri1998} Neri, R., Kahane, C., Lucas, R., et al. 1998, \aaps, 130, 1
\bibitem[Ochsenbein et al.(2000)]{Ochsenbein2000} Ochsenbein, F., Bauer, P., \& Marcout, J. 2000, \aaps, 143, 23
\bibitem[Olivier et al.(2001)]{Olivier2001} Olivier, E. A., Whitelock, P., \& Marang, F. 2001, \mnras, 326, 490
\bibitem[Olofsson et al.(1998)]{Olofsson1998} Olofsson, H., Lindqvist, M., Nyman, L.-\AA., \& Winnberg, A. 1998, \aap, 329, 1059
\bibitem[Olofsson et al.(2002)]{Olofsson2002} Olofsson, H., Gonz\'alez Delgado, D., Kerschbaum, F., \& Sch\"oier, F. L. 2002, \aap, 391, 1053
\bibitem[Pardo et al.(2022)]{Pardo2022} Pardo, J. R., Cernicharo, J., Tercero, B., et al. 2022, \aap, 658, A39
\bibitem[Pattillo et al.(2018)]{Pattillo2018} Pattillo, R. J., Cieszewski, R., Stancil, P. C., et al. 2018, \apj, 858, 10
\bibitem[Perrin et al.(1999)]{Perrin1999} Perrin, G., Coud\'e du Foresto, V., Ridgway, S. T., et al. 1999, \aap, 345, 221
\bibitem[Pety(2005)]{Pety2005} Pety, J. 2005, in SF2A-2005: Semaine de l'Astrophysique Francaise, ed. F. Casoli et al. (Les Ulis: EDP-Sciences), 721
\bibitem[Piñeiro et al.(1987)]{Pineiro1987} Piñeiro, A. L., Tipping, R. H., \& Chackerian, C., 1987, \jms, 125, 91
\bibitem[Price et al.(2001)]{Price2001} Price, S. D., Egan, M. P., Carey, S. J., et al. 2001, \aj, 121, 2819
\bibitem[Price et al.(2010)]{Price2010} Price, S. D. Smith, B. J., Kuchar, T. A., et al. 2010, \apjs, 190, 203
\bibitem[Ramstedt et al.(2008)]{Ramstedt2008} Ramstedt, S., Sch\"oier, F. L., Olofsson, H., \& Lundgren, A. A. 2008, \aap, 487, 645
\bibitem[Ramstedt \& Olofsson(2014)]{Ramstedt2014} Ramstedt, S. \& Olofsson, H. 2014, \aap, 566, A145
\bibitem[Ramstedt et al.(2020)]{Ramstedt2020} Ramstedt, S., Vlemmings, W. H. T., Doan, L., et al. 2020, \aap, 640, A133
\bibitem[Raymonda et al.(1970)]{Raymonda1970} Raymonda, J. W., Muenter, J. S., \& Klemperer, W. A. 1970, \jcp, 52, 3458
\bibitem[Ridgway \& Keady(1988)]{Ridgway1988} Ridgway, S. T. \& Keady, J. J. 1988, \apj, 326, 843
\bibitem[Rothman et al.(2005)]{Rothman2005} Rothman, L. S., Jacquemart, D., Barbe, A., et al. 2005, \jqsrt, 96, 139
\bibitem[Saberi et al.(2019)]{Saberi2019} Saberi, M., Vlemmings, W. H. T., De Beck, E. 2019, \aap, 625, A81
\bibitem[Sahai \& Bieging(1993)]{Sahai1993} Sahai, R. \& Bieging, J. H. 1993, \aj, 105, 595
\bibitem[Sanz et al.(2003)]{Sanz2003} Sanz, M. E., McCarthy, M. C., \& Thaddeus, P. 2003, \jcp, 119, 11715
\bibitem[Sch\"oier \& Olofsson(2001)]{Schoier2001} Sch\"oier, F. L. \& Olofsson, H. 2001, \aap, 368, 969
\bibitem[Sch\"oier et al.(2002)]{Schoier2002} Sch\"oier, F. L., Ryde, N. \& Olofsson, H. 2002, \aap, 391, 577
\bibitem[Sch\"oier et al.(2004)]{Schoier2004} Sch\"oier, F. L., Olofsson, H., Wong, T., et al. 2004, \aap, 422, 651
\bibitem[Sch\"oier et al.(2006a)]{Schoier2006a} Sch\"oier, F. L., Olofsson, H., \& Lundgren, A. A. 2006a, \aap, 454, 247
\bibitem[Sch\"oier et al.(2006b)]{Schoier2006b} Sch\"oier, F. L., Fong, D., Olofsson, H., et al. 2006b, \apj, 649, 965
\bibitem[Sch\"oier et al.(2007)]{Schoier2007} Sch\"oier, F. L., Bast, J., Olofsson, H., \& Lindqvist, M. 2007, \aap, 473, 871
\bibitem[Sch\"oier et al.(2013)]{Schoier2013} Sch\"oier, F. L., Ramstedt, S., Olofsson, H., et al. 2013, \aap, 550, A78
\bibitem[Scicluna et al.(2022)]{Scicluna2022} Scicluna, P., Kemper, F., McDonald, I., et al. 2022, \mnras, 512, 1091
\bibitem[Skrutskie et al.(2006)]{Skrutskie2006} Skrutskie, M. F., Cutri, R. M., Stiening, R., et al. 2006, \aj, 131, 1163
\bibitem[Smith et al.(2004)]{Smith2004} Smith, B. J., Price, S. D. \& Baker, R. I. 2004, \apjs, 154, 673
\bibitem[Stark et al.(1987)]{Stark1987} Stark, G., Yoshino, K., \& Smith, P. L. 1987, \jms, 124, 420
\bibitem[Suh(1999)]{Suh1999} Suh, K.-W. 1999, \mnras, 304, 389
\bibitem[Suh(2000)]{Suh2000} Suh, K.-W. 2000, \mnras, 315, 740
\bibitem[Tercero et al.(2021)]{Tercero2021} Tercero, F., L\'opez-P\'erez, J. A., Gallego, J. D., et al. 2021, \aap, 645, A37
\bibitem[Tobo{\l}a et al.(2008)]{Tobola2008} Tobo{\l}a, R., Lique, F., K{\l}os, J., \& Cha{\l}asi\'nski, G. 2008, \jpb, 41, 155702
\bibitem[Tsuji(1973)]{Tsuji1973} Tsuji, T. 1973, \aap, 23, 411
\bibitem[van Dishoeck(1988)]{vanDishoeck1988} van Dishoeck, E. F. 1988, Astrophys. Space Sci. Lib., 146, 49
\bibitem[van Dishoeck et al.(2006)]{vanDishoeck2006} van Dishoeck, E. F., Jonkheid, B., \& van Hemert, M. C. 2006, \fdis, 133, 231
\bibitem[van Loon et al.(2005)]{vanLoon2005} van Loon, J. Th., Cioni, M.-R. L., Zijlstra, A. A., \& Loup, C. 2005, \aap, 438, 273
\bibitem[Velilla-Prieto et al.(2017)]{Velilla-Prieto2017} Velilla-Prieto, L., S\'anchez-Contreras, C., Cernicharo, J., et al. 2017, \aap, 597, A25
\bibitem[Velilla-Prieto et al.(2019)]{Velilla-Prieto2019} Velilla-Prieto, L., Cernicharo, J., Ag\'undez, M., et al. 2019, \aap, 629, A146
\bibitem[Verbena et al.(2019)]{Verbena2019} Verbena, J. L., Bujarrabal, V., Alcolea, J., et al. 2019, \aap, 624, A107
\bibitem[Whitelock et al.(1994)]{Whitelock1994} Whitelock, P., Menzies, J., Feast, M., et al. 1994, \mnras, 267, 711
\bibitem[Whitelock et al.(2008)]{Whitelock2008} Whitelock, P. A., Feast, M. W., \& van Leeuwen, F. 2008, \mnras, 386, 313
\bibitem[Winnewiser \& Cook(1968)]{Winnewiser1968} Winnewiser, G. \& Cook, R. L. 1968, \jms, 28, 266
\bibitem[Winnewiser et al.(1997)]{Winnewiser1997} Winnewiser, G., Belov, S. P., Klaus, Th., \& Schieder, R. 1997, \jms, 184, 468
\bibitem[Wirsich(1994)]{Wirsich1994} Wirsich, J. 1994, \apj, 424, 370
\bibitem[Woods et al.(2003)]{Woods2003} Woods, P. M., Sch\"oier, F. L., Nyman, L.-\AA., \& Olofsson, H. 2003, \aap, 402, 617
\bibitem[Wright et al.(2010)]{Wright2010} Wright, E. L., Eisenhardt, P. R. M., Mainzer, A., et al. 2010, \aj, 140, 1868
\bibitem[Xu et al.(2019)]{Xu2019} Xu, Z., Luo, N., Federman, S. R., et al. 2019, \apj, 882, 86
\bibitem[Yang et al.(2010)]{Yang2010} Yang, B., Stancil, P. C., Balakrishnan, N., \& Forrey, R. C. 2010, \apj, 718, 1062

\end{thebibliography}
\end{document}